\documentclass[a4paper,11pt]{article}
\usepackage{jheppub} 
\pdfoutput=1
\usepackage{graphicx,array}
\usepackage[dvipsnames]{xcolor}
\usepackage{amstext}
\usepackage{amssymb}
\usepackage{slashed}
\usepackage{cancel}
\usepackage{subcaption}
\usepackage{booktabs}

\usepackage{pifont}

\definecolor{orange}{rgb}{1,0.5,0}
\definecolor{brown}{rgb}{0.59, 0.29, 0.0}
\definecolor{note_fontcolor}{rgb}{0.80078125, 0.80078125, 0.80078125}

\newcommand{\order}[1]{ \mathcal{O}(#1)}

\newcommand{\Luv}{\Lambda_\text{UV}}

\newcommand{\Lir}{\Lambda_\text{IR}}

\newcommand{\LDC}{\Lambda_{DC}}

\newcommand{\SM}{\text{SM}}

\newcommand{\kappaO}{\kappa_{\mathcal{O}}}
\newcommand{\DeltaO}{\Delta_{\mathcal{O}}}
\newcommand{\cO}{c_{\mathcal O}}
\newcommand{\MPl}{M_\text{Pl}}

\newcommand{\n}{\langle n\rangle}
\newcommand{\g}{\langle \gamma \rangle}
\newcommand{\E}{\langle E \rangle}

\usepackage{tikz}
\usetikzlibrary{arrows,shapes.misc,decorations.markings,decorations.pathmorphing,positioning,intersections}

\usepackage{feynmp}
\usepackage{graphicx}
\usepackage{slashed}
\usepackage{tikz-feynman,contour}
\usepackage{tikz}
\usetikzlibrary{arrows,shapes.misc,decorations.markings,decorations.pathmorphing,positioning,intersections,patterns}

\tikzset{scalar/.style={dashed, > = latex, thick,
        decoration={markings,
            mark= at position 0.5 with {\arrow{#1}} ,
        },
        postaction={decorate}
    	},
    	gaugeboson/.style={decorate, decoration={snake},thick, segment length=3.2mm},
    	cross/.style={cross out, draw, 
         minimum size=2*(#1-\pgflinewidth), 
         inner sep=0pt, outer sep=0pt},
         gluon/.style={decorate, draw=black,
    decoration={coil,amplitude=4pt, segment length=5pt}}
}

\tikzfeynmanset{
  eftvert/.style={
    /tikzfeynman/dot,
    /tikz/minimum size=7pt,
  },
	every unparticle/.style={> = latex, double distance=2pt, decoration={markings,mark= at position 0.5 with {\arrow{#1}},},postaction={decorate}},
}

\tikzset{
	scalar/.style={dashed, > = latex, thick,decoration={markings,mark= at position 0.5 with {\arrow{#1}},},postaction={decorate}},
	gaugeboson/.style={decorate, decoration={snake},thick, segment length=3.2mm},
    fermion/.style={> = latex, thick,decoration={markings,mark= at position 0.5 with {\arrow{#1}},},postaction={decorate}},
    unparticle/.style={> = latex, double distance=2pt, thick,decoration={markings,mark= at position 0.5 with {\arrow{#1}},},postaction={decorate}}
}

\usepackage{subcaption}

\usepackage[bottom]{footmisc} 

\title{Searching for Elusive Dark Sectors with Terrestrial and Celestial Observations}

\author[a,b]{Roberto Contino}
\author[a,b]{, Kevin Max}
\author[a,b,c]{, and Rashmish K.~Mishra}

\affiliation[a]{Scuola Normale Superiore, Piazza dei Cavalieri 7, 56126 Pisa, Italy}
\affiliation[b]{INFN Sezione di Pisa, Largo Bruno Pontecorvo 3, 56127 Pisa, Italy}
\affiliation[c]{Harvard University, 17 Oxford Street, Cambridge, MA, 02139, USA}

\emailAdd{roberto.contino@sns.it}
\emailAdd{kevin.max@sns.it} 
\emailAdd{rashmishmishra@fas.harvard.edu} 

\abstract{%
We consider the possible existence of a SM-neutral and light dark sector coupled to the visible sector through irrelevant portal interactions. Scenarios of this kind  are motivated by dark matter and arise in various extensions of the Standard Model. We characterize the dark dynamics in terms of one ultraviolet scale $\Luv$, at which the exchange of heavy mediator fields generates the portal operators, and by one infrared scale~$\Lir$, setting the mass gap. At energies $\Lir \ll E \ll \Luv$ the dark sector behaves like a conformal field theory and its phenomenology can be studied model independently. We derive the constraints set on this scenario by high- and low-energy laboratory experiments and by astrophysical observations. Our results are conservative and serve as a minimum requirement that must be fulfilled by the broad class of models satisfying our assumptions, of which we give several examples. The experimental constraints are derived in a manner consistent with the validity of the effective field theory used to define the portal interactions. We find that high-energy colliders give the strongest bounds and exclude UV scales up to a few TeV, but only in specific ranges of the IR scale. The picture emerging from current searches can be taken as a starting point to design a future experimental strategy with broader sensitivity.}

\begin{document} 
\maketitle
\flushbottom

\section{Introduction and Motivations}
\label{sec:introduction}

New physics beyond the Standard Model (SM) of particle physics is well motivated from several considerations, one of the most appealing being the requirement of a Dark Matter (DM) candidate. 
It is possible that some of the new fields reside in a light and neutral `dark' sector, coupled to the SM only through portal interactions formed by the product of one SM and one dark singlet operator.
Scenarios of this kind are predicted in various extensions of the SM and have been intensively studied under the assumption that the portal operators have dimension 4 or less, see for example~\cite{Essig:2013lka,Alekhin:2015byh,Alexander:2016aln,Beacham:2019nyx} and references therein.
In this work we analyze the more \textit{elusive} dark sectors where the portal operators are higher-dimensional and are generated at some ultraviolet (UV) scale~$\Luv$ by heavy mediator fields. The DM candidate might reside in the dark sector (DS) or be part of the UV dynamics. Given the constraints on new dynamics charged under the SM, set by current and past experiments, we assume that the UV scale is larger than the electroweak scale, $\Luv \gtrsim 100\,\text{GeV}$, although some of our results apply to theories with a lower UV scale as well when allowed. The portal interactions can thus be written in terms of $SU(3)_c \times SU(2)_L \times U(1)_Y$ invariant operators.

We will adopt a broad characterization of the dark dynamics in terms of one infrared (IR) scale, $\Lir$, setting its mass gap. We assume, for simplicity, that no other parametrically different scale exists in the theory. At energies between $\Luv$ and $\Lir$ the new dynamics is approximately conformal and flows slowly (i.e.~logarithmically) in the vicinity of a fixed point of its renormalization group. 
The fixed point can be free (if the dark dynamics is asymptotically free), weakly or strongly coupled. When probed at energies $\Lir \ll E \ll \Luv$ the dark dynamics can be thus described as a conformal field theory (CFT) in terms of its composite operators. Having a sufficiently large hierarchy $\Lir \ll \Luv$ is the working hypothesis of our analysis. Notice that it is also a prerequisite to explain the stability of the DM candidate as accidental, if the DM is part of the DS.

It is important at this point to ask what is the minimal structure that must be present in the dark sector. At energies well above $\Lir$, this corresponds to identifying the set of lowest-dimensional gauge-invariant operators which define the CFT. Clearly, the CFT must at least contain some relevant deformation 
\begin{align}
{\cal L}_\text{def} = c_{\cal O}\, \frac{\mathcal{O}}{\Luv^{\DeltaO-4}} 
\label{eq:cft-deformation}
\end{align}
to break the conformal invariance in the IR and generate the hierarchy between $\Lir$ and $\Luv$. A {\it natural} hierarchy, as we will assume in the following, implies that, in absence of a symmetry protection, the operator $\mathcal{O}$ must be slightly relevant, i.e. must have a scaling dimension $\Delta_\mathcal{O} = 4 -\epsilon$ with $\epsilon \ll 1$. Alternatively, one can also have $\Delta_\mathcal{O} \lesssim 4$ if the coefficient $c_{\cal O}$ is the (only) spurion of a global symmetry and has a value $c_{\cal O} \simeq (\Lir/\Luv)^{4-\Delta_\mathcal{O}}$ at $\Luv$. Clearly, no scalar singlet operators with dimension much smaller than 4 can exist in a natural dark sector since they would destabilize the hierarchy. 

Our analysis will be restricted, for simplicity, to dark sectors that are \textit{unitary} and \textit{local} CFTs.~\footnote{The flow near complex, non-unitary CFTs has been conjectured in Ref.~\cite{Gorbenko:2018ncu} to correspond to the Walking Technicolor regime, see also Ref.~\cite{Kaplan:2009kr}. It would be interesting to investigate how our analysis gets modified when the theory flows near one such complex CFT.} This implies that there must necessarily exist also a local stress-energy tensor operator, $T_{\mu\nu}^{DS}$, with scaling dimension equal to 4. Furthermore, if the dark sector has additional global symmetries, the list of CFT operators will include the corresponding conserved currents, $J^{DS}_\mu$, with dimension equal to 3. The CFT spectrum may contain other relevant operators, depending on the specific underlying dark dynamics. Their presence, however, is not a robust feature implied by our general assumptions or by symmetry arguments.

Any of the above CFT operators can appear in a portal interaction multiplied by one SM gauge-singlet operator. The lowest-lying SM operators are listed in Table~\ref{tab:SMoperators}. The first three have dimension smaller than 4 and can give rise to the well-studied marginal or relevant portals. The others necessarily appear in irrelevant portals.
\begin{table}
\begin{center}
\begin{tabular}{cc}
Operator & Dimension \\
\hline
$H^\dagger H$ & 2 \\[0.05cm]
$B_{\mu\nu}$ & 2 \\[0.05cm]
$\ell H$ & 5/2 \\[0.05cm]
$J_{\mu}^{SM} = \bar \psi \gamma^\mu \psi, \, H^\dagger i \!\overleftrightarrow{D}_{\!\!\mu} H$ & 3 \\[0.1cm]
$O_{\mu\nu}^{\SM} = F_{\mu\alpha}^i F_\nu^{\alpha\, i}, \, D_\mu H^\dagger D_\nu H, \, \bar\psi \gamma_\mu D_\nu \psi$ & 4 \\[0.1cm]
$O_{SM} = \bar\psi i\! \not\!\! D \psi, \, D_\mu H^\dagger D^\mu H, \, F_{\mu\nu} F^{\mu\nu}, \, F_{\mu\nu} \tilde F^{\mu\nu}, \, \bar \psi_L H \psi_R, \, (H^\dagger H)^2$ & 4
\end{tabular}
\caption{\small List of the SM gauge-singlet operators with (classical) dimension equal or smaller than~4. Here $\psi$ and $F_{\mu\nu}$ stand respectively for any SM fermion and any SM gauge field strength.}
\label{tab:SMoperators}
\end{center}
\end{table}
We will focus on the portals that can be constructed with the CFT operators ${\cal O}, J_\mu^{DS}, T_{\mu\nu}^{DS}$ and those of Table~\ref{tab:SMoperators}. These are:~\footnote{The portals $J_\mu^{DS} \partial_\nu B_{\mu\nu}$ and $\partial_\mu {\cal O} J_{SM}^\mu$ can be rewritten in terms of respectively $J_\mu^{DS} J^{\mu}_{SM}$ and $ {\cal O} O_{SM}$ by using the SM equations of motion.}
\begin{equation}
{\cal O} H^\dagger H, \quad  {\cal O} O_{SM}, \quad  J_\mu^{DS} J^{\mu}_{SM}, \quad T_{\mu\nu}^{DS} O^{\mu\nu}_{SM}\, .
\end{equation}
Dimensional analysis suggests that the portal ${\cal O} O_{SM}$ is less important than the Higgs portal ${\cal O} H^\dagger H$. One can consider UV theories where ${\cal O} H^\dagger H$ is generated with a suppressed coefficient, though notice that, in general, ${\cal O} H^\dagger H$ is radiatively induced from ${\cal O} O_{SM}$ at the 1-loop level, so the relative suppression cannot be smaller than a SM loop factor. This might be enough for ${\cal O} H^\dagger H$ to still give the leading effects.
An important exception is when ${\cal O}$ is an axion field with an associated Peccei-Quinn shift symmetry and $O_{SM} = G_{\mu\nu} \tilde G^{\mu\nu}$. In the case of the QCD axion, neither ${\cal O} H^\dagger H$ nor any potential for ${\cal O}$ is generated above the QCD scale. A hierarchy $\Lir\sim \Lambda_{QCD}^2/\Luv$ is instead generated by ${\cal O}^2$ after QCD confinement. Depending on the UV dynamics, additional portals of the type ${\cal O} O_{SM}$ can be present, with $O_{SM} = F_{\mu\nu} \tilde F^{\mu\nu}$ or $\bar \psi_L H \psi_R$. Apart from the special and thoroughly studied axion case, the portal ${\cal O} O_{SM}$ usually plays a subleading role compared to ${\cal O} H^\dagger H$.~\footnote{One exception arises if ${\cal O}$ can singly excite a CP-odd resonance, whose decay will proceed through the ${\cal O}  F_{\mu\nu} \tilde F^{\mu\nu}$ portal and not through ${\cal O} H^\dagger H$. This is the case of CP-odd glueballs in a pure-YM dark sector; we thank Alessandro Podo for pointing this out. Notice that if, as in the previous example, ${\cal O}$ has dimension~4, then the constraints on ${\cal O} O_{SM}$ are expected to be similar to those on $T_{\mu\nu}^{DS} O^{\mu\nu}_{SM}$ discussed in this work.}  We will neglect it in the following and focus on the remaining three portals.

Notice that, while portals involving the Higgs boson, the $Z$ or the top quark require values of $\Luv$ larger than the EW scale to be consistently defined, those featuring only light quarks and leptons can in principle be generated at much smaller scales provided the UV mediators do not have $O(1)$ SM charges and elude current experimental searches. This implies that some of the bounds we will derive are of interest even though they probe values of $\Luv$ well  below the EW scale.

We define our portal Lagrangian between the dark and SM sectors schematically as:
\begin{equation}
\label{eq:portal}
{\cal L}_{\text{portal}} = \frac{\kappa_{\mathcal O}}{\Luv^{\Delta_\mathcal{O} -2}} \mathcal{O}\, H^\dagger H + \frac{\kappa_{J}}{\Luv^2} J_\mu^{DS} J^{\mu}_{SM} + \frac{\kappa_T}{\Luv^4} T^{\mu\nu}_{DS} O_{\mu\nu}^{SM} \, ,
\end{equation}
where $\kappa_{\mathcal O}$, $\kappa_{J}$ and $\kappa_{T}$ are dimensionless coefficients. Our notation here is schematic since, as discussed later, different couplings may be introduced for different SM operators $O^{SM}_{\mu\nu}$ and $J^{\mu}_{SM}$.

The coefficient $\kappa_{\mathcal O}$ cannot be too large otherwise the hierarchy would be destabilized. Indeed, by contracting the two Higgs fields in a loop, the Higgs portal in Eq.~(\ref{eq:portal}) induces a radiative UV correction to the relevant deformation ${\cal L}_\text{def}$ in Eq.~\eqref{eq:cft-deformation}. The hierarchy does not get destabilized provided that
\begin{equation}
\label{eq:stabilitybound}
\kappa_{\mathcal O}\lesssim 16\pi^2 \left(\frac{\Lir}{\Luv}\right)^{4-\Delta_{\mathcal O}} \qquad \text{(UV threshold)}\, .
\end{equation}
An additional contribution to ${\cal L}_\text{def}$ in Eq.~\eqref{eq:cft-deformation} is generated at the electroweak scale, i.e.~when $H$ acquires a vev $v$; this leads to the condition
\begin{equation}
\label{eq:EWstabilitybound}
\kappa_{\mathcal O}\lesssim \frac{\Lir^2}{v^2} \left(\frac{\Lir}{\Luv}\right)^{2-\Delta_{\mathcal O}} \qquad \text{(EW contribution)}\, .
\end{equation}
In most of the parameter space (i.e.~for $\Luv > 4\pi v$), this constraint is weaker than that of Eq.~(\ref{eq:stabilitybound}), although the latter may be avoided if some UV mechanism is at work which tunes $c_{\mathcal O}$ to be small at $\Luv$.
Similar considerations apply to the coefficient of ${\cal O} O_{SM}$, which is subject to a bound analog to Eq.~(\ref{eq:stabilitybound}). Furthermore, if $O_{SM} = \bar q_L H q_R$, the portal ${\cal O} O_{SM}$ gives an additional contribution to ${\cal L}_\text{def}$ at the QCD scale from the quark condensate. One can also envisage a scenario, as done in Ref.~\cite{Hong:2019nwd}, where $\kappa_{\mathcal O}$ (or the coefficient of ${\cal O} O_{SM}$) saturates its upper bound, and the hierarchy is generated by the portal interactions themselves.~\footnote{Ref.~\cite{Hong:2019nwd} studied the cosmology of dark sectors where the hierarchy is generated by ${\cal O} H^\dagger H$ or ${\cal O} \bar q_L H q_R$. The 1-loop UV corrections to~$\Delta{\cal L}$ was neglected.}

In this paper we focus on elusive dark sectors that feature the portals of Eq.~(\ref{eq:portal}). These are minimal scenarios as, in general, additional portals may be present. We derive general constraints on these theories from laboratory experiments and astrophysical data by making use only of the general features of the dark dynamics, without relying on its specific details. More explicitly, our analysis will exploit the high-energy conformal regime and the fact that the lightest dark state has mass of order $\Lir$ (as implied by the absence of other infrared scales in the dark sector).
Our results will be conservative and can be improved if a full theory is defined explicitly. Indeed, knowing the IR behavior of the dark dynamics allows one to perform complete rather than just approximate calculations of rates and cross sections, and thus to derive stronger constraints.
Furthermore, as discussed in section~\ref{sec:Strategy}, effective operators generated by the exchange of UV degrees of freedom and made of SM fields alone can lead to constraints on $\Luv$ that are stronger than those obtained from our analysis (but are opaque about the details of the underlying DS). These effects have been thoroughly studied in the literature and several systematic analyses have been performed. In this work we will provide a conservative characterization of these constraints by estimating the smallest value of the effective coefficients compatible with the existence of our portal interactions. 

Our approach is not entirely new and in fact has some overlap with previous studies on Hidden Valleys and on the phenomenology of conformal field theories. The scenarios that are referred to as Hidden Valleys are similar to those we consider in this study: new confining dynamics with low mass scale is assumed to couple to the SM through some irrelevant portal, generated for example by heavy mediators~\cite{Strassler:2006im}. This possibility was envisaged before the beginning of the LHC operation, pointing out that the energy increase provided by the LHC could have been enough to climb over the barrier separating us from the Hidden Valley if the mediators have mass of order a few TeV. In that case, the mediators can be produced on shell and decay copiously to the hidden hadrons with spectacular experimental signatures. The LHC data collected at Run1 and Run2 have discovered no new particles and suggest that, if realized at all in nature, these scenarios must be hidden from us through a higher barrier. In this work we thus assume that the mediators are sufficiently heavy to be out of the direct reach of the LHC, and ask if we can test the existence of the dark sector, i.e.~the hidden sector with low mass scale. Hence, while the theories studied in this paper have a large overlap with Hidden Valleys (though, notice, we do not assume the dark sector to be necessarily strongly coupled and confining), our approach and assumptions are different. 

On the front of the phenomenology of conformal field theories, there is a vast literature on `unparticle' physics where similar experimental data were used to set constraints on the theoretical parameter space. The question originally motivating the study of unparticles is whether new dynamics can first manifest itself and be discovered at colliders in its conformal regime~\cite{Georgi:2007ek}.  We differ from those works for the choice of the portals in Eq.~(\ref{eq:portal}), our thorough inclusion of experimental bounds, and for our self-consistent use of effective field theory techniques. Furthermore, while unparticle studies assume that the CFT degrees of freedom are stable on distances relevant for the analysis, we have also considered the constraints that arise when these CFT excitations decay inside the detector with displaced vertices. 

Previous studies of the phenomenology of dark sectors coupled to the SM through irrelevant portals include Refs.~\cite{Juknevich:2009ji,Juknevich:2009gg,Batell:2009di,Falkowski:2009yz,Kanemura:2010sh,Djouadi:2011aa,Djouadi:2012zc,Greljo:2013wja,Fedderke:2014wda,Freitas:2015hsa,Fedderke:2015txa,Katz:2015zba,Fichet:2017bng,Brax:2017xho,Costantino:2019ixl,Cheng:2019yai,Darme:2020ral,Banks:2020gpu}. While these papers have some aspects in common with our work and some of their assumptions are similar to ours, we believe that our approach is original and our analysis extends previous results. We will focus on laboratory experiments and astrophysical observations that can test and set limits on elusive dark sectors. An additional important probe comes from cosmology, and a study in this direction has been performed in Ref.~\cite{Hong:2019nwd}. 

The outline of the paper is as follows. In section~\ref{sec:Examples} we illustrate some examples of elusive dark sectors, exhibiting their UV completion. Section~\ref{sec:Strategy} explains our strategy and estimates the effects from DS virtual effects and DS production. Three possible experimental manifestations of the DS excitations, in the form of missing energy, displaced decays and prompt decays, are discussed, and the validity of the effective field theory is analyzed. The bounds from terrestrial experiments and celestial observations are derived in section~\ref{sec:Bounds}. We analyze: resonant and non-resonant DS production at high-energy colliders; high-intensity experiments; stellar evolution and supernova energy loss; positronium decays; fifth-force experiments; and electroweak precision tests. We draw our summary and conclusions in section~\ref{sec:Conclusion}. The appendix includes useful formulas on two-point correlators (\ref{app:2pt}), additional details on a 5D Randall-Sundrum dark sector (\ref{app:RSmodel}), and formulas for the probabilities used to compute the rate of displaced decays (\ref{sec:probabilities}).

\section{Examples of Elusive Dark Sectors}
\label{sec:Examples}

Although our analysis will be model independent and will not make reference to the underlying dark dynamics, it is useful to discuss a few specific models that can serve as benchmark examples. In this section we will thus consider four different kinds of dark sectors and specify the mediator fields that generate their portal interactions.

\subsection{Pure Yang-Mills dark sector}
\label{sec:YM}

One of the simplest and most motivated example of dark sectors is pure Yang-Mills (YM) dynamics. Models of this kind have been considered in the context of glueball DM~\cite{Boddy:2014yra}, and can arise as the low-energy limit of theories of accidental DM with dark fermions heavier than the dynamical scale~\cite{Mitridate:2017oky}. Their mass gap is generated dynamically at dark confinement and the lightest states in the spectrum are the dark glueballs. Consider as an example the $L\oplus N$ model of Ref.~\cite{Mitridate:2017oky}, defined in terms of one Dirac fermion $L$ and one Majorana fermion $N$ transforming as fundamental representations of an $SO(N_{DC})$ dark color group. Under the SM gauge symmetry, $N$ is a singlet while $L$ transforms as a $2_{-1/2}$ of $SU(2)_{EW} \times U(1)_Y$.
The Lagrangian (in 4-component notation) is:
\begin{equation}
\label{eq:LNmodel}
\begin{split}
\Delta {\cal L} = & -\frac{1}{4g_{DC}^2} {\cal G}_{\mu\nu}{\cal G}^{\mu\nu} + \bar L (i \!\!\not\!\! D - m_L) L +  \frac{1}{2} \bar N (i \!\!\not\!\! D - m_N) N \\[0.1cm]
& - \left( y_L \,\bar N P_L L H + y_R \, \bar N P_R L H + h.c.\right),
\end{split}
\end{equation}
where ${\cal G}$ is the dark gluon field and $P_{L,R}$ are left and right projectors. 
The theory has an accidental dark baryon parity that makes the lightest baryon cosmologically stable and a potential DM candidate~\cite{Mitridate:2017oky}. If both $m_L$ and $m_N$ are larger than the dark dynamical scale $\LDC$, then the low-energy dark sector consists of a pure YM dynamics, while the DM candidate resides in the UV sector.~\footnote{For example, if $m_L > m_N > \LDC$ then the lightest dark baryon, i.e.~the DM candidate, is a bound state of~$N$ with spin $N_{DC}/2$ and mass $\sim N_{DC} m_N$~\cite{Mitridate:2017oky}.} Integrating out the heavy fermions at 1-loop generates the dim-6 and dim-8 operators
\begin{align}
 & {\cal G}_{\mu\nu}{\cal G}^{\mu\nu} H^\dagger H && \kappaO \sim \frac{\alpha_{DC}(\Luv)}{4\pi} (|y_L|^2 + |y_R|^2) \\[0.2cm]
\label{eq:d6YMportal}
& {\cal G}_{\mu\nu}{\cal G}^{\mu\nu} W_{\alpha\beta} W^{\alpha\beta} , \, {\cal G}_{\mu\alpha}{\cal G}^{\alpha}_\nu W^\mu_{\beta} W^{\beta\nu} &&
\kappa_T \sim \alpha_{DC}(\Luv) \alpha_2(\Luv)
\end{align}
where $\Luv\sim m_L, m_N$.
There are two kinds of light states in this model: CP-odd and CP-even glueballs. While the latter can decay through the dim-6 portal, CP-odd glueballs can only decay through the dim-8 one and their lifetime is longer. 

As another example of a theory that leads to a pure YM dark sector, consider an $SU(N_{DC})$ theory with massive fermions $\psi$ transforming as the adjoint representation of dark color and as a $3_0$ of $SU(2)_{EW} \times U(1)_Y$~\cite{Contino:2018crt}. Since $\psi$ does not have Yukawa couplings to the Higgs, integrating it out does not lead to any dim-6 operator at 1-loop. Therefore, this theory has only the dim-8 portal of Eq.~(\ref{eq:d6YMportal}). The DM candidate in this case is the gluequark, a bound state made of one dark quark and dark glue. It is cosmologically stable due to an accidental dark parity, has mass of order $m_\psi \gg \Lambda_{DC}$ and thus resides in the UV sector.

\subsection{Strongly coupled dark sector}
\label{sec:SCDS}

Another interesting limit of the theory defined by Eq.~(\ref{eq:LNmodel}) is when the doublet is heavy, $m_L \gg \Lambda_{DC}$, while the singlet is light with mass of order of the dynamical scale, $m_N \lesssim \Lambda_{DC}$. In this case the dark sector is a strongly coupled $SO(N_{DC})$ theory with one Majorana fermion in the fundamental representation. The spectrum of lowest-lying states contains dark baryons (the lightest of which is accidentally stable and thus a DM candidate) and mesons. Integrating out the heavy doublet at tree level generates dim-5 and dim-6 portals ($\Luv \sim m_L$):
\begin{align}
\label{eq:SCDS-Hportal}
& \bar N P_L N H^\dagger H + h.c.  && \kappaO \sim y_L y_R^* \\[0.2cm]
\label{eq:SCDS-Zportal}
& \bar N \gamma^\mu \gamma^5 N  H^\dagger i \!\overleftrightarrow{D}_{\!\!\mu} H && \kappa_J \sim (|y_L|^2 - |y_R|^2) \, .
\end{align}
The dark current appearing in the dim-6 portal is purely axial, as a consequence of $N$ being a Majorana fermion. Equation~(\ref{eq:SCDS-Zportal}) thus gives $N$ an axial coupling to the $Z$ boson. A similar model with $SU(N_{DC})$ dark color group and a vectorlike (complex) representation for $N$ would give an additional portal with a vectorial current, hence a vectorial coupling to the $Z$. Such vectorial coupling is strongly constrained by direct detection experiments if dark baryons made of $N$ are the DM (see for example Ref.~\cite{Mitridate:2017oky}). In the model of Eq.~(\ref{eq:LNmodel}), the scattering of DM off nuclei via $Z$ exchange has a spin-dependent cross section, as a consequence of the axial coupling. The corresponding bounds are weaker, though not negligible (see~\cite{Aprile:2019dbj}). The strongest constraint holds for DM masses in the range $10-100\,$GeV and requires $m_L$ to be larger than a few TeV for Yukawas of order 1. For lower DM masses, the bound becomes much weaker and sizable Yukawas are allowed for $m_L$ above the weak scale. The DM can scatter also via a Higgs exchange, with a spin-independent cross section. The corresponding bounds are slightly stronger than those from the $Z$ exchange, but also disappear for DM masses smaller than $\sim 10\,$GeV (see~\cite{Aprile:2018dbl}). They can be evaded for any value of the DM mass if one of the Yukawa couplings vanishes or is very small. This would still allow for a large $\kappa_J$ in Eq.~(\ref{eq:SCDS-Zportal}).

\subsection{Dark sector with free fermions}
\label{sec:freefermion}

Another interesting example of dark sector is a theory of free fermions. As a first UV completion, consider a theory where $(B-L)$ is gauged by $X_\mu$ and spontaneously broken at high scale by a scalar field $\phi$ with $(B-L)$ charge $-2$. To make $(B-L)$ anomaly free we introduce three left-handed neutrinos $N_i$ with $(B-L)$ charge $-1$. We impose a $Z_2$ symmetry under which the $N_i$ are odd in order to forbid their Yukawa couplings to the Higgs field and make them stable. In two-component notation, the Lagrangian for the new fields reads
\begin{equation}
\begin{split}
\Delta {\cal L} = & -\frac{1}{4g_X^2} X_{\mu\nu} X^{\mu\nu} + \sum_{i=1}^3 N_i^\dagger i(\partial_\mu - i X_\mu) \bar\sigma^\mu N_i + |D_\mu \phi|^2 \\
&+ \sum_{\psi_{SM}} q_{B-L}^{[\psi_{SM}]} \psi_{SM}^\dagger X_\mu \bar \sigma^\mu \psi_{SM}  - \sum_{i=1}^3 (y_i N_i N_i \phi + h.c.)
- \lambda_\phi (\phi^\dagger \phi - v_\phi^2)^2\, ,
\end{split}
\end{equation}
where $\psi_{SM}$ are the SM fields and $q_{B-L}^{[\psi_{SM}]}$ is their charge under $(B-L)$. When \mbox{$(B-L)$} gets spontaneously broken, all new fields acquire mass ($m_{N_i} = y_i v_\phi$, $m_\phi = 4\sqrt{\lambda_\phi}v_\phi$, $m_X=2\sqrt{2} g_X v_\phi$). We assume that the $N_i$ are much lighter than $X_\mu$ and $\phi$, and thus take $y_i \ll g_X, \sqrt{\lambda_\phi}$. Integrating out $X_\mu$ at tree level generates the dim-6 portal ($\Luv\sim m_X$)
\begin{equation}
\label{eq:JJportalforFF}
\bar\psi_{SM} \gamma_\mu \psi_{SM} \sum_i \bar\psi_{N_i}^\dagger \gamma^\mu \gamma^5\psi_{N_i} \qquad\quad \kappa_J \sim q_{B-L}^{[\psi_{SM}]}  g_X^2\, ,
\end{equation}
where $\psi_{N_i}$ are Majorana fermions in 4-component notation. Searches performed at the LHC for a $Z'$ decaying into leptons and jets set rather stringent lower bounds on the mass of the mediator $X_\mu$, of order $1-5\,$TeV for $O(1)$ couplings $g_X$~\cite{CMS-PAS-EXO-19-019,Sirunyan:2019vgj,Aad:2019fac,Aad:2019hjw}.

As another example, consider a theory with one SM-neutral Majorana fermion $\chi$ and one scalar $\phi$ with hypercharge $-1$. If $\chi$ and $\phi$ are odd under an exact dark parity, the Lagrangian is
\begin{equation}
\Delta {\cal L} = \left(D_\mu \phi\right)^\dagger \!\left(D^\mu \phi\right) + \frac{1}{2} \bar\chi (i \!\!\not\!\partial - m_\chi) \chi + \left( y \,\bar e_R \phi \chi + h.c. \right) - m_\phi^2 \phi^\dagger \phi - \lambda_\phi (\phi^\dagger\phi)^2\, .
\end{equation}
We take $m_\phi \gg m_\chi$, so that integrating out $\phi$ at tree level generates the dim-6 portal
\begin{equation}
\label{eq:JJportalforFF2}
\bar e_R \gamma^\mu e_R \, \bar\chi \gamma_\mu \gamma^5\chi\qquad\quad \kappa_J \sim y^2\, . 
\end{equation}
Thanks to dark parity, $\chi$ is absolutely stable, while $\phi$ decays to $e_R \bar\chi$ through its Yukawa coupling. This theory is similar to a simplified supersymmetric model with neutralino and selectron, where $\chi$ plays the role of the neutralino and $\phi$ of the selectron. This suggests that searches for supersymmetry at LEP can set limits on the mass of the mediator $\phi$, in particular those looking for slepton pair production followed by the decay to electron plus neutralino (see~\cite{LEPSUSYWG/04-01.1} and references therein). The lower bound on $m_\phi$ is expected to be of order $100\,\text{GeV}$ or smaller, depending on the mass of $\chi$.

A final example of UV completion is a theory with a single Dirac fermion $\psi$ coupled to a real scalar field $S$, both neutral under the SM gauge group.~\footnote{See Refs.~\cite{Kim:2008pp,Baek:2011aa,LopezHonorez:2012kv,Fedderke:2015txa} for similar, though different, models.} The Lagrangian is assumed to be invariant under a chiral parity $\psi \to \gamma^5 \psi$, $S\to -S$, and it reads
\begin{equation}
\label{eq:Lfreefermion3}
{\cal L} = \bar \psi i\! \!\not\!\partial \psi +\frac{1}{2} (\partial_\mu S)^2 - y\, \bar\psi \psi S - \lambda_S (S^2 - v_S^2)^2 - \lambda_{SH} \, S^2 H^\dagger H\, .
\end{equation}
The scalar potential gives $S$ a vev and breaks the chiral parity spontaneously. Assuming $m_S = 4\sqrt{\lambda_S}v_S \gg m_\psi = y v_S$ implies at low energy a dark sector with one free Dirac fermion. Integrating out $S$ at tree level generates a dim-5 Higgs portal
\begin{equation}
\label{eq:JJportalforFF3}
\bar \psi\psi H^\dagger H \qquad\quad \kappaO \sim \frac{\lambda_{SH} y v_s}{m_S} = \lambda_{SH} \frac{m_\psi}{m_S}\, ,
\end{equation}
as well as the operator $O_H = [\partial_\mu (H^\dagger H)]^2$ with coefficient $c_H \sim \lambda_{SH}^2 v_s^2/m_S^4$.
The value of $\kappaO$ satisfies the naturalness bounds (\ref{eq:stabilitybound}),(\ref{eq:EWstabilitybound}) as long as $\lambda_{SH} < \text{min}(16\pi^2, m_S^2/v^2)$. Differently from the strongly-coupled dark sector discussed above, in this theory there is only one spurion (i.e.~$m_\psi$) breaking the chiral parity, and $\kappaO$ automatically bears a suppressing factor $m_\psi/m_S \sim \Lir/\Luv$.
The operator $O_H$ implies a universal shift in the Higgs couplings of order $\delta g/g \sim (\lambda_{SH}/\lambda_S)^2 (v/v_S)^2$, which can be sufficiently small if $\lambda_{SH} \ll \lambda_S$ and/or $v\ll v_S$. Notice that the parity transformation $\psi \to -\psi$ also leaves Eq.~(\ref{eq:Lfreefermion3}) invariant and is not broken spontaneously; as a consequence, $\psi$ is absolutely stable.

In all the models discussed in this section, the fermions in the dark sector are stable and can be considered as potential DM candidates.

\subsection{5D Randall-Sundrum Dark sector}
\label{sec:5DRS}

Finally, let us discuss a 5-dimensional example of dark sector that is dual to a strongly-coupled 4-dimensional theory. Consider a Randall-Sundrum theory~\cite{Randall:1999ee}  where the full SM sector is localized on the UV brane and the only fields propagating in the bulk and on the IR brane are gravity and the fields required to stabilize the extra dimension, such as a Goldberger-Wise scalar~\cite{Goldberger:1999uk} or a gauge field~\cite{Garriga:2002vf}.
We add the following boundary action on the UV brane:
\begin{equation}
\label{eq:UVbraneaction}
\int d^4 x \sqrt{-g} \left( M_0^2 R + \frac{1}{\Luv^2} R_{\mu 5\nu 5} T^{\mu\nu}_{SM} \right)\, .
\end{equation}
The first term, with $M_0 \sim \MPl$, sets the strength of the gravitational interaction at low energy, so that one can assume $\Luv \sim M_5, k \ll \MPl$, where $M_5$ and $k$ are respectively the 5-dimensional Planck mass and the AdS curvature.  The UV brane gives an effective description of the dynamics at energies larger than $\Luv$, and in fact the model can be thought of as the low-energy effective limit of a multi-brane RS theory~\cite{Agashe:2016rle}. The dynamics in the bulk and on the IR brane, dual to the 4D CFT, play the role of the dark sector. The second term of Eq.~(\ref{eq:UVbraneaction}) induces a dim-8 portal interaction between the SM and the CFT in the dual theory, 
\begin{equation}
\label{eq:dim8RSportal}
\frac{\kappa_T}{\Luv^4} T_{\mu\nu}^{DS} T^{\mu\nu}_{SM} \, , \qquad \text{with} \quad \kappa_T \sim \frac {k^3}{M_5^3}\, ,
\end{equation}
as discussed in Appendix~\ref{app:RSmodel}. Other portals can be also generated by the interactions between the SM fields and those stabilizing the extra dimension. For example, a mixed interaction term on the UV brane between a Goldberger-Wise scalar and the SM Higgs field generates a dim-6 portal ${\cal O}H^\dagger H$ in the dual theory.

\subsection{Summary}

The models discussed above provide concrete realizations of dark sectors with portal interactions of the type considered in Eq.~\eqref{eq:portal}.
They will serve as benchmarks in Sec.~\ref{sec:openproduction} and in our final discussion of Sec.~\ref{sec:Conclusion}, where different constraints are analyzed and compared. 
Depending on the model, a DM candidate might reside in the dark sector or be part of the UV dynamics, and its abundance may be thermal or arise from a different production mechanism. 

In the pure Yang-Mills DS models, the DM candidate is one of the UV~states and its thermal abundance reproduces the DM experimental density for $\Luv \gtrsim 30\,$TeV~\cite{Mitridate:2017oky,Contino:2018crt}, which is too large a value to be probed with terrestrial experiments. Similarly, in the strongly-coupled $L\oplus N$ model the thermal density of dark baryons reproduces the observed DM abundance for $\Lir \sim 100\,$TeV~\cite{Antipin:2015xia}, which is again beyond the reach of current and future terrestrial experiments. Hence, in the region of parameter space that is probed with our analysis, the DM candidate of all these strongly-coupled models has either a non-thermal density or does not account for the (whole) DM abundance.

The models of Sec.~\ref{sec:freefermion}, on the other hand, are weakly coupled and the thermal density of their dark fermions can reproduce the observed DM abundance for lower masses, of order $\Lir \sim 1-100\,$GeV.
This is in the range accessible by the terrestrial experiments analized in Sec.~\ref{sec:Bounds}. A detailed study of the DM phenomenology of these models is beyond the scope of our paper, although it is reasonable to expect that it will not differ much from the one studied in Refs.~\cite{Kim:2008pp,Baek:2011aa,LopezHonorez:2012kv} in the context of similar theories.

Finally, the 5D Randall-Sundrum model of Sec.~\ref{sec:5DRS} does not have any obvious DM candidate, although the lightest Kaluza-Klein resonance might potentially play this role in the limit in which it becomes very light (compared to $\Luv$) and long lived. It would be interesting to analize this possibility in presence of a non-thermal production mechanism.

Table~\ref{tab:models} summarizes our benchmark models, indicating the DS content, its possible UV completions and the leading portals to the SM. For additional models see for example Refs.~\cite{LopezHonorez:2012kv,Greljo:2013wja,Freitas:2015hsa,Fedderke:2015txa}.
\begin{table}[h!]
\centering
\begin{tabular}{|l|c|c|}
\hline
Dark Sector & UV completion & Portals \\
\hline && \\[-0.3cm]
Pure $SO(N_{DC})$ Yang-Mills & L+N model & $\kappaO, \kappa_T$ \\
                                              & V model & $\kappa_T$ \\[0.15cm]
$SO(N_{DC})$ + 1 Majorana fermion & L+N model & $\kappaO, \kappa_J$ \\[0.15cm]
Strongly coupled CFT with only $T_{\mu\nu}^{DS}$ & 5D RS model & $\kappa_T$ \\[0.15cm]
Free Fermions: && \\
3 Majorana $N_i$ & gauged $U(1)_{B-L}$ model & $\kappa_J$ \\
1 Majorana $\chi$ & `slepton + neutralino' model & $\kappa_J$ \\
1 Dirac $\psi$ & model with real scalar mediator $S$ & $\kappaO$ \\[0.05cm]
\hline
\end{tabular}
\caption{\small Summary of benchmark models that serve as examples of dark sectors with irrelevant portals interactions.}
\label{tab:models}
\end{table}

\section{Strategy}
\label{sec:Strategy}

In this section we discuss how the dark dynamics can be probed using processes at energies $\sqrt{s} < \Luv$. We can envisage three different situations, sketched in Fig.~\ref{fig:luv-lir-s-hierarchy}, depending on the value of $\sqrt{s}$. Furthermore, one can consider two broad classes of effects:
\begin{figure}
\centering
\includegraphics[scale=.75]{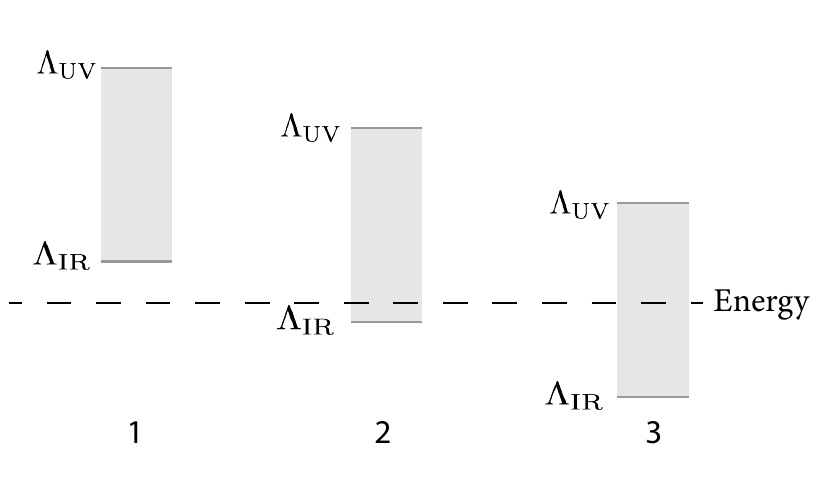}
\caption{\small Cartoon of the three possible situations characterizing the energy $\sqrt{s}$, at which the dark sector is probed, compared to the scales $\Luv,\Lir$.}
\label{fig:luv-lir-s-hierarchy}
\end{figure}
%
\begin{itemize}
\item Indirect contributions to SM processes from virtual exchange of DS or UV states
\item Production of DS states.
\end{itemize}
Indirect effects are the only ones that can occur if $\Lir >\sqrt{s}$, as in situation 1 of Fig.~\ref{fig:luv-lir-s-hierarchy}. Production of DS states, on the other hand, occurs differently in situations 2 and~3 of Fig.~\ref{fig:luv-lir-s-hierarchy}. One can imagine discovering the dark sector through the production of a few new states upon crossing the IR energy threshold. This is situation 2 of Fig.~\ref{fig:luv-lir-s-hierarchy}. On the other hand, a dark sector with low mass gap and feeble interactions with the SM could be first observed directly in its conformal regime if one reaches a minimum luminosity. Discovery in this case is not limited by energy, and the new states can be produced well above threshold (situation~3 of Fig.~\ref{fig:luv-lir-s-hierarchy}). In what follows we estimate the relative importance of indirect effects and DS production, and try to highlight the best strategy to probe the dark dynamics. 
As we will see, whenever the energy relevant for the physical observable is much higher than the IR threshold scale, like in the situation 3 of Fig.~\ref{fig:luv-lir-s-hierarchy}, bounds on the dark sector can be set in a model-independent way. The rate of production of dark states near threshold, like in situation 2, depends instead on the details of the dark dynamics and cannot be predicted on general grounds. 


\subsection{Indirect (virtual) effects}
\label{sec:indirect}

\begin{figure}
\centering
\includegraphics[scale=.75]{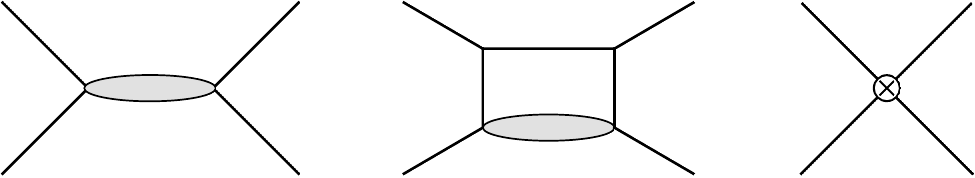}
\caption{\small Contributions to processes involving SM external states: virtual exchange of DS states at tree-level and 1-loop (first two diagrams from the left), contact interaction from UV-generated dim-6 operators (diagram on the right). Solid lines denote SM particles, the gray blob stands for a DS propagator.}
\label{fig:indirect}
\end{figure}

The DS degrees of freedom can be exchanged virtually in processes involving SM external states. This requires (at least) two insertions of the portal interactions, either at tree-level or at loop-level, depending on the process and the portal involved. Physical amplitudes are thus written in terms of two-point correlators of the DS operators appearing in the portal interactions. These have the form
\begin{align}
\langle O_{DS}(p) O_{DS}(-p) \rangle \sim \frac{c}{16\pi^2} \left( p^{2\Delta-4} + p^{2\Delta-6} \Lir^2 + \dots + \Lir^{2\Delta-4} \right) + \text{divergent terms}\, ,
\label{eq:DS2point}
\end{align}
for a generic DS operator $O_{DS}$ with dimension $\Delta$, where~$c$ accounts for the multiplicity of DS states. An additional contribution to the same process comes from the exchange of UV states. This is a local effect and can be encoded by a single insertion of dim-6 operators generated at the UV scale. The different contributions are illustrated in Fig.~\ref{fig:indirect}.
The dim-6 (as well as higher-dimensional) operators are in fact required as counterterms to cancel the power-law divergences that arise  for $D>5$, where $D$ is the overall dimension of the portal, in the two-point correlator of Eq.~(\ref{eq:DS2point}).
In the spirit of effective field theory, this is a UV threshold correction arising at the scale $\Luv$.  
For example, a tree-level diagram with two insertions of $J^{DS}_\mu H^\dagger i \!\overleftrightarrow{D}^{\!\mu} H$ requires a counterterm $O_T = (H^\dagger i \!\overleftrightarrow{D}^{\!\mu} H)^2$ to remove the quadratic divergence 
of the two-point correlator $\langle J^{DS}_\mu J^{DS}_\nu \rangle$.~\footnote{Such quadratic divergence arises if the invariance associated to the conserved current $J^{DS}_\mu$ is broken by the UV dynamics. As an example, consider the $L\oplus N$ model of Sec.~\ref{sec:SCDS}, where the axial $U(1)$ acting on the singlet $N$ is broken by the Yukawa couplings.}
We can thus estimate a minimum value of the coefficient of a generic dim-6 operator made of $n$ SM fields, compatible with the existence of the portal interactions: 
\begin{equation}
\label{eq:UVthreshold}
\Delta c_6(\Luv) \sim g_{SM}^{n-4}\frac{\kappa^2}{\Luv^2} \frac{c}{16\pi^2} \left(\frac{g^2_{SM}}{16\pi^2}\right)^{\!\ell}\quad\quad \text{(UV threshold)}\, .
\end{equation}
Here $g_{SM}$ is a generic SM coupling and $\ell$ is the number of loops at which $\Delta c_6$ is generated. For $D<5$, diagrams with two portal insertions can be made finite with counterterms already present in the SM Lagrangian, and they do not imply any UV threshold correction. While Eq.~(\ref{eq:UVthreshold}) corresponds to the minimum value of the dim-6 coefficients compatible with the existence of the portal interactions, an additional and possibly larger contribution can arise from the virtual exchange of just UV states. The size of such effect clearly depends on the type of UV physics and cannot be estimated on general grounds. For integer $D$, with $D\geq 5$, diagrams with two portal insertions
will also have a logarithmic divergence, which implies a renormalization of the dim-6 operators and a contribution to their RG evolution below $\Luv$. A naive estimate of such effect gives:
\begin{equation}
\label{eq:RGrunning}
\Delta c_6(\mu) \sim g_{SM}^{n-4}\frac{\kappa^2}{\Luv^2} \frac{c}{16\pi^2} \left(\frac{g^2_{SM}}{16\pi^2}\right)^{\!\ell}\left( \frac{\bar \Lambda^2}{\Luv^2}\right)^{D-5} \log\frac{\mu}{\Luv} \quad\quad \text{(RG running)} \, ,
\end{equation}
where $\bar\Lambda \equiv \max( \Lir, m_H)$ and $\mu$ is an RG scale below $\Luv$ and above $\Lir$. The degree of divergence can be lowered to zero (corresponding to a log divergence) by making insertions of the Higgs mass term (hence $\bar\Lambda = m_H$) if the diagram features Higgs propagators, or by making use of the subleading terms in the DS correlator of Eq.~(\ref{eq:DS2point}) (hence $\bar\Lambda = \Lir$). For example, for integer $D \geq 5$ the operator $O_H = [\partial_\mu (H^\dagger H)]^2$ will be renormalized at tree level by ${\cal O} H^\dagger H$.

For a given process with SM external states, the DS gives an additional contribution, not associated with divergences, that takes a different form depending on whether the energy $\sqrt{s}$ is above or below the IR scale $\Lir$.
If $\sqrt{s} < \Lir$, then the DS dynamics can be integrated out at $\Lir$ and generates (for any $D$) an IR threshold correction to dim-6 operators. We estimate in this case
\begin{equation}
\label{eq:IRthreshold}
\Delta c_6(\Lir) \sim g_{SM}^{n-4}\frac{\kappa^2}{\Luv^2} \frac{c}{16\pi^2} \left(\frac{g^2_{SM}}{16\pi^2}\right)^{\!\ell} \left( \frac{\bar \Lambda^2}{\Luv^2}\right)^{D-5}  \quad\quad \text{(IR threshold)} \, .
\end{equation}
This is smaller than Eq.~(\ref{eq:RGrunning}) by a log factor. 
If $\sqrt{s} > \Lir$, then the exchange of DS states will induce a long-distance contribution to the rate of events $R$ of order
\begin{equation}
\label{eq:longdistance}
\frac{\Delta R}{R} \sim \frac{\kappa^2 c}{16\pi^2 g_{SM}^2} \left(\frac{g^2_{SM}}{16\pi^2}\right)^{\!\ell} \left(\frac{s}{\Luv^2}\right)^{D-4} \quad\quad \text{(Long-Distance)} \, , 
\end{equation}
arising through the interference with the SM amplitude. This should be compared with the correction from the interference of the SM amplitude with diagrams featuring one insertion of a dim-6 operator, $\Delta R/R \sim c_6/g_{SM}^{n-2} (s/\Luv^2)$.

We can, at this point, establish the relative importance of the various virtual effects in Eqs.~(\ref{eq:UVthreshold}),(\ref{eq:RGrunning}),(\ref{eq:IRthreshold}) and (\ref{eq:longdistance}). 
In the case $\sqrt{s} < \Lir$ (situation 1 of Fig.~\ref{fig:luv-lir-s-hierarchy}), the contributions from both DS and UV states are local and parametrized by dim-6 operators. As such, they are qualitatively indistinguishable at low energy. Furthermore, for $D\geq 5$ the UV threshold correction is always larger than the RG running, which in turn dominates (for $D$ even) over the IR thresholds. For $4 < D < 5$, instead, the DS exchange gives only an IR threshold contribution, which 
can (depending on the UV dynamics) be larger than the one generated by heavy mediators at $\Luv$. 

If $\sqrt{s} > \Lir$ (situations 2 and 3 of Fig.~\ref{fig:luv-lir-s-hierarchy}), then for $D\geq 5$ the UV threshold corrections are larger than the long-distance effects, which in turn are larger than the RG running. In principle, one could distinguish experimentally the long-distance from local effects, since the former induce a non-analytic dependence of the cross section on the energy~\cite{Georgi:2007si} (see also the discussion in Sec.~\ref{sec:fifthforce}). For $4 < D < 5$, the DS exchange generates only a long-distance contribution, which can win over the UV effect induced by heavy mediators.

To summarize, UV thresholds are expected to give the most important virtual effects for $D\geq 5$; portals with $4 < D < 5$, instead, generate only long-distance (for $\sqrt{s} > \Lir$) or IR threshold (for $\sqrt{s} < \Lir$) corrections, and can give the largest indirect contribution.

\subsection{Production of DS states}
\label{sec:openproduction}

The rate of production of DS states scales as $(1/\Luv^2)^{D-4}$, and is clearly suppressed for large portal dimensions $D$. On the other hand, the experimental significance of the new physics events strongly depends on the kind of signature and on the size of the SM background. Depending on the lifetime of the lightest DS particle(s) (LDSP), one can have processes at colliders with missing energy, displaced vertices or prompt DS decays. In the rest of this subsection we will estimate the lifetime of the LDSP, explain our strategy to quantify the yield of events with respectively missing energy and displaced vertices, and discuss the validity of the effective field theory approach.

\subsubsection*{Lifetime of the Lightest DS Particle}

At energies $\sqrt{s} \gg \Lir$ (situation 3 of Fig.~\ref{fig:luv-lir-s-hierarchy}), the DS operator will excite a CFT state made of DS degrees of freedom whose evolution depends on the underlying dark dynamics. In strongly-coupled dark dynamics, there will be a phase of parton showering followed by dark hadronization, at the end of which many DS particles are produced. Weakly-coupled dark dynamics, on the other hand, will lead to few particles. In either case, these states will generally decay among themselves through fast transitions, and eventually decay to the LDSP $\psi$. Metastable or stable particles can also exist as a consequence of symmetries or kinematic suppressions. The LDSP itself might be stable if charged under some dark symmetry preserved by the portals. Generically, $\psi$ will decay to SM states through the portal interactions. The rate for this transition is expected to be much smaller than that characterizing inter-DS decays, especially in the case of strongly-coupled dynamics. Hence, the general expectation is that in a given process with dark excitations in the final state, these will promptly decay to $\psi$ and to stable particles (if present), and at later times $\psi$ decays back to the SM.

If the LDSP decays through a portal with dimension $D$ and is heavier than the EW scale, its lifetime can be naively estimated to be
\begin{equation}
\label{eq:lifetime}
\tau_{\psi} \sim \left[ \Lir \frac{\kappa^2}{8\pi} \left(\frac{f^2}{\Lir^2}\right) \left(\frac{\Lir^2}{\Luv^2}\right)^{D-4}\right]^{-1}\, ,
\end{equation}
where $f$ is a decay constant defined by $\langle 0 | {\cal O}|\psi\rangle = \hat A \, f\,  \Lir^{\Delta-2}$, and $\hat A$ is a dimensionless tensor that depends on the quantum numbers of ${\cal O}$ and $\psi$. For example, if the DS operator is a conserved current, then $\hat A$ is proportional to the polarization vector $\epsilon_\mu$ of $\psi$ if the latter is a massive spin-1 state, and to $p_\mu/\Lir$ if $\psi$ has spin 0 (as for a Nambu-Goldstone boson). For strongly-coupled dark dynamics, one expects the decay constant to scale as $f\sim \sqrt{c}$ in the limit of large $c$, where $c$ is proportional to the number of degrees of freedom of the DS (see Eq.~\eqref{eq:DS2point}). 
The LDSP can decay through one of the minimal portals of Eq.~(\ref{eq:portal}) or through operators with different quantum numbers and a larger dimension. 
The value of $\tau_\psi$ can differ from the estimate of 
Eq.~(\ref{eq:lifetime}) if $\psi$ is lighter than the EW scale and its main decay channel requires EW symmetry breaking. This occurs for example when $\psi$ mixes with the Higgs boson or the $Z$, respectively through the ${\cal O}H^\dagger H$ or $J_\mu^{DS} H^\dagger i \!\overleftrightarrow{D}^{\!\mu} H$ portal. In this case, one can compute $\tau_\psi$ (for $m_\psi < m_{Z,h}$) as
\begin{equation}
\label{eq:lifetimewithmixing}
\tau_\psi = \left( \Gamma_{i} \sin^2\theta_i \right)^{-1}, 
\qquad\quad \tan 2\theta_i = \frac{2 \delta_i}{m_\psi^2 - m_i^2}\, ,
\qquad  i=Z,h \, ,
\end{equation}
where $\Gamma_{Z,h}$ are the total decay widths of the $Z$ and $h$  (defined as the sum of the partial decay widths into the accessible SM final states) evaluated at $m_{Z,h} = m_\psi$. The mixing angle $\theta_{Z,h}$ is computed from the mass mixing terms
\begin{equation}
\delta_h = \kappa_{\cal O} v f \left(\frac{\Lir}{\Luv}\right)^{\Delta-2}\, , \qquad
\delta_Z = \kappa_J \, v f\,  \frac{m_Z\Lir}{\Luv^2}\, ,
\end{equation}
where we assumed that $\psi$ has spin 1 when it mixes with the $Z$. If the decay proceeds through the mixing with the Higgs boson, the value of $\tau_\psi$ from Eq.~(\ref{eq:lifetimewithmixing}) is parametrically larger than the estimate (\ref{eq:lifetime}) by a factor $m_h^2/\Lir^2$. In the case of mixing with the $Z$, on the other hand, the lifetime is parametrically similar to that induced by a generic $D=6$ portal.

\subsubsection*{Missing Energy Events}

In the limit of a large hierarchy, i.e.~for $\Lir/\Luv$ small enough, the LDSPs produced in high-energy collisions will decay outside the detector, and manifest themselves as missing energy. We will classify an event as a missing energy one if all of its LDSPs emerging from the primary collision decay outside the detector. The probability for one LDSP to decay within a distance $x$ from the primary vertex is $\exp(-x/c\tau_\psi\gamma)$, where $\gamma$ is the boost factor of the LDSP. We will thus estimate the probability for an event to be a missing-energy one as
\begin{equation}\label{eq:PmissE}
\text{P}[all > L_d] = \exp\left(- \frac{\n L_d}{c\tau_\psi \g}\right) \, ,
\end{equation}
where $L_d$ is the detector length, $\n$ is the average number of DS particles per event, and $\g$ is the average boost factor. As already mentioned, the average number of DS particles depends on the type of dark dynamics. We will consider two benchmark values: the first, $\n =2$, is representative of weakly-coupled dark sectors; in the second we set
\begin{gather}
\label{eq:naverage}
\n = A  \left(\frac{1}{\log(\E^2/\bar \Lambda^2)}\right)^B  \exp\!\left( C\sqrt{\log(\E^2/\bar \Lambda^2)}\right),
\hspace{0.6cm}
\begin{split}
A &= 0.06 \\
B &= 0.5 
\end{split}
\hspace{0.6cm}
\begin{split}
C &= 1.8 \\
\bar\Lambda &= 0.1\, \Lir
\end{split}
\end{gather}
to characterize the behavior of $\n$ in strongly-coupled dark sectors in terms of the energy $\E$ of the DS system. The functional dependence of Eq.~(\ref{eq:naverage}) corresponds to the leading-order theoretical prediction in QCD~\cite{Ellis:1991qj}, according to which $\langle n\rangle \propto \alpha_s^b \exp(c/\sqrt{\alpha_s})$, where $b$ and $c$ are known constants.
The values of the numerical coefficients in Eq.~(\ref{eq:naverage}) well approximate those of QCD with 5 flavors, except for the overall normalization $A$ that cannot be computed perturbatively in QCD and has been fixed so that $\n=2$ for $\E=2\Lir$.~\footnote{This normalization gives a smaller average number of dark hadrons at $\E/\Lir$ compared to the QCD prediction at $\E/\Lambda_{QCD}$. This is in fact reasonable given that the QCD spectrum includes particles (i.e.~the pions and other pseudo Nambu-Goldstone bosons) that are parametrically lighter than other resonances.} We take Eq.~(\ref{eq:naverage}) as representative of strongly-coupled dark sectors near a fixed point where couplings evolve (nearly) logarithmically like in QCD.~\footnote{Notice however that in gauge theories with large 't Hooft coupling $\lambda$ one has $\langle n(Q)\rangle \propto Q^{1-3/2\sqrt{\lambda}}$~\cite{Hatta:2008tn}.} Finally, we will estimate the average boost factor in Eq.~(\ref{eq:PmissE}) as
\begin{equation}
\label{eq:gammaaverage}
\g = \frac{\E}{\n \Lir}\, .
\end{equation}

\begin{figure}
\centering
\includegraphics[scale=0.50]{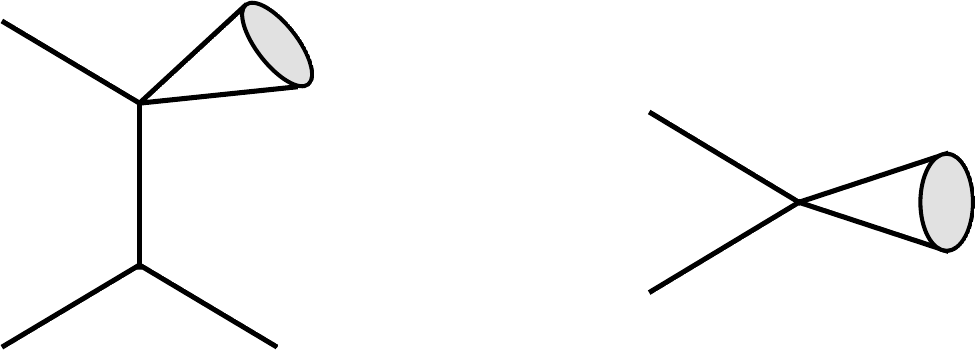}
\caption{\small Prototype Feynman diagrams for DS production: associated production $SM+SM\to DS+SM$ (left), and single production $SM+SM\to DS$ (right). Solid lines denote SM particles, the gray blob stands for a DS state.}
\label{fig:directproduction}
\end{figure}

Triggering on missing-energy events requires (at least) one SM tagging object in the final state, and the prototype Feynman diagram for DS production in this case is that on the left of Fig.~\ref{fig:directproduction}.
For $\Lir \ll \sqrt{\hat s} \ll \Luv$, where $\hat s = p_{DS}^2$ is the squared momentum of the DS system, the inclusive cross section can be predicted independent of the low-energy details of the dark dynamics by exploiting its conformal behavior. From the optical theorem it follows
\begin{equation}
\label{eq:opticaltheorem}
\sum_n \int\! d\Phi_{DS} |\langle 0 | O_{DS} | n\rangle|^2 = 2\, \text{Im} \left[ i\langle 0 | T\!\left\{ O_{DS} O_{DS} \right\} \! | 0\rangle \right]
\end{equation}
for a generic operator $O_{DS}$ that interpolates the dark state $|n\rangle$ from the vacuum, denoting the dark sector phase space with $d\Phi_{DS}$.
Since conformal invariance determines the two-point function of $O_{DS}$ in terms of its dimension and up to an overall constant, the inclusive cross section well above threshold can be predicted in a model-independent way.
As an analogy, consider for example the production of QCD hadrons in $e^+ e^-$ collisions: near threshold the inclusive cross section exhibits a complicated pattern of resonances, but at energies $\sqrt{\hat s} \gg \Lambda_{QCD}$ its behavior is determined by the asymptotic freedom of QCD, and depends only on the number of colors and the fact that the photon couples to a conserved quark current. In this regime, resumming the contributions of all the hadronic states reproduces the much simpler quark contribution (quark-hadron duality), as dictated by perturbativity. Notice, however, that the universal behavior of the inclusive cross section stems from the fact that the theory is nearly conformal, and having a free fixed point is not crucial. Similar results, therefore, hold also for a strongly-coupled dark dynamics in its conformal regime.

\begin{figure}[t]
\centering
\includegraphics[width=0.485\textwidth]{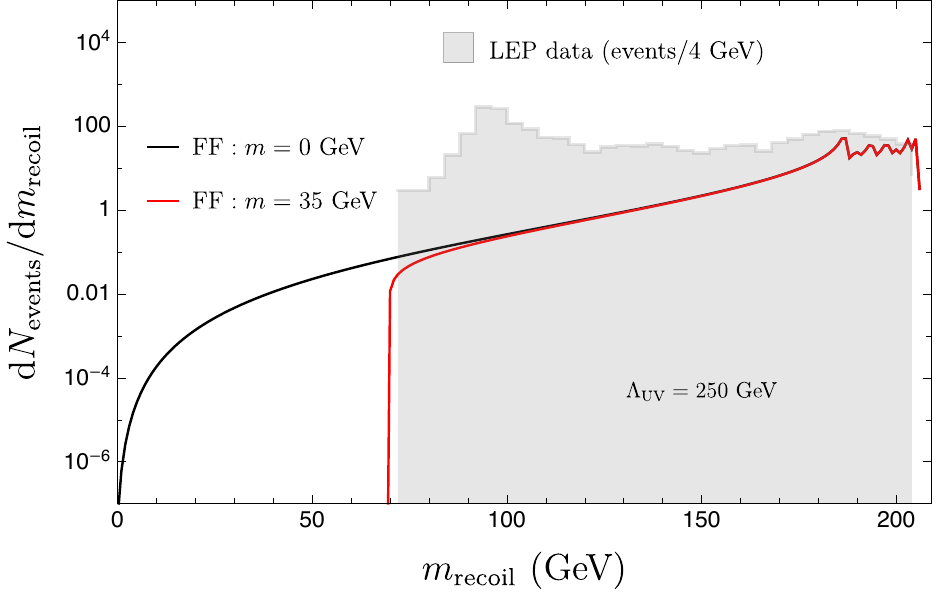}
\hspace{0.2cm}
\includegraphics[width=0.485\textwidth]{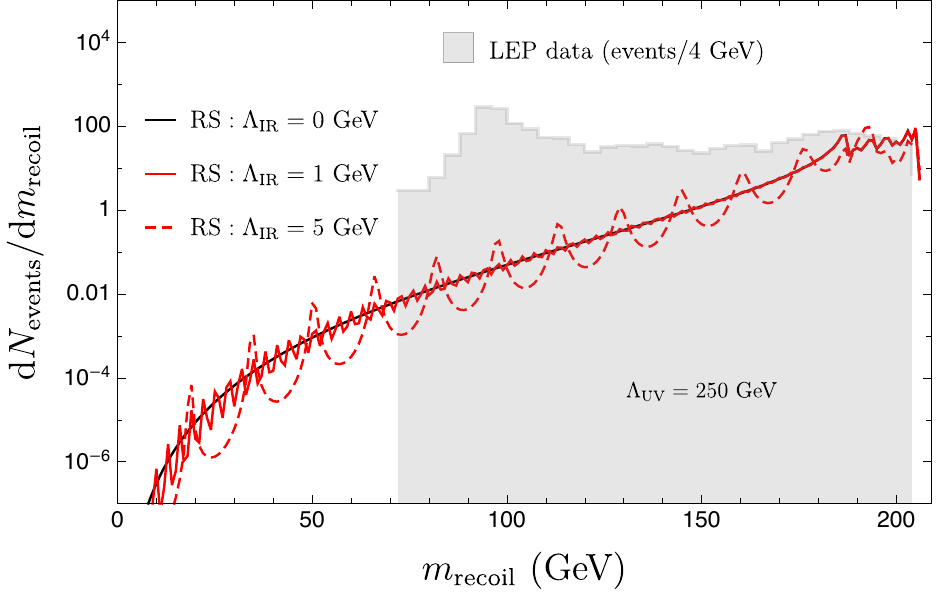}
\caption{\small Differential number of events for $e^+ e^- \to DS + \gamma$ as a function of the recoil mass (equal to the DS invariant mass) at LEP. The black and red curves in the left panel correspond to the prediction of the second model of Sec.~\ref{sec:freefermion}, which leads to a DS with one Majorana fermion coupled to the SM through the $D=6$ portal of Eq.~(\ref{eq:JJportalforFF2}). We have set $\kappa_J =1$, $\Luv = m_\phi = 250\,$GeV, and the fermion mass $m_\chi = m$ to the value indicated in the plot.
Similarly, the curves in the right panel show the prediction of the RS model of Sec.~\ref{sec:5DRS}, where the DS couples through the $D=8$ portal of Eq.~(\ref{eq:dim8RSportal}). We have set $(N_{CFT}^2-1) =10$ (corresponding to $c_T = 400$, see Eq.~(\ref{eq:cTholo})), $\kappa_T =1$ and $\Luv =250\,$GeV. The gray region shows the number of events measured at LEP by the L3 Collaboration~\cite{Achard:2003tx}.}
\label{fig:threshold}
\end{figure}

In our analysis we will approximate the inclusive cross section for DS production by including only the contribution from the conformal
regime and by using the optical theorem as in Eq.~(\ref{eq:opticaltheorem}). We will thus neglect the events produced near threshold (in practice, we will impose a lower cut on $p_{DS}^2$). Including them obviously increases the total cross section and leads to more stringent constraints. Our results will be thus conservative. The importance of such threshold contribution depends on the dimensionality of the portal responsible for the DS production and on the energy range probed by the collider. In the case of irrelevant portals, the (partonic) cross section usually grows with the energy; larger dimensions of the DS operator lead to faster growths at high energy and thus enhance the contribution away from threshold. As a consequence, the bulk of events can be produced in the deep conformal regime, where our approximation is accurate. To illustrate this point, we show in Fig.~\ref{fig:threshold} the number of events predicted at LEP for the process $e^+e^- \to DS + \gamma$ as a function of the recoil mass (i.e.~the invariant mass of the DS). 
This process has been measured by L3~\cite{Achard:2003tx} and OPAL~\cite{Abbiendi:2000hh} and sets constraints on elusive DS, as discussed in Sec.~\ref{sec:non-resonant}. 
The plots of Fig.~\ref{fig:threshold} report the theoretical predictions for two benchmark dark sectors:
the case of a free Majorana fermion coupled through the $D=6$ portal of Eq.~(\ref{eq:JJportalforFF2}), and the 5D Randall-Sundrum theory with the $D=8$ portal of Eq.~(\ref{eq:dim8RSportal}).
They show clearly that in those cases the bulk of the events are created away from threshold, in the regime where the DS dynamics is conformal.
This situation should be contrasted with the case of relevant or marginal portals, where threshold events are more important and could first lead to discovery~\cite{Strassler:2008bv}. 
Notice that the $D=8$ portal in the RS theory has the proper quantum numbers to singly excite the radion and spin-2 resonances, and that these appear as resonant peaks in the right panel of Fig.~\ref{fig:threshold}. We have used a modified expression of the two-point form factor as in Eq.~(\ref{eq:modifiedFF}), with a radion mass $m_\phi = \Lir$. In the free-fermion case, the $D=6$ portal excites pairs of fermions, and for this reason no resonant peak appears in the left panel of Fig.~\ref{fig:threshold}.

\subsubsection*{Validity of the Effective Field Theory Description}

An additional aspect of our calculation is the validity of the Effective Field Theory (EFT) approximation. The form of the portal interactions considered in this work, between the dark and SM sectors, arises by integrating out mediators of mass close to $\Luv$. If the momentum at which this interaction is probed exceeds $\Luv$, it is no longer a good approximation to describe it as a contact interaction mediated by a local operator. As we consider various experimental bounds, the validity of the EFT approximation must be enforced for internal self-consistency. 
The experimental data are usually presented in terms of a differential distribution of the number of events or cross section as a function of some kinematic variables (e.g. 3-momentum, transverse momentum or recoil mass) relative to one or more of the visible particles. From momentum conservation, these kinematic variables are related to the momentum $p_{DS}$ that flows into the contact interaction. Assuming an $s$-channel exchange of the mediator field, the EFT expansion is controlled by $p_{DS}^2/\Luv^2$; consistency requires $p_{DS}^2/\Luv^2 \ll 1$, which translates into a condition on the kinematic variables. If the data are presented as a histogram, the condition in general varies bin by bin.

Taking this into consideration, we will adopt the approach advocated for example in Ref.~\cite{Contino:2016jqw}, and use the subset of events that allows us to derive a self-consistent bound on $\Luv$. To see how this works in practice, let us consider a scattering process with a mono-X final state plus missing momentum, as in the left diagram of Fig.~\ref{fig:directproduction}. In this case
\begin{equation}
p_{DS}^2 = \hat s - 2 \sqrt{\hat s} \, {\not}{p} \:,
\end{equation}
where we take the final SM state to be massless, and ${\not}{p} \equiv |{\not}{\vec{p}}|$ is the magnitude of its 3-momentum. Requiring $p_{DS}^2/\Luv^2 < \xi$, where $\xi$ is some value smaller than 1, translates into a condition on $\Luv$:
\begin{equation}
\Luv \gtrsim \frac{1}{\xi^{1/2}} \sqrt{\hat s-2\sqrt{\hat s}\,{\not}{p}} \geq \frac{1}{\xi^{1/2}} \sqrt{\hat s-2\sqrt{\hat s}\,{\not}{p_T}} \, ,
\label{eq:eft_validity_bins}
\end{equation}
where in the last step we have used ${\not}{p_T} \equiv {\not}{p} \, \sin\theta < {\not}{p}$. Here ${\not}{p_T}$ is the transverse missing momentum carried by the DS (equal to the transverse momentum of the SM final state). The EFT is within its validity as long as the missing momentum is sufficiently large. One can thus exclude bins with low ${\not}{p_T}$ to extend the validity of the analysis to smaller values of $\Luv$. Including a given range of bins, consistency with EFT implies a lower and an upper bound in the range of $\Luv$. Removing progressively bins of low ${\not}{p_T}$ and finally taking the union of the excluded regions, we obtain the overall bound. The advantage is that while taking all the data may not result in a valid exclusion region at all,  discarding data in some bins gives a self-consistent, though weaker, bound. In the following, when applying this procedure, we will fix $\xi = 0.1$.

\subsubsection*{Events with displaced decays}

Besides events with missing energy, the production of DS states can lead to displaced vertices (DV) if some of the LDSPs decay inside the detector far from the interaction region. For a fixed value of $\Luv$, this occurs in a range of IR scales $\Lir$ that varies with the portal dimensionality $D$. Depending on the experimental analysis, events are selected by requiring a minimum number of decays in specific regions of the detector (inner detector, calorimeters, muon spectrometer). To analyze those data we construct a probability for each event to pass the required conditions as explained in Appendix~\ref{sec:probabilities}. This probability is maximized and close to 1 for  lifetimes $\tau_\psi$ in a certain interval, which in turn corresponds to an interval of $\Lir$ values at fixed $\Luv$.
Events with displaced vertices can be triggered on and reconstructed without the need of a SM tagging object. The leading production diagram is thus the one on the right of Fig.~\ref{fig:directproduction}. As for missing energy events, the rate of DV events can be computed conservatively by including only the contribution from the conformal regime $\sqrt{\hat s} \gg \Lir$, but in this case the result depends on additional quantities whose value is model dependent. For example, the LDSPs from strongly-coupled dark sectors will be produced with energies and angular distributions determined by the showering and hadronization processes. This leads to an acceptance efficiency in the reconstruction of the displaced vertices that depends on the type of dark dynamics. Since the goal of our analysis is to assess the importance of DV searches in testing our theories, we will estimate the event rate by using the two benchmark values of $\n$ described above and by making reasonable assumptions to average out any further model dependency.

\subsubsection*{Events with prompt decay}

Finally, for small hierarchy of scales, i.e.~$\Lir/\Luv$ not too small, the LDSPs produced in a DS event will decay promptly. The significance of these events strongly depends on the details of the underlying DS dynamics and cannot be assessed in a model-independent way. An analysis of this kind goes beyond the scope of this work, and we will not consider the region of the parameter space where only prompt decays occur.

\section{Terrestrial and Astrophysical Bounds}
\label{sec:Bounds}

In this section we present our analysis of the terrestrial and astrophysical processes that can probe the dark sector dynamics. Following the strategy outlined in the previous section, we will derive constraints on the scales $\Lir$ and $\Luv$, for given coupling $\kappa$ and dark multiplicity~$c$, by considering individually each of the portals in Eq.~(\ref{eq:portal}). We have analyzed both processes with production of DS states and processes where these are virtually exchanged. The complete list is reported in Table~\ref{tab:processes}. 
\begin{table}
\centering
\begin{minipage}[t]{0.53\textwidth}
\underline{Probes of DS production} \\[-0.4cm]
\begin{itemize}
\setlength{\itemindent}{-0.18in}
\setlength\itemsep{0in}
\item $Z$ and Higgs boson decays 
\item Non-resonant production at LEP and LHC
\item High-intensity experiments
\item Supernova and stellar cooling
\item Positronium lifetime
\end{itemize} 
\end{minipage} \hspace{0.4cm}
\begin{minipage}[t]{0.4\textwidth}
\underline{Probes of DS virtual exchange} \\[-0.4cm]
\begin{itemize}
\setlength{\itemindent}{-0.18in}
\setlength\itemsep{0in}
\item Fifth-force experiments
\item EW precision tests
\end{itemize} 
\end{minipage}
\caption{\small List of processes and experiments analyzed in this work that probe the dark sector dynamics.}
\label{tab:processes}
\end{table}
It includes searches at high-energy colliders, where DS excitations can manifest themselves as missing energy, displaced vertices or in precision observables, and fixed-target and beam dump experiments, which probe the DS-SM interaction at energies of order $10-100\,$GeV. Complementary to these, there is another class of experiments that probe the DS-SM interaction at much lower energies. They study the effect of the DS on long-range forces or precision observables like the ortho-positronium lifetime. Finally, there are celestial constraints coming from astrophysical observations, which probe the DS-SM interaction at MeV and keV energies.

In the following we discuss each process starting from those with DS production.

\subsection{DS production from $Z$ and Higgs boson decays}

When $\Lir$ is smaller than the EW scale, one of the most efficient ways to produce DS states at colliders is through the decay of the $Z$ and Higgs bosons. Such resonant production proceeds respectively through ${\cal O}H^\dagger H$ (Higgs portal) and $J^\mu_{DS} H^\dagger i \!\overleftrightarrow{D}_{\!\!\mu} H$ ($Z$ portal). The rate of DS events can be computed, in the narrow width approximation, as the SM cross section for Higgs or $Z$ production times the branching ratio for their decay into DS states. No issue arises in this case with the validity of the effective field theory description, since the energy characterizing the production of DS states is that of the $Z$ or Higgs boson mass, while $\Luv$ is required to be larger. One can extract the inclusive decay width of the Higgs or $Z$ boson into DS states from the imaginary part of the 2-point correlator of the DS operator in the portal. For example, working at leading order in the Higgs portal interaction, the pole residue and width of the Higgs boson propagator are corrected by:
\begin{align}
\label{eq:Zh}
Z_h \equiv 1+ \delta Z_h & = 1 + \frac{\kappaO^2 v^2}{\Luv^{2\Delta_\mathcal{O}-4}}\ \frac{d}{d p^2} \text{Re } i\left<\mathcal{O}(-p)\mathcal{O}(p)\right> |_{p^2 = m_h^2} \\
\label{eq:GammahtoDS}
\Gamma_{h\to DS} &= -\frac{1}{m_h} \frac{\kappaO^2 v^2}{\Luv^{2\Delta_\mathcal{O}-4}}\ \text{Im } i\left<\mathcal{O}(-p)\mathcal{O}(p)\right> |_{p^2 = m_h^2} \, .
\end{align}
We approximate the imaginary part of the 2-point correlator at $m_h \gg \Lir$ by using its conformal expression in Eq.~(\ref{eq:imO}) of Appendix~\ref{app:2pt}, and obtain
\begin{equation}
\label{eq:GammahtoDSexplicit}
\Gamma_{h\to DS} = \frac{\kappa_\mathcal{O}^2\,\cO}{\pi^{3/2}} \, \frac{\Gamma(\DeltaO + 1/2)}{\Gamma(\DeltaO-1)\Gamma(2\DeltaO)}
\: \frac{v^2\, m_h^{2\DeltaO-5}}{\Luv^{2\DeltaO-4}} \, .
\end{equation}
Similar steps in the case of the $Z$ portal, and the use of Eq.~(\ref{eq:imJ}) in Appendix~\ref{app:2pt}, lead to the $Z\to DS$ decay width:
\begin{equation}
\label{eq:ZtoDSwidth}
\Gamma_{Z\to DS} =-\frac{\kappa_J^2 v^2 m_Z}{3\Luv^4}\sum_{i=1,2,3} \!\epsilon_\mu^i \epsilon_\nu^{*i}\,\text{ Im } i \langle J^{\mu}_\text{DS}(p)J^\nu_\text{DS}(-p) \rangle\big|_{p^2=m_Z^2} = \frac{\kappa_J^2 v^2 m_Z^3}{\Luv^4}\frac{c_J}{96\pi}\, .
\end{equation}

The total width of the $Z$ boson has been measured accurately by the LEP experiments, which put an upper bound on beyond-the-SM contributions $\Delta \Gamma_Z < 2.0 \text{ MeV}$ at $95\%$ confidence level~\cite{ALEPH:2005ab}. Using this result and Eq.~(\ref{eq:ZtoDSwidth}) leads to the constraint
\begin{align}
\label{eq:invZwidth}
\Luv > 525\,\text{GeV} \times (\kappa_J^2 c_J)^{1/4}  \, . 
\end{align}
A similarly inclusive bound can be obtained through a global fit to data from Higgs searches at the LHC. The correction to the residue of the Higgs propagator due to the DS exchange implies a universal shift of the Higgs couplings by a factor $Z_h^{n/2}$, where $n$ is the number of Higgs bosons in the vertex.
Neglecting possible modifications to the couplings from the UV dynamics, the common signal strength modifier used by LHC collaborations can be expressed as
\begin{equation}
\mu \equiv \frac{\sigma \times BR}{\sigma_{SM} \times BR_{SM}} 
\simeq 1 + \delta Z_h - \frac{\Gamma_{h\to DS}}{\Gamma^{SM}_h}\, ,
\end{equation}
where $\Gamma^{SM}_h$ is the SM Higgs boson total decay width, $\Gamma^{SM}_h=4.07\,$MeV~\cite{PhysRevD.98.030001}. From Eq.~(\ref{eq:Zh}), (\ref{eq:GammahtoDS}) and Eqs.~(\ref{eq:2pO}), (\ref{eq:imO}) it follows that $\delta Z_h$ is smaller than $\Gamma_{h\to DS}/\Gamma^{SM}_h$ by a factor $\Gamma^{SM}_h/m_h \ll 1$, and can be thus neglected. Using Eq.~(\ref{eq:GammahtoDS}) and the measurement $\mu = 1.17 \pm 0.10$ made by CMS with $13\,$TeV data~\cite{Sirunyan:2018koj} gives the constraint
\begin{equation}
\label{eq:mufit}
\Luv > m_h \times \left(1.9 \times 10^{5}\, \kappaO^2 \cO\,  \frac{\Gamma(\DeltaO + 1/2)}{\Gamma(\DeltaO-1)\Gamma(2\DeltaO)} \right)^\frac{1}{2\DeltaO-4} \, ,
\end{equation}
at~95~\%~probability.~\footnote{This bound is derived by constructing a posterior probability as a function of $\delta = 1-\mu$ in terms of the likelihood $\exp[(1-\delta - 1.17)^2/0.02]$ and a flat prior. We find $\delta = \Gamma_{h\to DS}/\Gamma^{SM}_h < 0.11$ with 95~\%~probability, which in turn implies Eq.~(\ref{eq:mufit}).}

\begin{figure}[t]
\centering
\includegraphics[width=0.48\textwidth]{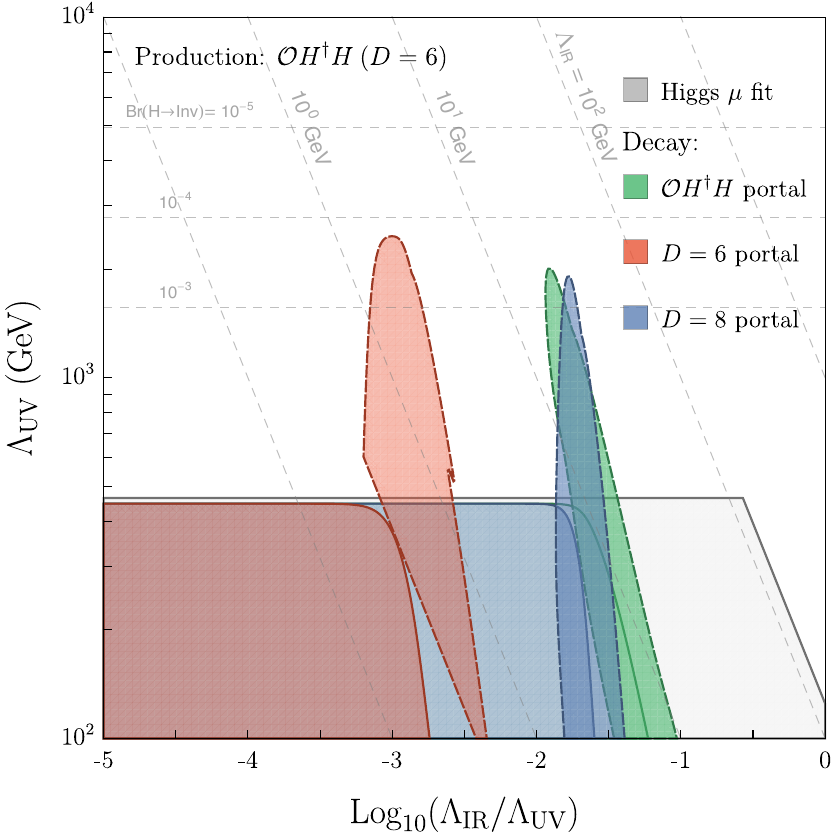}
\hspace{0.2cm}
\includegraphics[width=0.48\textwidth]{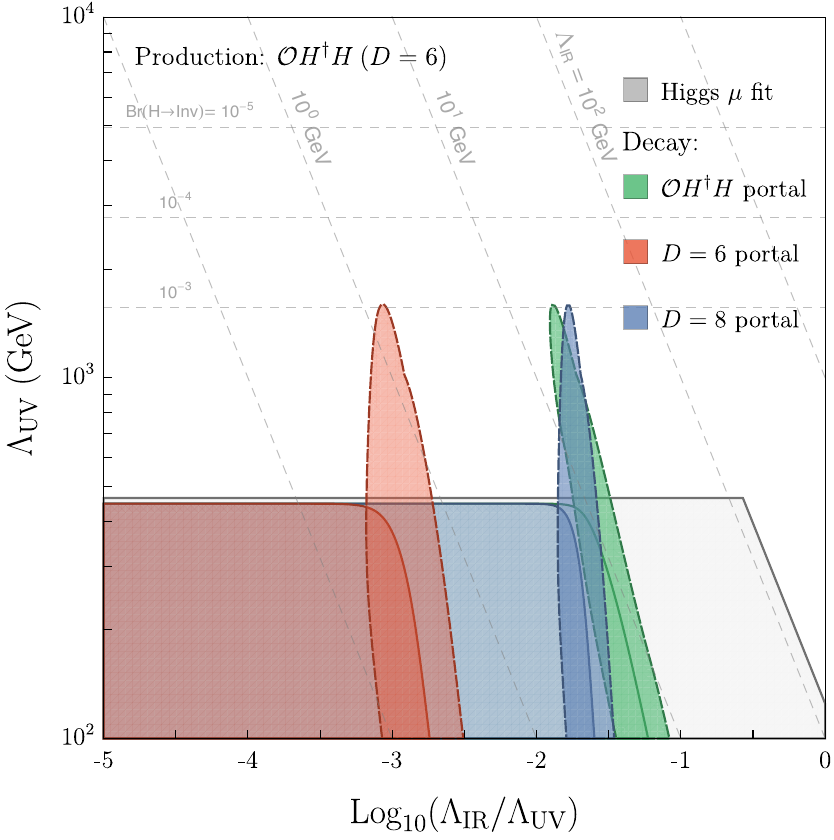} \\[0.15cm]
\includegraphics[width=0.48\textwidth]{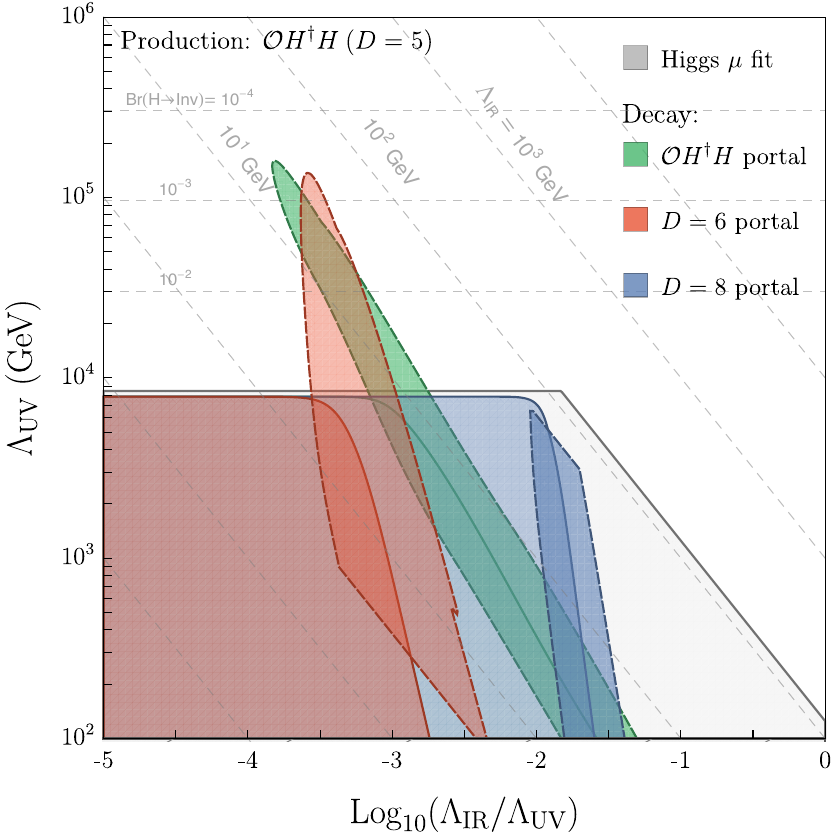}
\hspace{0.2cm}
\includegraphics[width=0.48\textwidth]{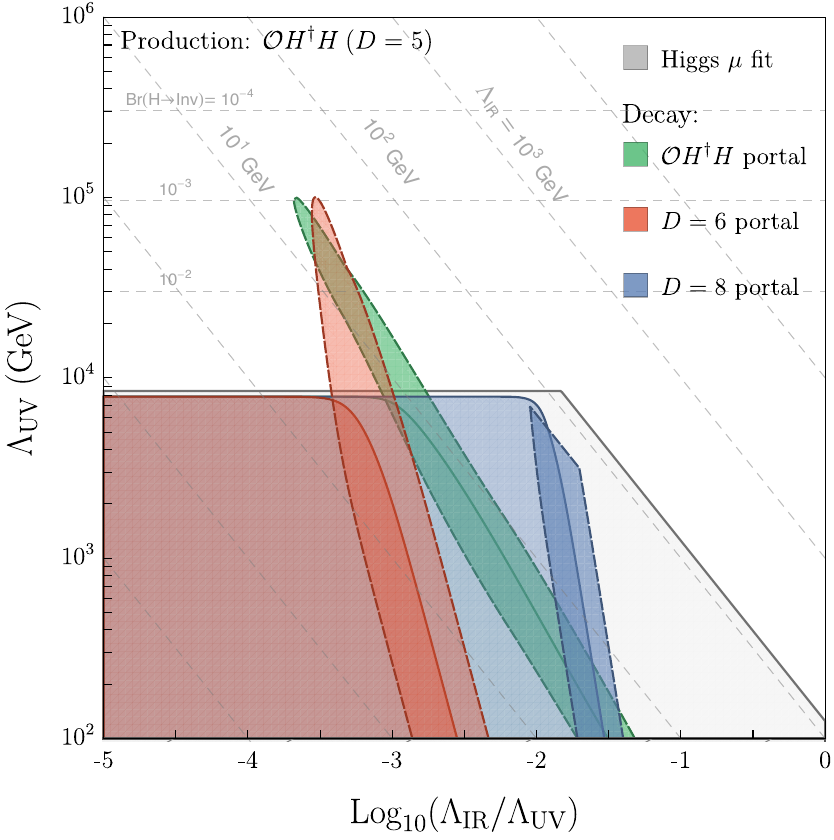}
\caption{\small Constraints on resonant DS production through a $D=6$ (upper panels) and \mbox{$D=5$} (lower panels) Higgs portal. Plots on the left (on the right) assume a strongly (weakly) coupled dark sector. Exclusion regions from the bound on the Higgs invisible width of Ref.~\cite{ATLAS:2020cjb} (solid contours) and the searches for displaced vertices of Refs.~\cite{Aaboud:2018aqj, Aad:2019xav} (dashed contours) are shown for three different types of LDSP decay portal: the same Higgs portal responsible for the production (green), generic $D=6$ (red) and $D=8$ (blue) portals. Also shown in gray is the exclusion from the LHC fit to Higgs data of Eq.~(\ref{eq:mufit}). All the plots assume $\kappa_i^2 c_i=1$ for the various portals.}
\label{fig:resonantHiggs}
\end{figure}

\begin{figure}[t]
\centering
\includegraphics[width=0.48\textwidth]{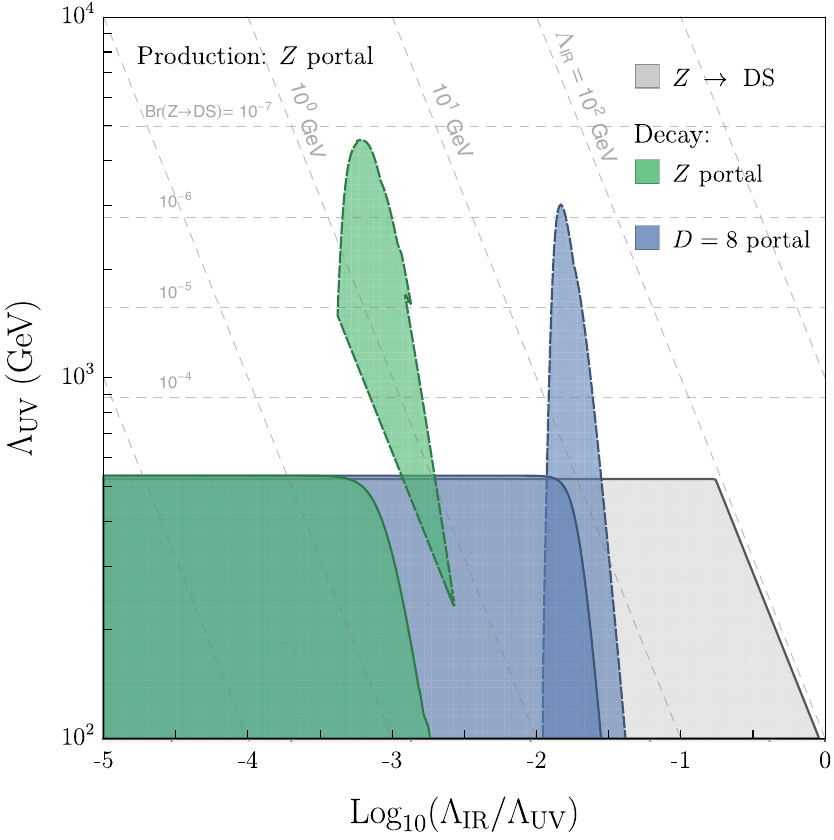}
\hspace{0.2cm}
\includegraphics[width=0.48\textwidth]{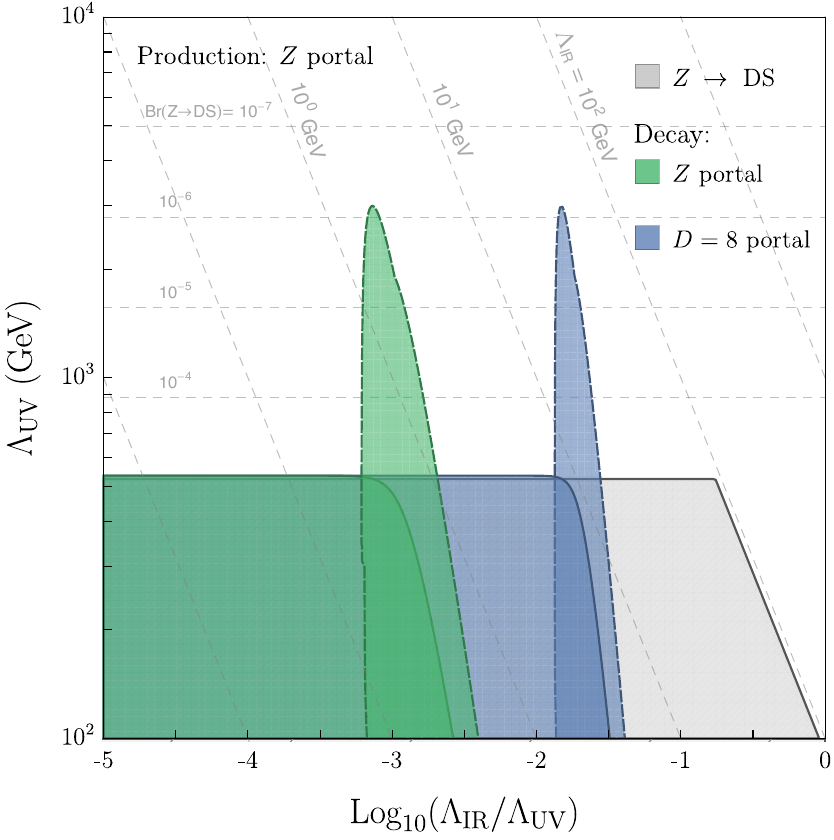}
\caption{\small Constraints on resonant DS production through the $Z$ portal. The plot on the left (on the right) assumes a strongly (weakly) coupled dark sector. Exclusion regions from the mono-jet search of Ref.~\cite{Aaboud:2017phn} (solid contours) and the searches for displaced vertices of Refs.~\cite{Aaboud:2018aqj, Aad:2019xav} (dashed contours) are shown for two different types of LDSP decay portal: the same $Z$ portal responsible for the production (green), and a generic $D=8$ portal (blue). The exclusion from the invisible width measurement at LEP of Eq.~(\ref{eq:invZwidth}) is shown in gray. Both plots assume $\kappa_i^2 c_i=1$ for the various portals.}
\label{fig:resonantZ}
\end{figure}

Measuring precisely the decay rates of the $Z$ and of the Higgs boson into SM particles leads to the indirect constraints on the production of DS states discussed above. Searches at high-energy colliders, however, also look for non-standard decay modes in a variety of final states. The bound on the Higgs invisible branching ratio set by the LHC experiments, for example, constrains the region of parameter space where the LDSP decays outside the detector. We use the recent result obtained by the ATLAS collaboration, $BR_{inv} < 0.13$ at 95\% CL~\cite{ATLAS-CONF-2020-008}, and estimate the number of missing-energy events through Eqs.~(\ref{eq:PmissE}), (\ref{eq:GammahtoDSexplicit}). 
For small values of $\Lir$ the probability of Eq.~(\ref{eq:PmissE}) is approximatively~1 and the bound turns out to be very similar to that of Eq.~(\ref{eq:mufit}). Conversely, for $\Lir$ large enough the majority of LDSPs decay inside the detector and the probability of Eq.~(\ref{eq:PmissE}) goes to zero.
The corresponding exclusion region is shown in Fig.~\ref{fig:resonantHiggs} (solid contours) for three different LDSP decay portals: the same Higgs portal responsible for Higgs-resonant production, and generic $D=6$ and $D=8$ portals.
%
%
Similar constraints come from \mbox{mono-X} searches sensitive to the resonant production of a $Z$ boson followed by its decay into invisible final states. We have analyzed missing-energy searches performed at LEP2 by the L3 collaboration (at a centre-of-mass energy between 189 GeV and 209 GeV) in association with a photon~\cite{Achard:2003tx} and a $Z$ boson~\cite{Achard:2004cf}, and by the OPAL collaboration (at $\sqrt{s}=189\,$GeV) with single photon events~\cite{Abbiendi:2000hh}. The corresponding bounds turn out to be weaker than the inclusive one from the $Z$ decay width and will not be discussed.  From the LHC Run2 at $\sqrt{s}=13\,$TeV we have analyzed the ATLAS mono-jet~\cite{Aaboud:2017phn}, mono-photon~\cite{Aaboud:2017dor} and mono-$Z$~\cite{Aaboud:2017bja} searches. From Run1 at $\sqrt{s}=8\,$TeV we considered the mono-jet search of Ref.~\cite{Aad:2015zva}. All of these studies have found signals consistent with a pure SM background and set constraints on the resonant production of DS states through the $Z$ portal. The strongest bound comes from the mono-jet analysis at $13\,$TeV and is comparable to that from the $Z$ width at LEP. The corresponding exclusion region is shown in Fig.~\ref{fig:resonantZ} (solid contours) for two choices of the LDSP decay portal: the same $Z$ portal, and a generic $D=8$ portal.
%
%

For values of $\Lir$ not too small, some of the LDSP produced in the event can decay inside the detector, far from the primary vertex. Signatures of this kind are searched for by the LHC collaborations in a variety of final states. A nice overview of searches for long-lived particles can be found in a recent document written by the LHC LLP Community~\cite{Alimena:2019zri}. Recasting all the existing experimental bounds into our theoretical parameter space is beyond the scope of this work. An idea of their effectiveness can be however obtained by considering the searches performed by ATLAS for displaced hadronic jets in the muon spectrometer (MS)~\cite{Aaboud:2018aqj} or in both the MS and the inner detector (ID)~\cite{Aad:2019xav}. These are particularly optimized since they make use of dedicated trigger and vertex algorithms to analyze jets in the MS, with relatively low thresholds. Among the search strategies pursued in Refs.~\cite{Aaboud:2018aqj, Aad:2019xav}, the simplest ones require no additional prompt decays and are inclusive of any other activity in the event. Specifically, we will make use of the analysis in Ref.~\cite{Aaboud:2018aqj} that searches for events with at least two displaced hadronic vertices in the MS, and the analysis of Ref.~\cite{Aad:2019xav} where events with (at least) one decay in the MS and one in the ID are selected. We model the probability that a given event gives rise to such signatures as explained in Appendix~\ref{sec:probabilities}, and assume an overall efficiency for triggering and reconstructing an event equal to $0.01$. The bounds obtained for $Z$ and Higgs resonant production are shown respectively in Figs.~\ref{fig:resonantHiggs} and~\ref{fig:resonantZ} (dashed contours). They are stronger than those set on the rate of DS events by missing-energy searches by $2-3$ orders of magnitude, and can probe branching ratios into DS states of order $10^{-6}$ for the $Z$ and a few$\,\times 10^{-4}$ for the Higgs boson. 

Other searches for long-lived particles performed by ATLAS and CMS typically require extra prompt activity or missing energy in addition to the displaced vertices. Since the request of prompt and energetic SM particles reduces the production rate, these analyses are naively expected to be less effective in constraining the dark sector theories considered in this work. A possible important exception is the case where the DM is part of the DS and produced together with the LDSP. Events of this kind always contain (possibly a large amount of) missing energy, which can be used to trigger the event and reduce the background. It would be therefore interesting to assess the constraints imposed on elusive dark sectors by searches that require displaced vertices in association with large missing energy, like those with displaced photons or jets. We leave this study to a future work. Finally, searches for long-lived particles at LHCb also make use of dedicated triggers for displaced vertices. These are however required to be inside the tracker, i.e.~within a distance of $200\,$mm ($30\,$mm) from the primary vertex in the beam (transverse) direction. This limits the sensitivity to short LDSP lifetimes. Considering that typically hard kinematic cuts are imposed to reduce the background and that LHCb has a smaller integrated luminosity than ATLAS and CMS, the effectiveness of these analyses is expected to be smaller than that of the searches considered above.

\subsection{Non-resonant DS production at LEP and LHC}
\label{sec:non-resonant}

\begin{figure}[t]
\centering
\includegraphics[width=0.48\textwidth]{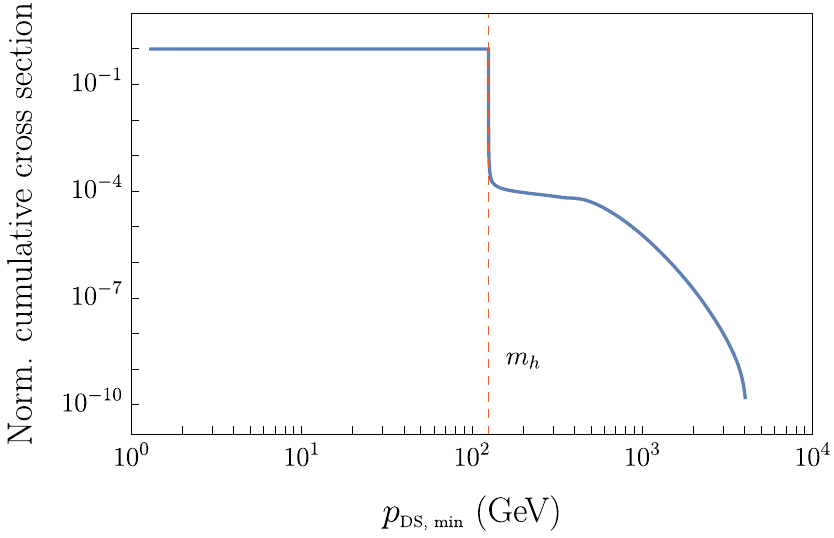}
\hspace{0.2cm}
\includegraphics[width=0.48\textwidth]{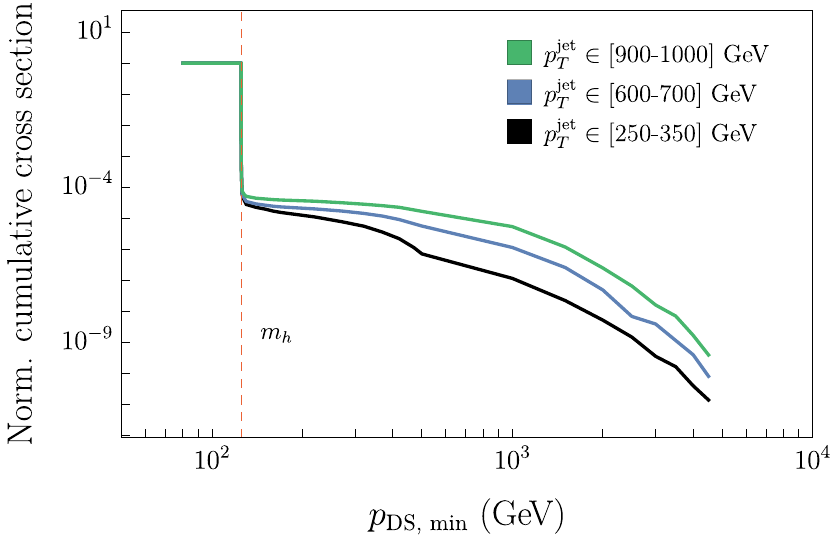}
\caption{\small Normalized cumulative cross section for $pp\to DS$ (left plot) and $pp\to DS + j$ (right plot) at the $13\,$TeV LHC for a $D=6$ Higgs portal, shown as a function of the lower limit of integration over the invariant mass of the DS system (obtained by analytic calculation of matrix elements, convolved with the PDFs). In the right plot three curves are shown corresponding to events in three different bins of the jet transverse momentum.}
\label{fig:resonantvsnonresonant}
\end{figure}

In presence of (unsuppressed) ${\cal O} H^\dagger H$ and $J^\mu_{DS} H^\dagger i \!\overleftrightarrow{D}_{\!\!\mu} H$ portals, and for $\Lir$ below the electroweak scale, the strongest constraints on the dark sector come from $Z$ and Higgs decays, as discussed in the previous subsection. Values of $\Lir$ larger than the electroweak scale are more difficult to probe since in that case the production of DS states is non-resonant and has a smaller rate. The relative importance of resonant vs non-resonant DS production at the LHC is illustrated in Fig.~\ref{fig:resonantvsnonresonant} for a $D=6$ Higgs portal. 
%
%
The cumulative cross section for the processes $pp\to DS$ and $pp\to DS+j$ drops sharply when the lower limit of integration over the invariant mass of the DS system is raised above the Higgs mass threshold. Non-resonant production is thus expected to give weaker bounds than those from resonant processes. Furthermore, for $\Lir$ above the electroweak scale and not too large $\Luv$, the LDSP decays promptly. As already discussed, prompt decays at colliders give model-dependent signatures whose analysis is beyond the scope of this work. On the other hand, the DS could interact with the SM through portals different than ${\cal O} H^\dagger H$ and $J^\mu_{DS} H^\dagger i \!\overleftrightarrow{D}_{\!\!\mu} H$ (alternatively, these latter could be generated with a suppressed coefficient). In this case, for any value of $\Lir$, one needs to analyze non-resonant processes to assess the current bounds on the DS dynamics.

\begin{figure}[t]
\centering
\includegraphics[width=0.48\textwidth]{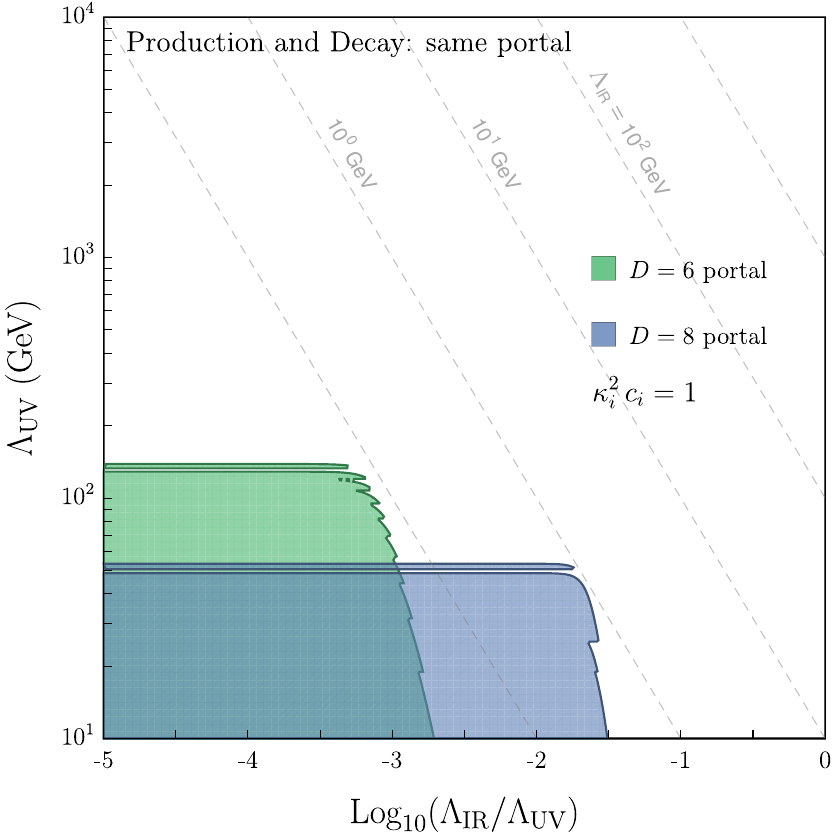}
\hspace{0.2cm}
\includegraphics[width=0.48\textwidth]{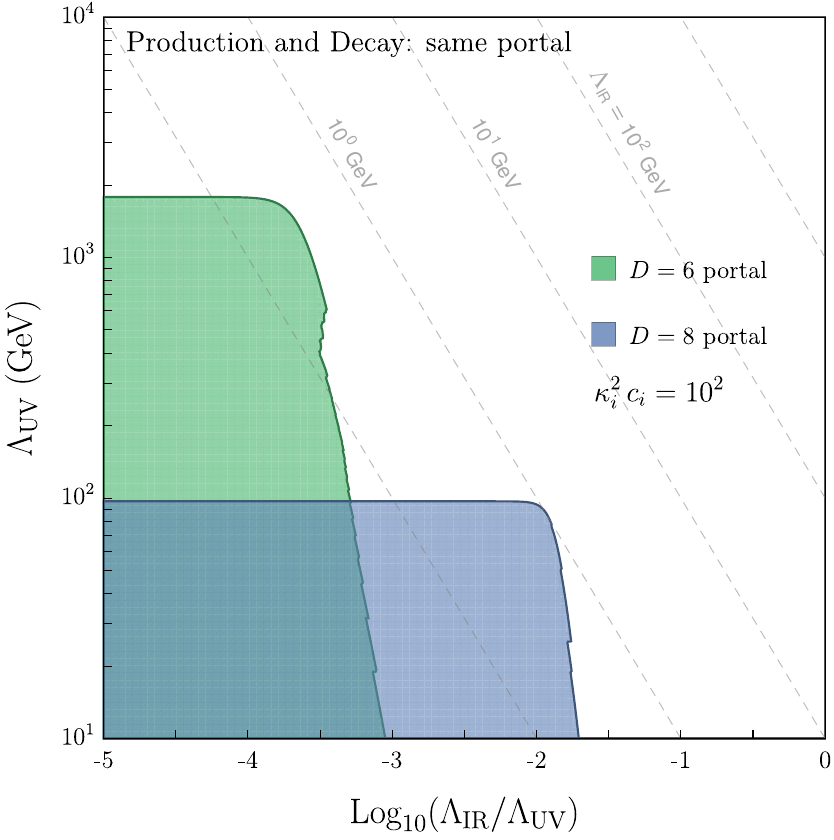}
\caption{\small Constraints on non-resonant DS production from mono-photon searches at LEP. Two choices of portals are shown: $J_\mu^{DS} \bar e\gamma^\mu e$ ($D=6$) and $T_{\mu\nu}^{DS} (F_\alpha^\mu F^{\alpha\nu} + \bar e \gamma^\mu D^\nu e)$ ($D=8$). It is assumed that the same portal is responsible for both the DS production and the LDSP decay, and that the dark sector is strongly coupled. The value of $\kappa_i^2 c_i$ is fixed to $1$ in the plot on the left and to $10^2$ in the plot on the right. 
}
\label{fig:nonresonant}
\end{figure}

Assuming a non-resonant DS production, for small $\Lir$ the LDSP decays outside the detector and the constraints are set by missing-energy searches. We have analyzed the mono-X searches performed at LEP2~\cite{Achard:2003tx, Achard:2004cf, Abbiendi:2000hh} and LHC~\cite{Aaboud:2017phn,Aaboud:2017dor,Aaboud:2017bja, Aad:2015zva} discussed previously for $Z$ decays. Dark sector production proceeds through the prototype diagram on the left of Fig.~\ref{fig:directproduction}, where the SM tagging particle can be an electron, photon, $Z$ boson or a jet stemming from a quark or gluon.\footnote{The corresponding Feynman rules have been generated with {\ttfamily{FeynRules}}~2.3~\cite{Degrande:2011ua,Alloul:2013bka}, using a model file based on~\cite{Das:2016pbk}. The squared matrix elements have been computed with {\ttfamily{FeynArts}}~3.10 \cite{Hahn:2000kx}.}~Their yield has been computed in bins of missing momentum by assigning each event a weight given by Eq.~(\ref{eq:PmissE}). For each data set, we make use of different combinations of bins in missing energy in order to increase the EFT validity, as explained in Sec.~\ref{sec:openproduction}. 
We find that the strongest bounds come from mono-photon searches at LEP~\cite{Achard:2003tx,Abbiendi:2000hh}, while the impact of LHC searches is limited by the request of the EFT validity, since the corresponding analyses make use of events at higher energies or invariant masses. Figure~\ref{fig:nonresonant} shows the exclusion regions that we have obtained from LEP data 
for the following two portals involving electrons and photons: $J_\mu^{DS} \bar e\gamma^\mu e$ ($D=6$) and $T_{\mu\nu}^{DS} (F_\alpha^\mu F^{\alpha\nu} + \bar e \gamma^\mu D^\nu e)$ ($D=8$).
%
%
As expected, the constraints are much weaker than those from resonant DS production if the comparison is done for the same value of $\kappa_i^2 c_i$. 

For large enough $\Lir$, the LDSP can give rise to displaced decays inside the detector. As for the case of resonant DS production, we focused on the searches for displaced jets made by ATLAS in Refs.~\cite{Aaboud:2018aqj, Aad:2019xav}, and computed the signal yield by assigning each event a weight through the probabilities reported in Appendix~\ref{sec:probabilities}. We find that no bound compatible with the validity of the effective field theory can be set in this case unless $c_i \kappa_i^2$ has a very large value, $c_i \kappa_i^2 \gtrsim 10^3$.

\subsection{Constraints from High-Intensity Experiments}
\label{subsec:FixTargetConstraints}

Dark sectors with sufficiently low IR scale can be probed by high-intensity experiments operating at center-of-mass energies smaller than those reached at modern high-energy colliders. In this case the strategy is that of producing the DS particles by pushing the intensity, rather than the energy, frontier. Simple dimensional analysis suggests that this approach can probe most effectively dark sectors that couple through relevant or marginal portals~\cite{Batell:2009di}. As a prototype of high-intensity experiments consider those where an intense proton or electron beam hits a fixed target or a beam dump. Dark sector particles can be produced directly in the hard scattering between the incident beam particle and the target, or originate from the decay of QCD hadrons produced in the collision. The cross section for direct DS production naively scales as $\sigma \sim (c\kappa^2/E^2) (E/\Luv)^{2(D-4)}$, where $E$ is the beam energy. Then, a very naive estimate of the ratio of the numbers of DS events produced at a collider and a fixed-target experiment is~\cite{Batell:2009di}
\begin{equation}
\label{eq:colliderVStarget}
\frac{N_{collider}}{N_{target}} = \frac{\sigma_{collider} \, {\cal L}_{collider}}{\sigma_{target}\, {\cal L}_{target}} \sim
10^{-3} \left(\frac{E_{collider}}{E_{target}}\right)^{2D-10} \left(\frac{{\cal L}_{collider}}{100\,\text{fb}^{-1}}\right) \left(\frac{10^{20}}{N_\text{POT}}\right)\, ,
\end{equation}
where the integrated luminosity at the fixed target experiment, ${\cal L}_{target} = N_\text{POT} \ell \rho$, depends on the total number of incident particles (protons or electrons) delivered on target, $N_\text{POT}$, the length $\ell$ of the target and its atomic density $\rho$. To derive Eq.~(\ref{eq:colliderVStarget}) we have assumed $\ell = 10\,$cm and $\rho = 10^{23}\,\text{cm}^{-3}$. This estimate suggests that portals with $D \leq 5$ can be effectively probed at fixed-target experiments with high luminosity, while high-energy colliders are parametrically more efficient for $D>5$. 
Clearly, a quantitatively more accurate estimate should take into account the effect of the parton distribution functions at hadron colliders, the finite mass of the target nucleus in fixed-target experiments, as well as the geometric acceptance of the detector in each case. However, the qualitative conclusion that can be drawn from Eq.~(\ref{eq:colliderVStarget}), i.e. that direct DS production through higher-dimensional portals can be best probed by pushing the energy frontier, is generally correct and in agreement with the results of our analysis reported in this section.
An estimate similar to (\ref{eq:colliderVStarget}) can be derived to compare the rates of DS particles produced in the decay of QCD hadrons at colliders and fixed target experiments. Such rate scales naively as $\sim \sigma_{incl}(E) c\kappa^2 (M/\Luv)^{2(D-4)}$, where $M$ is the mass of the decaying hadron and $\sigma_{incl} $ is an inclusive QCD cross section. The relatively mild increase of the latter with the c.o.m. energy (see for example Ref.~\cite{Pancheri:2016yel}) is not sufficient to make colliders competitive with high-intensity experiments in this case. Decays of QCD hadrons to DS particles will be thus most effectively probed by dedicated low-energy experiments with large integrated luminosity.

In this section we will study the sensitivity that high-intensity experiments have on elusive dark sectors analyzing both of the possible production modes. Let us consider first the production that occurs in the hard scattering between an intense proton or electron beam and a fixed target. 

\subsubsection*{Direct DS production from the hard scattering}

There are two broad experimental strategies that have been adopted to detect the DS particles. A first class of experiments makes use of a shield or active detector regions to block or veto any particle emerging from the collision, with the exception of neutrinos and DS states. These can reach a detector placed downstream of the shield where they decay in flight to SM states or scatter with the detector material. Neutrino experiments, such as CHARM~\cite{Bergsma:1984ff}, LSND~\cite{Athanassopoulos:1996ds}, NuTeV~\cite{Adams:2001ska}, MINOS~\cite{Adamson:2007gu} and MiniBooNE~\cite{Aguilar-Arevalo:2017mqx}, belong to this class. They utilize very intense proton beams (with up to $10^{20} - 10^{23}$ protons delivered on target) and may include a decay volume where neutrinos are produced by the in-flight decay of pions and kaons. Other experiments, such as E137 and E141 at SLAC~\cite{Bjorken:1988as,Riordan:1987aw} and E774 at Fermilab~\cite{Bross:1989mp}, utilized an electron beam and were dedicated to the search for new long-lived neutral particles. 
A second class of experiments, such as NA64 at CERN~\cite{Andreas:2013lya} and the proposed LDMX~\cite{Akesson:2018vlm}, are designed to measure the energy (and possibly the momentum) of the electron beam before and after the collision with the target. Calorimetry is then used to veto any significant hadronic activity following the collision. Long-lived dark sector particles can either decay outside the detector and thus give rise to events with missing energy or momentum, or lead to displaced decays inside the detector. 

A complete analysis of all these experiments is clearly beyond the scope of this work. We will thus focus on two of them, one in the first experimental class and one in the second class, and use them to illustrate the sensitivity that fixed-target experiments have on elusive dark sectors. We will consider, in particular, theories with a $D=6$ portal of the form $J^{em}_{\mu} J^{\mu}_{DS}$, where $J_{em}^\mu = \bar e \gamma^\mu e$ is the SM electron current and $J^{\mu}_{DS}$ is a DS current.

\begin{figure}[t]
\centering
\includegraphics[width=0.7\textwidth]{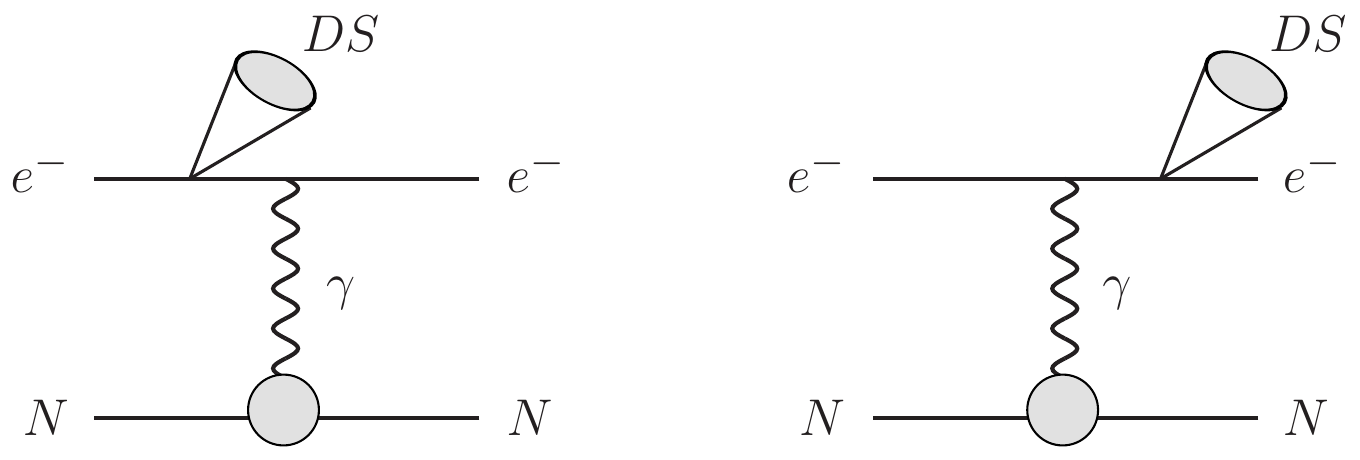} 
\caption{\small Feynman diagrams characterizing direct DS production at fixed target experiments with electron beams like NA64 and E137. The DS particles can be radiated off the initial or final electron line through the $D=6$ portal $(\bar e \gamma^\mu e) J^{\mu}_{DS}$.}
\label{fig:DSbremsstrahlung}
\end{figure}

\begin{figure}[tp]
\centering
\includegraphics[width=0.65\textwidth]{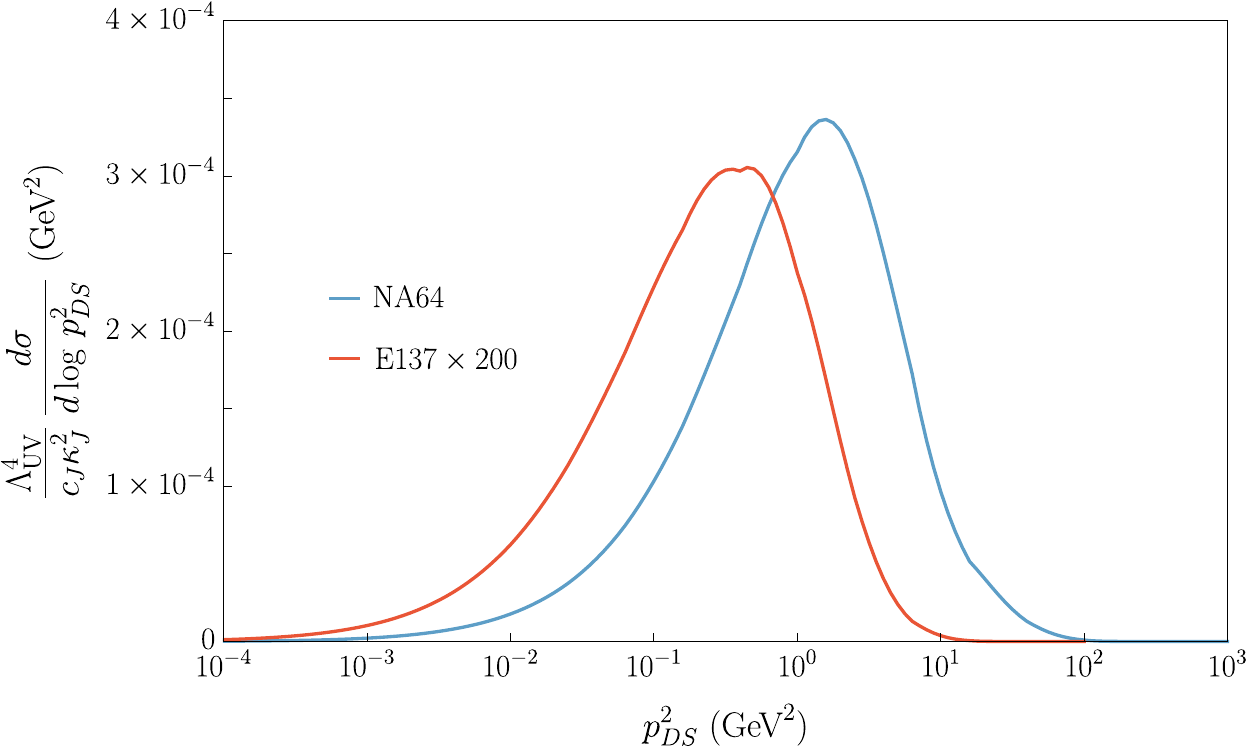} 
\caption{\small Differential cross section for DS production as a function of the DS invariant mass squared at NA64 and E137. The NA64 curve is obtained by imposing the cut $E_\text{miss} \geq 50\,$GeV.}
\label{fig:diffxsec-NA64-E137}
\end{figure}

Among the experiments that can search for missing energy we consider NA64. It features a high-intensity electron beam with energy $E_0 = 100\,$GeV  hitting an active lead target (the ECAL). Dark sector excitations can be emitted through dark bremsstrahlung in the scattering of the incident electron with the target nucleus, see Fig.~\ref{fig:DSbremsstrahlung}, and decay outside the detector (the HCAL) if sufficiently long lived.
The HCAL itself is used to veto any hadronic activity that follows a deep inelastic scattering where the nucleus breaks apart. The analysis of Ref.~\cite{NA64:2019imj} in particular, selects events that are characterized in their final state by one electron with energy $E'$ plus missing energy $E_\text{miss} \equiv E_0 - E'$, without further activity. SM backgrounds are removed by requiring $E_\text{miss} \geq 50\,$GeV. Using a dataset corresponding to $2.84 \times 10^{11}$ electrons on target, no event is found which passes all the cuts, with an estimated background of 0.53 events. This result is interpreted to set constraints on dark photon models where the dark photon is radiated off the electron line and decays to DM particles which escape detection. These models are particular examples of a dark sector where the invariant mass of the DS system is fixed (for a small dark photon decay width) to the dark photon mass, $p_{DS}^2 = m_{\gamma_D}^2$. More in general, the DS system will consist of several particles and have arbitrary invariant mass, compatible with phase space constraints. It is convenient to reduce this general situation to the case of a dark photon with varying mass by factorizing the Lorentz invariant phase space as $d\Phi_{2+n}  = (2\pi)^{-1} dp_{DS}^2 d\Phi_{3}d\Phi^{DS}_n$. Here $d\Phi^{DS}_n$ denotes the $n$-body phase space of the DS system with total momentum $p_{DS}$; $d\Phi_{3}$ is instead the 3-body phase space obtained by replacing the entire dark sector with a single particle of momentum $p_{DS}$ and mass $p_{DS}^2$. The integration over the DS phase space can be performed easily by using the optical theorem; the result is written in terms of the imaginary part of the 2-point correlator of the DS operator $J^\mu_{DS}$:
\begin{equation}
\label{eq:NA64_cs}
\begin{split}
\sigma(e N \to e N + DS) = & \, \frac{\kappa_J^2}{\Luv^{4}} \frac{1}{4 E_0 m_N} \frac{1}{2\pi} \int \! dp_{DS}^2 \! \int \! d\Phi_3 \,
 \mathcal{M}_\mu \mathcal{M}_{\nu}^*\, G(t) \\[0.1cm]
& \times 2 \,\text{Im}\! \left[i \langle 0|T(J^\mu_{DS}(p_{DS}) J^\nu_{DS}(-p_{DS}))|0 \rangle \right]\, ,
\end{split}
\end{equation}
where ${\cal M}_\mu$ is the matrix element with one insertion of the portal interaction, and $G(t)$ is a form factor that parametrizes atomic and nuclear scatterings. Here $t = (p_N' -p_N)^2$ is the momentum transfer, and $p_N$, $p_N'$ are respectively the initial and final \mbox{4-momenta} of the nucleus $N$, whose mass is denoted by $m_N$. We set $G(t) = G_{2, el}(t) + G_{2, in}(t)$, where  $G_{2, el}(t)$ and $G_{2, in}(t)$ are respectively the elastic and inelastic contributions to the form factor, as defined by Eqs.~(A18) and (A19) of Ref.~\cite{Bjorken:2009mm}, see also Refs.~\cite{Kim:1973he,Tsai:1973py}. 
The production of a dark photon of mass $m_{\gamma_D}$ is characterized by a small emission angle $\theta_{\gamma_D} \lesssim \max[(m_{\gamma_D}m_e/E_0^2)^{1/2},(m_{\gamma_D}/E_0)^{3/2}]$ and by a spectrum of momentum transfer peaked at $t_\text{min}$, where $-t_\text{min}\approx m_{\gamma_D}^4/E_0^2$ if $m_{\gamma_D}\gtrsim m_e$~\cite{Bjorken:2009mm,Kim:1973he,Tsai:1973py}. 
Formula~(\ref{eq:NA64_cs}) applies in that case as well if one replaces $(\kappa_J/\Luv^2) J^\mu_{DS} \to (\varepsilon e) A^\mu_D$, where $A^\mu_D$ is the dark photon field and $\varepsilon$ its kinetic mixing parameter. The imaginary part of the 2-point correlator in this case gives $\pi \delta(p_{DS}^2-m_{\gamma_D}^2)\sum_i \epsilon_i^\mu \epsilon_i^{\nu *}$, where $\epsilon_i^\mu$ is the polarization vector of the dark photon. Using the imaginary part of the 2-point correlator given by Eq.~(\ref{eq:imJ}), one can thus express the DS cross section in terms of the cross section for the production of a dark photon; we obtain
\begin{equation}
\label{eq:DP-to-DS}
\frac{d\sigma}{dp_{DS}^2}(e N \to e N + DS) = \frac{\kappa_J^2 c_J}{\Luv^{4}} \frac{p_{DS}^2}{96\pi^2} \frac{\sigma(e N \to e N + A_D)}{(\varepsilon e)^2}\, ,
\end{equation}
where the dark photon cross section on the right-hand side has to be evaluated for $m_{\gamma_D} = (p_{DS}^2)^{1/2}$. Using the exact tree-level calculation of Ref.~\cite{Liu:2017htz} (see also~\cite{Gninenko:2017yus}) to compute the dark photon cross section, and performing the cut $E_\text{miss} \geq 50\,$GeV,  from Eq.~(\ref{eq:DP-to-DS}) we obtained the differential cross section shown in Fig.~\ref{fig:diffxsec-NA64-E137}.
The production of an elusive dark sector with a current-current portal at NA64 is thus equivalent to a convolution of dark photon theories with mass spectrum in the range $\sim 0.1-10\,$GeV, which corresponds to a minimum momentum transfer $-t_\text{min} \sim 10^{-8} - 1\,\text{GeV}^2$. This suggests that most of the incident electrons at NA64 scatter off the target atom or nucleus, well above the electron screening regime and below the onset of deep inelastic scattering.~\footnote{Processes with $d \lesssim -t \lesssim 4m_p^2$, where $m_p$ is the proton mass and $d = 0.164\,\text{GeV}^2 A^{-2/3}$ is the inverse nuclear size squared, are characterized by the scattering of the incident electron off the target nucleus. Scatterings off the target atom take place when $1/a^2 \lesssim -t \lesssim d$, where $a = 111\, Z^{-1/3}/m_e$ is the atomic radius, whereas for $-t \ll 1/a^2$ the atomic electrons screen the charge of the nucleus and the form factor dies off. In the opposite limit of very large momentum transfer, $-t \gg 4m_p^2$, the process occurs in the regime of deep inelastic scattering, where the incident electron scatters off the constituents quarks. In this case the final state is characterized by an intense hadronic activity. See Refs.~\cite{Kim:1973he,Tsai:1973py}.}
By integrating the differential cross section of Fig.~\ref{fig:diffxsec-NA64-E137}, we obtain the total cross section at NA64:
\begin{equation}
\sigma(e N \to e N + DS) = 0.8 \times 10^{-41} \text{cm}^2 \, \left(c_J \kappa_J^2\right) \left(\frac{500\,\text{GeV}}{\Luv}\right)^4 .
\end{equation}

\begin{figure}[tp]
\centering
\includegraphics[width=0.48\textwidth]{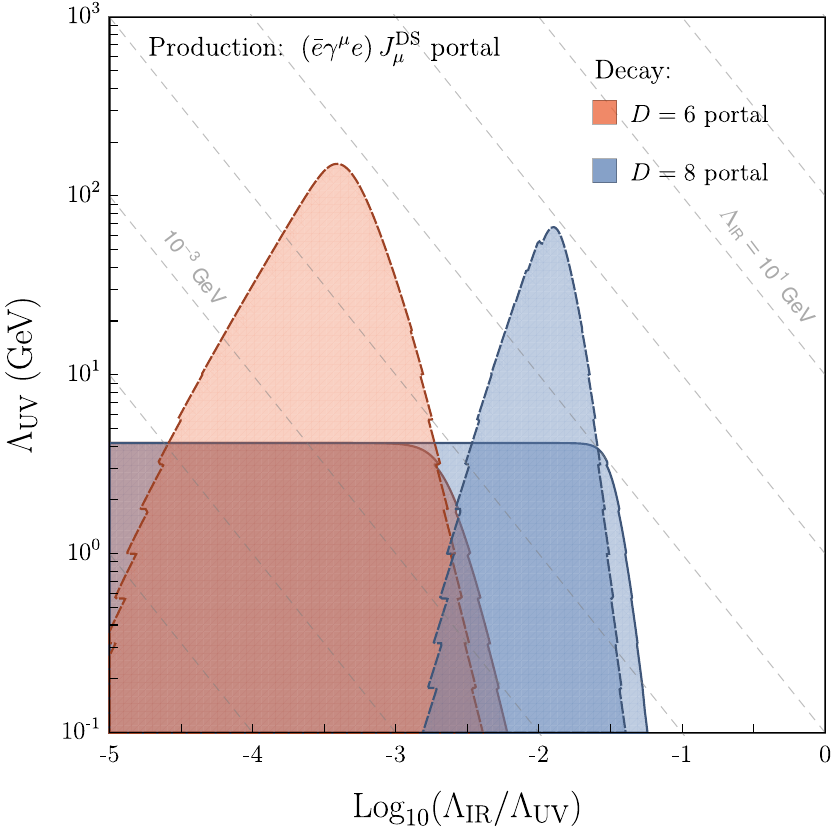} 
\caption{\small Constraints from NA64 (solid contours) and E137 (dashed contours) on elusive dark sectors with portal $(\bar e \gamma^\mu e) J_{\mu}^{DS}$. The plot assumes a strongly-coupled DS dynamics and two possible portals mediating the LDSP decay, respectively with dimension $D=6$ (blue region) or $D=8$ (red region). For both portals, $\kappa^2 c$ is set to 1.}
\label{fig:exclusionNA64+E137}
\end{figure}

Using Fig.~\ref{fig:diffxsec-NA64-E137} and assuming a total luminosity ${\cal L} = 5 \times 10^{33}\,\text{cm}^{-2}$,~\footnote{This is obtained as ${\cal L} = N_\text{EOT} \rho_\text{Pb} \ell$, where $N_\text{EOT} = 2.84 \times 10^{11}$, $\rho_\text{Pb} = 0.3 \times 10^{23}\,\text{cm}^{-3}$ and we set the thickness of the detector to 1 radiation length, $\ell = 0.56\,\text{cm}$.} we derived the bound that missing-energy searches at NA64 set on elusive dark sectors. For sufficiently low $\Lir$, all LDSPs decay outside the detector and the constraint is independent of the IR scale. In this limit we find
\begin{equation}
\Luv > 4\,\text{GeV} \times (c_J \kappa_J^2)^{1/4} \quad \text{for} \quad 
\begin{cases}
\Lir\ll  9\, \text{MeV}\left(c_J \kappa_J^2\right)^{-1/6}   &(D=6)\\[0.2cm]
\Lir\ll  120\,\text{MeV} \left(c_J \kappa_J^2\right)^{-1/10}  &(D=8)\, ,
\end{cases}
\end{equation}
where $D$ is the dimension of the decay portal. For larger values of $\Lir$, fewer DS events give rise to missing energy and the constraint gets weaker. The corresponding exclusion curve is shown in Fig.~\ref{fig:exclusionNA64+E137} (solid contours) for a strongly-coupled DS dynamics and two possible portals mediating the LDSP decay, respectively with dimension $D=6$ (blue region) or $D=8$ (red region). 
Very similar results hold for weakly-coupled dynamics. Compared to those arising from high-energy collider searches, this bound is rather weak and does not constrain values of $\Luv$ above the electroweak scale. To derive it, we implemented the procedure explained in Sec.~\ref{sec:openproduction} to enforce the EFT validity, i.e. we restricted the integration of the differential distribution of Fig.~\ref{fig:diffxsec-NA64-E137} to values below the UV scale.

A stronger bound comes from the E137 experiment performed at SLAC. The experimental setup is as follows: an incident electron beam with energy $E_0 = 20\,$GeV hits a beam dump target made of aluminium plates interlaced with cooling water. The particles produced by the collision must traverse a hill of $179\,$m in thickness before reaching a $204\,$m-long open region followed by a detector. Bounds can be placed on long-lived dark particles that decay in the open region or rescatter with the material in the detector. 
No signal events were observed after two runs during which $\sim 30\,$C of electrons (respectively $10\,$C in Run~1 and $20\,$C in Run~2, corresponding to a total of $\sim 2\times 10^{20}$ electrons) were delivered on target. An interpretation of this result in terms of dark photon theories was given in Refs.~\cite{Bjorken:2009mm,Batell:2014mga}. 
We used it to derive a bound on elusive dark sectors with portals $(\bar e \gamma^\mu e) J_{\mu}^{DS}$ as follows.
First, we computed the differential cross section for atomic and nuclear scatterings of the incident electrons off the target using Eq.~(\ref{eq:DP-to-DS}). The result is shown in Fig.~\ref{fig:diffxsec-NA64-E137}. The invariant mass spectrum peaks in the range $0.03-3\,$GeV, which corresponds to values of the minimum momentum transfer $-t_\text{min} \sim 10^{-11}-10^{-3}\,$GeV. Most of the incident electrons at E137 thus scatter off the target atom. Since no veto is imposed at E137 on the hadronic activity of the final state, we have explicitly computed the contribution of deep inelastic scatterings, finding that is small (it becomes important only at very large invariant masses $p_{DS}^2 \gtrsim 25\,\text{GeV}^2$) and safely negligible to derive the bounds described below. From Fig.~\ref{fig:diffxsec-NA64-E137} we obtain the total cross section at E137:
\begin{equation}
\sigma(eN\to eN +DS) = 0.4 \times 10^{-43} \text{cm}^2 \, \left(c_J \kappa_J^2\right) \left(\frac{500\,\text{GeV}}{\Luv}\right)^4 \, .
\end{equation}

Using the differential cross section of Fig.~\ref{fig:diffxsec-NA64-E137}, we computed the rate of LDSP decays that occur in the open region and are seen by the detector. To this aim, we estimated the geometric acceptance simply as the fraction of particles from each LDSP decay that passes through the front area of the detector. We approximated as collinear the emission of the DS excitation through bremsstrahlung (this is a reasonably good approximation for light dark photons, see for example Ref.~\cite{Bjorken:2009mm}), and assumed an isotropic distribution of the decay products from the LDSP decay in its center-of-mass frame. Finally, we have used an integrated luminosity ${\cal L} = 3.4\times 10^{43}\,\text{cm}^{-2}$ for Run1 and twice as much for Run2.~\footnote{Here we have used $\rho_\text{Al} = 0.6 \times 10^{23}\,\text{cm}^{-3}$ and set the thickness of the detector to 1 radiation length, $\ell = 8.9\,\text{cm}$.} The exclusion region that we obtained is shown in Fig.~\ref{fig:exclusionNA64+E137} for $D=6$ and $D=8$ portals mediating the LDSP decay. In the relevant range of hierarchies, the exclusion on $\Luv$ extends up to $\sim 150\,$GeV and is much stronger than the one set by NA64, despite the smaller cross section, thanks to the vastly larger number of electrons delivered on target.

\subsection*{DS production from hadron decays}

The other way to produce DS particles at high-intensity experiments is through the decay of QCD hadrons. To achieve a good sensitivity on elusive dark sectors, very large samples of hadron decays are needed. These are obtained at experiments with particularly intense proton beams and at experiments dedicated to the study of rare decays. 
One can broadly identify two classes of decays: those where a parent QCD hadron annihilates into DS excitations, possibly emitting an additional photon (annihilation decays), and those where it decays to a lighter hadron plus DS excitations (radiative decays). We will assume for simplicity that the portal interaction conserves baryon number and flavor. One can thus further distinguish between flavor-conserving and flavor-violating decays; these latter proceed necessarily through a flavor-violating SM loop and are correspondingly suppressed.

Annihilation decays are mediated by portals whose SM operator has the appropriate quantum numbers to excite the parent meson from the vacuum, in particular by $J_\mu^{SM} J^\mu_{DS}$ portals where $J_\mu^{SM}$ is a vector or an axial quark current. Decays of interest are for example those of light unflavored pseudoscalar or vector mesons ($\pi^0$, $\eta$, $\eta'$, $\rho$, $\omega$, $\phi$, etc.), as well as those of flavored mesons ($K_L$, $D$ and $B$). These processes have been considered in previous studies and used to constrain specific dark sectors whose excitations are either long lived and escape detection, see for example Refs.~\cite{McElrath:2007sa,Badin:2010uh,Gninenko:2015mea,Barducci:2018rlx,Darme:2020ral}, or promptly decay back (at least partly) to the SM, see for example Ref.~\cite{Hostert:2020gou}. See also Ref.~\cite{Kamenik:2011vy} for a model-independent approach. To give an idea of how precisely one can probe elusive dark sectors through annihilation decays, we consider the decay of light vector mesons and assume that the DS excitations are sufficiently long lived to escape detection. In the case of the portal $J_\mu^{em} J^\mu_{DS}$, where $J_\mu^{em}$ is the SM electromagnetic current, we find that
\begin{equation}
\frac{BR(V\to DS)}{BR(V\to e^+ e^-)} = \frac{c_J\kappa_J^2}{128\pi^2} \frac{1}{\alpha_{em}^2} \frac{m_V^4}{\Luv^4}\, ,
\end{equation}
where $V$ denotes a light unflavored vector meson, and we used the optical theorem to compute the phase space integral over the DS system. The invisible decay of $V=\phi,\omega$ has been searched for by the BESSIII Collaboration~\cite{Ablikim:2018liz} through $J/\psi \to V \eta$. Using a sample of $1.3 \times 10^9$ $J/\psi$ events, they obtained $BR(\phi \to \text{invisible}) < 1.7 \times 10^{-4}$ and $BR(\omega \to \text{invisible}) < 7.3 \times 10^{-5}$ at 90\% confidence level.  The upper limit on $BR(\phi \to \text{invisible})$ implies
\begin{equation}
\Luv > 2.3\,\text{GeV} \left(c_J \kappa_J^2\right)^{1/4} \qquad \text{for } \Lir \lesssim 6\,\text{MeV} \left( \kappa_J^2 c_J\right)^{-0.18}\, ,
\end{equation}
where the bound on $\Lir$ assumes a strongly-coupled dark sector and ensures that the LDSPs are long lived and escape detection. The limit on $BR(\omega\to\text{invisible})$ gives a slightly weaker constraint. The decays of pseudoscalar mesons can also be used to probe elusive dark sectors, although their rate vanishes for $J_\mu^{SM} J^\mu_{DS}$ portals where the DS current is conserved. In the case of partially conserved DS currents, the decay rate depends on the scale of explicit breaking of the associated global symmetry and receives a contribution only from values of the DS invariant mass below the onset of the conformal regime. Its estimate is thus model dependent and will not be pursued here.

Radiative decays are also interesting and are tested by various experiments. For example, experiments with very intense proton beams such as LSND and MINOS are particularly suited to probe flavor-conserving decays of light mesons and baryons, such as: $\rho \to \pi + DS$, $K^* \to K + DS$, $\Delta\to N+DS$, etc. All these decays are expected to occur, for example, through $J_\mu^{SM} J^\mu_{DS}$ portals where $J_\mu^{SM}$ is a quark vector current. This strategy has been applied for example in Refs.~\cite{Batell:2009di,Darme:2020ral} to set constraints on dark sectors. Flavor-changing meson decays can be best probed, instead, at dedicated experiments. Here we focus, in particular, on the decays \mbox{$B^+\to K^+ + DS$} and \mbox{$K^+\to \pi^+ + DS$}, where the DS particles decay outside the detector and thus lead to missing energy. See for example Refs.~\cite{Bird:2004ts,Bird:2006jd,Darme:2020ral,Kamenik:2011vy} for previous related studies of this kind of processes.

\begin{figure}[t]
\centering
\includegraphics[width=0.90\textwidth]{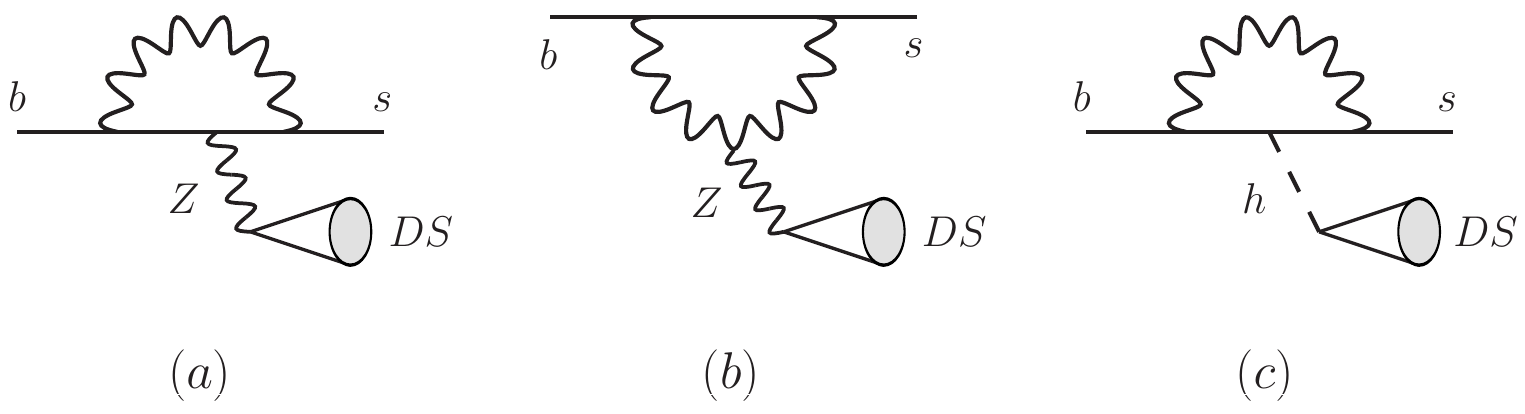} 
\caption{\small Feynman diagrams contributing to $B^+ \to K^+ + DS$ in theories with a $Z$-portal (diagrams $(a)$ and $(b)$) and a Higgs portal (diagram $(c)$).}
\label{fig:mesondecay}
\end{figure}

The decay $B^+\to K^+ + DS$ can be mediated by a $J_\mu^{SM} J^\mu_{DS}$ portal where $J^\mu_{SM} = \bar t\gamma^\mu t$ or $iH^\dagger \overleftrightarrow{D^\mu} H$. In the case of a $Z$ portal, for example, the transition occurs via the \mbox{$Z$-penguin} diagrams of Fig.~\ref{fig:mesondecay}a-b, in analogy with the decay $B^+\to K^+ + \nu\bar\nu$ in the SM. 
In fact, the neutrinos themselves behave as a dark sector with very low mass scale (hence conformal at energies of order of the $B$ mass), which couples to the $Z$ through a conserved current: $(g/2\cos\theta_W) Z_\mu J^\mu_{(\nu)}$, where $J^\mu_{(\nu)}$ is the neutrino current. Our elusive dark sector couples through the $Z$ portal in very much the same way: $(m_Z v \kappa_J/\Luv^2) Z_\mu J^\mu_{DS}+\dots$. The decay rate of $B^+\to K^+ + DS$ can be thus computed by adapting the SM calculation of $B^+\to K^+ + \nu\bar\nu$ (see~\cite{Buchalla:2008jp} and references therein) by simply replacing the neutrino system with the DS one and omitting the box diagrams. From the upper limit $BR(B^+\to K^+ + \nu\bar\nu) < 3.7\times 10^{-5}$, obtained by the \textsc{BaBar} collaboration with a dataset of $\sim 10^8$ $B\bar B$ pairs~\cite{Lees:2013kla}, we find the constraint
\begin{align}
\Luv > 83\,\text{GeV} \left(c_J \, \kappa_J^2\right)^{1/4}  \quad\text{for}\quad \Lir \ll 90\,\text{MeV} \,(c_J \kappa_J^2)^{-0.17} \, .
\end{align}

The decay $K^+ \to \pi^+ + DS$ can be also used to constrain Higgs and current-current portals. In the case of the $Z$ portal, the transition occurs via penguin diagrams as in Fig.~\ref{fig:mesondecay}a-b, where both the top and charm quarks can circulate in the loop. The rate can be computed by adapting the SM calculation for $K^+\to \pi^+ + \nu\bar\nu$ (see Ref.~\cite{Buras:1998raa}), as discussed above for $B^+\to K^++DS$. We then use the upper limit $\text{BR}(K^+ \to \pi^+ + \nu\bar{\nu})<1.73\times 10^{-10}$, set by the E949 collaboration~\cite{PhysRevLett.101.191802} from a sample of $\sim 10^{12}$ $K^+$ decays, to constrain the $Z$ portal. We find:
\begin{equation}
\Luv > 80\,\text{GeV} \left(c_J \kappa_J^2 \right)^{1/4}\quad\text{for}\quad \Lir \ll 80\,\text{MeV} \,(c_J \kappa_J^2)^{-0.18}\, .
\end{equation}

The decay $B^+\to K^+ + DS$ can also proceed through the Higgs portal, as shown in Fig.~\ref{fig:mesondecay}c, via a loop with the top quark. The transition $b \to hs$ has been calculated in Refs.~\cite{PhysRevD.26.3086,Bird:2006jd} and expressed in terms of an effective coupling
\begin{equation}
C_{bs} \, \bar{s}_L b_R h + \text{h.c.}\, ,  \quad\qquad  C_{bs} = \frac{3 g_2^2 m_b m_t^2 V_{ts}^* V_{tb}}{64 \pi^2 m_W^2 v} \simeq 5.9 \times 10^{-6}\, .
\end{equation}
Using this result and the optical theorem it is straightforward to compute the decay rate into DS excitations; we find:
\begin{equation}
\label{eq:FCNC_phase_space}
\begin{split}
\Gamma(B^+ \to K^+ + DS) = & \, \frac{1}{2 M_B} \frac{\kappaO^2}{m_h^4}\,
\frac{v^2}{\Luv^{2\Delta_\mathcal{O}-4}} \int \frac{d^3 p_K}{(2\pi)^3} \frac{1}{2 E_K} |\mathcal{M}(B^+ \to K^+ h)|^2 \\[0.2cm]
& \times \, 2\,\text{Im}\, \langle\mathcal{O}(p_{DS})\mathcal{O}(-p_{DS})\rangle \, ,
\end{split}
\end{equation}
where $E_K = \sqrt{M_{K^+}^2 + |\vec{p}_K|^2}$, and $p_{DS} = p_B - p_K$. The matrix element $\mathcal{M}(B^+ \to K^+ h)$ is given by~\cite{Bailey:2015dka}
\begin{align}
|\mathcal{M}(B^+& \to K^+ h)|^2 =  |C_{bs}|^2 \left|f_0^{K}(p_{DS}^2)\right|^2 \left(\frac{M_B^2-M_{K^+}^2}{m_b-m_s}\right)^2\, ,
\end{align}
and we use the form factor reported in Ref.~\cite{Ball:2004ye}: $f_0^{K}(q^2)=0.33\, [1-q^2/(37.46\text{\,GeV})]^{-1}$. We then approximate the imaginary part of the DS correlator with its conformal limit given by Eq.~(\ref{eq:imO}), and use the experimental upper limit on $BR(B^+\to K^+ + \nu\bar\nu)$ to set constraints on the Higgs portal. We find:
\begin{equation}
\label{eq:BtoKHiggsportal}
\begin{aligned}
\Delta_\mathcal{O}  = 4: &\quad \Luv > 1.3\,\text{GeV} \left(\cO \kappaO^2\right)^{1/4} && \quad\text{for}\quad \Lir \ll 800\,\text{MeV} \,(\cO \kappaO^2)^{-0.1} 
\\[0.2cm]
\Delta_\mathcal{O}  = 3: &\quad  \Luv > 2.1\,\text{GeV} \left(\cO \kappaO^2\right)^{1/2} &&\quad \text{for}\quad \Lir \ll 750\,\text{MeV} \,(\cO \kappaO^2)^{-0.05}\, .
\end{aligned}
\end{equation}
The upper limit on $\Lir$ ensures that the LDSPs decay outside the detector and has been derived assuming a strongly-coupled dark dynamics. Notice that these results are compatible with the definition of the Higgs portal, and as such are consistent, only if the lower limit on $\Luv$ is larger than the EW scale. This would require $\cO \kappaO^2\sim 10^8 \, (10^4)$ for $\DeltaO =4$ ($\DeltaO =3$), values that are at least implausible to obtain from realistic UV completions.

\subsection{Celestial constraints}
\label{subsec:CelestialConstraints}

The presence of a dark sector can significantly impact the dynamics of stellar objects and astronomical events. In the case of axions or axion-like particles, the two largest effects on stellar evolution were found to be an accelerated energy loss of red giants before helium ignition, and a modified lifetime of horizontal branch stars~\cite{Raffelt:1999tx,PhysRevD.51.1495}. Another celestial signature can be a change of the energy loss in supernovae (SNe), if the DS particles are able to escape from the core.

Ample research on these phenomena has been performed in the literature, in particular on axion emission in stellar and SNe observations. Based on this groundwork, various studies have extended the phenomenology to models of dark photons and four-fermion portal interactions~\cite{An:2013yfc,Hardy:2016kme,Chang:2016ntp,Chang:2018rso,Zhang:2014wra,Dreiner:2013mua,Dreiner:2013tja,DeRocco:2019njg,DeRocco:2019jti}. Closely related to our case is the study performed by Freitas and Wyler in Ref.~\cite{Freitas:2007ip}, where an unparticle dark sector has been probed, much akin to our $D=6$ current portal. We will therefore be able to adapt the results found by these authors to our most relevant case, i.e.~the $J_\mu^{SM} J^\mu_{DS}$ portal, where $J_\mu^{SM}$ is a current of SM fermions.~\footnote{Notice that the relative size of terms in the unparticle vector propagator proposed in \cite{Georgi:2007ek,Georgi:2007si} and employed in \cite{Freitas:2007ip} needs to be corrected by a factor which depends on the operator dimension $\Delta$~\cite{Grinstein:2008qk}. For a $J_\mu^{SM} J^\mu_{DS}$ portal, $\Delta= 3$ and the correct relative size of the terms in the propagator agrees with the one used in \cite{Freitas:2007ip}. We are thus left with a different overall normalization factor, which we have chosen by defining $\left<J^\mu_{DS}\, J^\nu_{DS}\right>$ as in Eq.~\eqref{eq:2ptJ}. In practice, our normalization yields a multiplicative factor $16 \pi^2/3$ with respect to the results of Ref.~\cite{Freitas:2007ip}.}
For this portal, we will obtain bounds from the observations of SN1978A and horizontal branch stars. In the case of SN1978A, the bound will be based on the `Raffelt criterion' of energy loss (see Eq.~\eqref{eq:SN_QD}), which states that any new particle species should not lead to an energy loss in the SN progenitor which is more efficient than that of neutrinos. We point out that recent studies  have complemented this strategy by looking for DM produced in the SN cooling process through direct detection experiments~\cite{DeRocco:2019jti} and gamma-ray burst observatories~\cite{DeRocco:2019njg}. For simplicity, however, here we focus on the energy loss argument.

\subsubsection{SN1987A}
\label{subsec:SN1978A}

The impact of an additional conformal sector on the observation of the supernova SN1978A is a shortening of the neutrino burst.~\footnote{This method of constraining new physics through SN1978A relies on the modelling of the supernova by a core collapse and a neutrino-driven supernova explosion. Under other assumptions, no such bound is found~\cite{Blum:2016afe,Bar:2019ifz}.} Using the results derived in~\cite{Freitas:2007ip}, we will make a quantitative estimate on the DS emission rate. We will then compare with the bound on the energy loss rate $Q_\text{SN}$ derived in Refs.~\cite{Raffelt:1999tx,Hannestad:2007ys}.~\footnote{An improved analysis takes into account the profile of the collapsing star \cite{Lee:2018lcj}.}

Due to the high concentration of nucleons in the supernova core, the dominant process for energy loss is the production of DS excitations through the scattering of nucleons. As argued in~Refs.~\cite{Freitas:2007ip,PhysRevD.33.897}, the main contribution is given by the scattering $n \, n \to n \, n + DS$, as other channels are smaller in comparison: $p \, p \to p \, p + DS$ is suppressed due to lower proton density, $e \, n \to e \, n + DS$ and $e \, e \to e \, e+DS$ are negligible due to Coulomb screening effects in the supernova core plasma~\cite{Freitas:2007ip,PhysRevD.33.897}. Therefore, we only consider DS emission in the scattering of neutrons as the leading effect. 

Since the SN temperature is much smaller than the neutron mass, $T_\text{SN}\approx 30$~MeV, the scattering occurs non-relativistically, and the DS emission is a soft one. The dominant contribution thus turns out to be DS bremsstrahlung off a neutron leg, whose rate can be factorized into that for a neutron-neutron hard scattering times the probability for soft radiation. Consider for example the diagram 
\begin{center}
  \begin{tikzpicture}[x=2mm,y=2mm,baseline]
    	\node (phi1) at (-12,3) {$n \,(p_1)$};
    	\node (phi2) at (-12,-3) {$n \,(p_2)$};
    	\node (phi4) at (6,3) {$n \,(k_1)$};
    	\node (phi5) at (6,-3) {$n \,(k_2)$};
    	\node (phi6) at (4,8.5) {$DS \,(p_{DS})$};
        \draw[fermion] (phi1) -- (-3,3) node[midway, above] {$ $} node (V1) {\Huge.};
        \draw[fermion] (-3,3) -- (phi4) ;        
        \draw[fermion] (phi2) -- (-3,-3) node[midway, below] {$ $};
        \draw[fermion] (-3,-3) -- (phi5) node[midway, below] {$ $};   
        \draw[fermion] (0,3) -- (3.5,5.5) node[midway, below] {$ $}; 
        \draw[fermion] (0,3) -- (1.5,6.5) node[midway, below] {$ $}; 
        \draw [color=black, fill=gray!40, rotate around={65:(2.5,6)}] (2.5,6) ellipse (.15cm and .225cm);
        \draw [fill,white] (-3,0) ellipse (.25cm and .7cm);
        \draw [pattern=north west lines, pattern color=black] (-3,0) ellipse (.25cm and .7cm);
        \node (phi8) at (-5.8,0) {$\mathcal{A}$};
  \end{tikzpicture}
\end{center} 
\vspace{0.4cm}
the amplitude for which can be written as 
\begin{align}
 i \mathcal{M} =  i\frac{\kappa_J}{\Luv^2}  \big[ \bar{u}(k_2)\bar{u}(k_1) \gamma_\mu  \frac{i(\slashed{q}+m_n)}{q^2-m_n^2} (-i \mathcal{A}) \,u(p_1)u(p_2) \big]  \,\langle DS|J_{DS}^{\mu}(p_{DS})|0 \rangle \, ,
\end{align}
where $q=p_{DS}+k_1$ and $\mathcal{A}$ is defined such that
\begin{equation}
\mathcal{M}_{nn\to nn} \equiv \bar{u}(q)\bar{u}(k_2)\, \mathcal{A}\, u(p_1) u(p_2)
\end{equation}
corresponds to the amplitude for the $2 \to 2$ on-shell scattering of neutrons. Retaining only the lowest-order terms in $p_{DS}$, the matrix element acquires the factorized form
\begin{equation}
i \mathcal{M}(nn \to nn+DS)= i   \big[ \mathcal{M}_{nn\to nn}\big] 
\frac{\kappa_J}{\Luv^2} \frac{(k_1)_\mu}{p_{DS}\cdot k_1}  \,\langle DS|J_{DS}^{\mu}(p_{DS})|0 \rangle \, .
\end{equation}
The factor $1/(p_{DS}\cdot k_1)$ from the propagator is of order $1/T_\text{SN}$ and brings in the enhancement due to the soft emission. A similar factorization holds from the other bremsstrahlung diagrams. The rate of $n \, n \to n \, n + DS$ can thus be computed, at leading order in $T_\text{SN}/m_n$, in terms of the cross section for neutron-neutron scattering, which can be extracted from nuclear data and has a value $\sigma_0(nn\to nn) \approx 25 \times 10^{-27}\,$cm$^2$ at the relevant energy~\cite{Hanhart:2000er}.

Having specified the scattering process, we define the object that will be compared to observational data: the energy loss rate~\cite{Raffelt:1996wa}
\begin{equation}
\label{eq:SN_Q}
\begin{split}
Q_{DS} = \int \! & d\Phi_{DS} \,p_{DS}^0  \prod_{i=1,2} \int \!\frac{d^3p_i}{(2\pi)^3} \frac{1}{2p^0_i}\int \!\frac{d^3k_i}{(2\pi)^3} \frac{1}{2k^0_i} \\
          &\times f_{p_1} f_{p_2} (1-f_{k_1})(1-f_{k_2}) \, \langle  |\mathcal{M}(nn\to nn + DS)|^2 \rangle\,  .
\end{split}
\end{equation}
Here $f_{p_1,p_2}$ and $f_{k_1,k_2}$ are the Maxwell-Boltzmann distributions of respectively the initial and final state neutrons,
\begin{equation}
f_p =  \frac{n_n}{2} \Big( \frac{2\pi}{m_n T_\text{SN}} \Big)^{3/2} \exp\left( -\frac{|\vec{p}|^2}{2m_nT_\text{SN}} \right)\, ,
\end{equation}
and $n_n$ denotes the neutron number density. We describe the supernova core in the non-degenerate limit, where the Pauli blocking factors are neglected, i.e.~$(1-f_{k_{1,2}})\to 1$.

The energy loss rate of SN1978A is obtained by an integration of Eq.~\eqref{eq:SN_Q} using the parameters $T_\text{SN}=30\,$MeV, $\sigma_0(nn\to nn) = 25 \times 10^{-27}\,$cm$^2$ and the neutron density $\rho_n = 3\times 10^{14}\,\,\text{g}/\text{cm}^3$. This evaluation has been performed analytically for a vector unparticle in Ref.~\cite{Freitas:2007ip}, and we adapt that result for $\Delta=3$  as
\begin{align}
Q^{\text{SN},nn}_{DS}= 2.5 \times 10^8 \,\text{MeV}^5 \left(c_J \, (\kappa_J^{nn})^2\right) \left(\frac{1 \text{ MeV}}{\Luv} \right)^4\, ,
\end{align}
where $\kappa_J^{nn}$ is the coefficient of a neutron current portal $(\bar{n}\gamma^\mu n) \, J^{DS}_\mu$.~\footnote{This can be related to the coefficient of the quark current portal $(\bar{q}\gamma^\mu q) \, J^{DS}_\mu$, we expect $\kappa_J^{nn} \approx \kappa_J^{qq}$.} This needs to be compared to the estimated bound on the energy loss in SN1987A~\cite{Raffelt:1999tx,Hannestad:2007ys},
\begin{equation}
\label{eq:SN_QD}
Q_\text{SN} \lesssim 3 \times 10^{33} \text{ erg cm$^{-3}$ s$^{-1}$}\, ,
\end{equation}
which yields the constraint
\begin{align}
\Luv \gtrsim 400\,\text{GeV} \left( c_J (\kappa_J^{nn})^2\right)^{1/4} \quad \text{for} \quad 
\Lir \ll  \min\!\left\{ T_\text{SN} , 90 \,\text{MeV}\left( c_J (\kappa_J^{nn})^2\right)^{-0.19} \right\} \, .
\end{align}
The upper limit on $\Lir$ follows from two requirements: first, the IR scale must be much smaller than the SN temperature,  $\Lir \ll T_\text{SN}$, in order to be able to describe the DS as an approximately conformal dynamics; second, DS excitations must escape the radius of the neutron core of the supernova, which we estimate from the SN mass $3 \times 10^{33}\,$g and neutron density to be $\order{10 \text{ km}}$.\footnote{We neglect reabsorption effects of the DS particles within the SN. For marginal portals, e.g.~the dark photon scenario, this effect can lead to a drastic reduction of the bounds~\cite{Chang:2016ntp,Hardy:2016kme}.}. The limit due to this second requirement has been derived for a strongly-coupled DS by assuming that the LDSP decays through the neutron current portal.

\subsubsection{Stellar evolution}
\label{subsec:HBstars}

An additional bound can be obtained from a similar calculation of the energy loss in red giants before helium ignition, which would imply a decreased lifetime of horizontal branch stars. In Ref.~\cite{Freitas:2007ip}, such a bound was derived by comparing the emission rate $Q^{\text{HB}}_{DS}$ with the energy loss rate for axions, $Q^{\text{HB}}_\text{ax} $. This latter has been used in the literature to constrain the axion--electron coupling $g_{aee}$ through a numerical simulation of the stellar evolution~\cite{Dearborn:1985gp}.

As horizontal branch stars are composed of electrons, photons, $\text{H}^+$ and $\text{He}^{2+}$ nuclei, there exist multiple processes which can radiate DS excitations. Adapting the results derived in Ref.~\cite{Freitas:2007ip}, we find that for a $J_\mu^{SM} J^\mu_{DS}$ portal, the dominant process is Compton scattering, $e + \gamma \to e+DS$, if $J_\mu^{SM}$ contains the electron current. The corresponding energy loss rate is
\begin{align}
Q^{\text{HB},C}_{DS} = 2.2 \times 10^{-24}\,\text{MeV}^5 \left( c_J \kappa_J^2 \right) \left( \frac{1\,\text{MeV}}{\Luv} \right)^{4} \,.
\end{align}
We obtain a bound on the UV scale by comparing this with the energy loss rate for axions, $Q^{\text{HB},C}_\text{ax} =  3.8 g_{aee}^2 \times 10^{-18} \text{ MeV}^5$, in combination with the most stringent bound on the axion--electron coupling $g_{aee} \lesssim 2 \times 10^{-13}$ obtained for horizontal branch stars in Refs.~\cite{Dearborn:1985gp,Raffelt:1994ry}. This amounts to a bound on the $(\bar e\gamma^\mu e) J_\mu^{DS}$ portal
\begin{align}
\Luv > 62\,\text{GeV} \left( c_J (\kappa_J^{ee})^2 \right)^{1/4} \qquad \text{for} \quad \Lir \ll \min\!\left\{ T_\text{HB} , 10\,\text{MeV} \,(c_J (\kappa_J^{ee})^2)^{-0.23} \right\}\, .
\end{align}
Similarly to the supernova case, the upper limit on $\Lir$ follows from requiring that the IR scale be much smaller than the temperature of the star,  $T_\text{HB} \approx 8.6 \text{ keV}$, and that DS excitations escape the radius of the star, $r_\odot \sim 10^5\,$km. The limit due to this second requirement has been derived for a strongly-coupled DS by assuming that the LDSP decays through the electron current portal. It turns out to be always satisfied, for not too large values of $c_J (\kappa_J^{ee})^2$, as long as $\Lir \ll T_\text{HB}$.

\subsection{Positronium lifetime}

The $e^+\,e^-$ bound system, positronium, comes in two spin states: orthopositronium, o-Ps ($S=1$) and parapositronium p-Ps, ($S=0$). Due to $C$ conservation in electromagnetic interactions, the leading decay of o-Ps is to three photons, and its relatively long lifetime offers a good opportunity to test the presence of portal interactions to the dark sector. In particular, o-Ps could annihilate into the dark sector, or decay to one photon plus DS excitations. We will focus on the case in which the LDSP is long lived and results in missing energy. The experimental bounds on the invisible decay of o-Ps and its decay to one photon plus missing energy are
\begin{align}
\label{eq:oPsinvEXP}
\text{Br}(\text{o-Ps} \to \text{invisible}) & \leq 4.2 \times 10^{-7} && \text{\cite{Badertscher:2006fm}}\\[0.2cm]  
\label{eq:oPs1gaEXP}
\text{Br}(\text{o-Ps} \to \gamma + \!\not\!\! E) & \leq 1.1 \times 10^{-6} && \text{\cite{Asai:1991rd}}
\end{align}
at $90\%$ confidence level. The sensitivity of these constraints as probes of elusive dark sectors can be easily quantified by considering the SM rate of ortopositronium decays into neutrinos. As a matter of fact, neutrinos are a perfect prototype of dark sector coupled, at low energy, through $D=6$ portals (i.e. the four-fermion operators generated by the exchange of weak bosons). The SM predicts
\begin{align}
\text{Br}(\text{o-Ps} \to \nu\bar\nu) & = 6.2 \times 10^{-18}  && \text{\cite{Czarnecki:1999mt}} \\[0.2cm]  
\text{Br}(\text{o-Ps} \to \gamma + \nu\bar\nu) & = 1.7 \times 10^{-21} && \text{\cite{Pokraka:2016jgy}}\, .
\end{align}
These branching fractions are much smaller than the experimental limits and this suggests that the current experimental precision is not sufficient to probe elusive dark sectors that couple through $D\geq 6$ portals generated at UV scales larger than the EW scale. Bounds on lower-dimensional portals can be stronger, depending on the portal and the nature of the dark sector. In the rest of this section we will compute the decay widths of the processes $\text{o-Ps} \to DS$, $\text{o-Ps} \to \gamma + DS$ and derive the corresponding bounds assuming $C$-conserving portals to electrons and photons. Such bounds will be relevant for dark sector theories with a UV scale much lower than the EW scale. We have checked that the limits on Higgs portals with $D<6$ are not significantly stronger, since the virtual exchange of the Higgs boson implies an additional suppressing factor $(m_e/m_h)^4 \sim 10^{-22}$ in the rate. Hence, although they can have lower dimensionality, Higgs portals are not efficiently constrained by positronium decays.

At leading order, the decay rate of positronium into a generic final state $X$ can be expressed by means of a factorized formula~\cite{Wheeler:1946xth,Pirenne1947} as
\begin{equation}
\Gamma(\text{o-Ps}\to X) = \frac{1}{3} |\psi(0)|^2 \,\big[4\, v_\text{rel}\, \sigma(e^+e^- \to X)\big]_{v_\text{rel} \to 0}\, ,
\label{eq:o-Psfactorizeddecay}
\end{equation}
where $\psi(0)$ is the o-Ps wave function at the origin, $v_\text{rel}$ is the relative velocity of $e^-$ and $e^+$ in their center of mass frame, and the factor $1/3$ is due to the three polarisations of orthopositronium. We will use this formula and compute the cross section for $e^+e^-$ annihilation into DS and into DS plus one photon for the benchmark portals $J_\mu^{DS} \bar e\gamma^\mu e$ ($D=6$) and $T_{\mu\nu}^{DS} F_\alpha^\mu F^{\alpha\nu}$, $T_{\mu\nu}^{DS} (\bar e \gamma^\mu D^\nu e)$ ($D=8$) respectively.

\subsubsection{o-Ps annihilation to DS}

\begin{figure}[t]
\begin{center}
	\begin{tikzpicture}[x=1.8mm,y=1.8mm,baseline]
    	\node (phi1) at (-9.0,4.8) {$e^-$};
    	\node (phi2) at (-9.0,-4.8) {$e^+$};
    	\node (phi4) at (4.8,0) {$DS$};
        \draw[fermion] (phi1) -- (-3,0) node[midway, above] {$ $} node (V1) {\Huge.};
        \draw[fermion] (-3,0) -- (1.5,1.5);
        \draw[fermion] (-3,0) -- (1.5,-1.5);
        \draw[fermion] (phi2) -- (-3,0) node[midway, below] {$ $};
        \draw [color=black, fill=gray!40 ] (1.5,0) ellipse (.15cm and .27cm);
       \node (a) at (-3,-10) {(a)};
	\end{tikzpicture}	
	\hspace{0.2cm}
	\begin{tikzpicture}[x=1.8mm,y=1.8mm,baseline]
    	\node (phi1) at (-7.5,4) {$e^-$};
    	\node (phi2) at (-7.5,-4) {$e^+$};
    	\node (phi4) at (8.3,4) {$DS$};
    	\node (phi5) at (7.5,-4) {$\gamma$};
        \draw[fermion] (phi1) -- (0,4) node[midway, above] {$ $} node (V1) {\Huge.};
        \draw[fermion] (0,4) -- (5.,5.5);
       \draw[fermion] (0,4) -- (5.,2.5);
       \draw [color=black, fill=gray!40 ] (5,4) ellipse (.15cm and .27cm);
        \draw[fermion] (phi2) -- (0,-4) node[midway, below] {$ $};
        \draw[gaugeboson] (0,-4) -- (5.5,-4) node[midway, below] {$ $};   
        \draw[fermion] (0,4) -- (0,-4) ;        
       \node (b) at (0,-10) {(b)};
	\end{tikzpicture}
	\hspace{0.2cm}	
\begin{tikzpicture}[x=1.8mm,y=1.8mm,baseline]
    	\node (phi1) at (-9.0,4.8) {$e^-$};
    	\node (phi2) at (-9.0,-4.8) {$e^+$};
    	\node (phi5) at (1.8,-4.8) {$\gamma$};
        \draw[fermion] (phi1) -- (-3,0) node[midway, above] {$ $} node (V1) {\Huge.};
    	\node (phi4) at (2.8,5.5) {$DS$};
       \draw[fermion] (-3,0) -- (-0.5,4.0);
        \draw[fermion] (-3,0) -- (1.5,2.5);
        \draw [color=black, fill=gray!40, rotate around={60:(0.5,3.25)}] (0.5,3.25) ellipse (.15cm and .23cm);
      \draw[fermion] (phi2) -- (-3,0) node[midway, below] {$ $};
        \draw[gaugeboson] (-3,0) -- (phi5) node[midway, below] {$ $};   
       \node (d) at (-3,-10) {(c)};
	\end{tikzpicture}	
	\hspace{0.2cm}
	\begin{tikzpicture}[x=1.8mm,y=1.8mm,baseline]
    	\node (phi1) at (-9.0,4.8) {$e^-$};
    	\node (phi2) at (-9.0,-4.8) {$e^+$};
    	\node (phi4) at (8.8,5.5) {$DS$};
    	\node (phi5) at (9.0,-4.8) {$\gamma$};
        \draw[fermion] (phi1) -- (-3,0) node[midway, above] {$ $} node (V1) {\Huge.};
        \draw[fermion] (3,0) -- (5.5,4.0);
        \draw[fermion] (3,0) -- (7.5,2.5);
        \draw[fermion] (phi2) -- (-3,0) node[midway, below] {$ $};
        \draw[gaugeboson] (3,0) -- (phi5) node[midway, below] {$ $};   
        \draw[gaugeboson] (-3,0) -- (3,0) node[midway, above] {$\gamma,Z$} node (V1) {\Huge.};        
        \draw [color=black, fill=gray!40, rotate around={60:(6.5,3.25)}] (6.5,3.25) ellipse (.15cm and .23cm);
       \node (c) at (0,-10) {(d)};
	\end{tikzpicture}
\end{center}
\caption{\small Feynman diagrams for $\text{o-Ps}\to DS$ (diagram (a)), mediated by the $D=6$ portal $J_\mu^{DS} \bar e\gamma^\mu e$, and for $\text{o-Ps}\to \gamma+ DS$, mediated by the $D=8$ portals $T_{\mu\nu}^{DS} (\bar e \gamma^\mu D^\nu e)$ (diagram (b) plus its crossed and diagram (c)) and $T_{\mu\nu}^{DS} F_\alpha^\mu F^{\alpha\nu}$ (diagram (d)).}
\label{fig:oPs}
\end{figure}
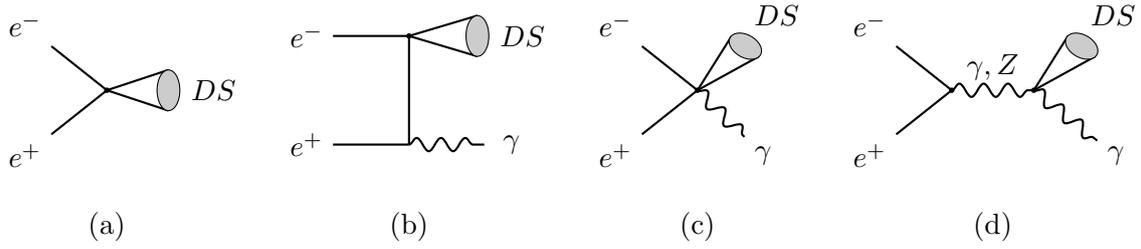

The $D=6$ portal $J_\mu^{DS} \bar e\gamma^\mu e$ can induce the annihilation of o-Ps into DS excitations through the diagram $(a)$ of Fig.~\ref{fig:oPs}.
By using the optical theorem to integrate over the DS phase space, the corresponding $e^+e^-$ annihilation cross section can be easily derived to be
\begin{equation}
\sigma(e^+e^- \to DS) = \frac{1}{(2 m_e)^2 \, v_\text{rel} } \frac{c_J\kappa_{J}^2}{2\pi}\frac{m_e^4}{\Luv^4}\, .
\end{equation}
Using Eq.~(\ref{eq:o-Psfactorizeddecay}), the leading order standard prediction for the decay rate into three photons, 
$\Gamma (\text{o-Ps} \rightarrow 3 \gamma ) = (4/3)^2 (\pi^2-9) (\alpha^3/m_e^2) |\psi(0)|^2$,
and the experimental limit (\ref{eq:oPsinvEXP}), we obtain the bound
\begin{equation}
\label{eq:oPsD6}
\Luv > 346\,\text{MeV} \left(\kappa_{J}^{2} c_J\right)^{1/4} \quad \text{for } \, \Lir \lesssim 3\,\text{MeV} \left( \kappa_J^2 c_J\right)^{-0.19}\, .
\end{equation}
The upper limit on $\Lir$ ensures that the LDSP be long lived and decay outside of the experimental apparatus, assuming a strongly-coupled DS (a very similar condition holds for weakly-coupled DS).

\subsubsection{o-Ps decay to one photon plus DS}

The $D=8$ portals $T_{\mu\nu}^{DS} F_\alpha^\mu F^{\alpha\nu}$ (where $F_{\mu\nu}$ is the photon field strength) and $T_{\mu\nu}^{DS} (\bar e \gamma^\mu D^\nu e)$ do not mediate o-Ps annihilations into the dark sector, but contribute to the decay $\text{o-Ps}\to \gamma+ DS$ via the diagrams (b),~(c),~(d) of Fig.~\ref{fig:oPs}. The corresponding $e^+ e^-$ annihilation cross section has the following form
\begin{align}
\sigma(e^+e^- \to \gamma +DS) & = \frac{1}{16 \pi^2 v_\text{rel}}\int_0^1 dx\, x\,\left<|\mathcal{M}(e^+e^- \rightarrow \gamma +DS)|^2\right>\, \\
\langle |\mathcal{M}(e^+e^- \rightarrow \gamma + DS)|^2 \rangle & =  \frac{\alpha}{15} c_T \left(\kappa_T^{ee}- \kappa_T^{\gamma\gamma} x \right)^2 \,(10-15x+6x^2) \frac{m_e^6}{\Luv^8}\, ,
\end{align}
where $x=E_\gamma/m_e$ and $\kappa_T^{ee}$, $\kappa_T^{\gamma\gamma}$ are the coefficients of the two portals. Here $\langle |\mathcal{M}(e^+e^- \rightarrow \gamma + DS)|^2 \rangle$ is the squared matrix element, summed/averaged over final/initial state polarizations and integrated over the DS phase space. Using the experimental limit (\ref{eq:oPs1gaEXP}) we obtain the bound
\begin{equation}
\begin{split}
\Luv &> 3.6\,\text{MeV} \times \left[ c_T \left( \frac{3}{2} (\kappa_T^{ee})^2-\frac{47}{30} \kappa_T^{ee} \kappa_T^{\gamma\gamma} +  \frac{1}{2} (\kappa_T^{\gamma\gamma})^2  \right)\right]^{1/8} \\[0.25cm]
\text{for } \, \Lir &\lesssim 0.4\,\text{MeV} \left( \kappa_T^2 c_T\right)^{-0.1}\, .
\end{split}
\end{equation}
Here again, the upper limit on $\Lir$ ensures that the LDSP decays outside the detector, and has been derived for a strongly-coupled DS and a $D=8$ decay portal with coefficient~$\kappa_T$. Both this bound and that of Eq.~(\ref{eq:oPsD6}) probe values of $\Luv$ well below the EW scale, but are still interesting and constrain theories where the portals $J_\mu^{DS} \bar e\gamma^\mu e$, $T_{\mu\nu}^{DS} F_\alpha^\mu F^{\alpha\nu}$ and $T_{\mu\nu}^{DS} (\bar e \gamma^\mu D^\nu e)$ are generated by very light UV mediators.

\subsection{Constraints from fifth-force experiments}
\label{sec:fifthforce}

So far we have analyzed the experimental constraints that arise from the production of DS excitations. Another way to test the dark sector is through processes involving the virtual exchange of DS degrees of freedom. As discussed in Sec.~\ref{sec:Strategy}, effects from \mbox{dimension-6} SM operators generated at the UV scale are naively expected to dominate over those induced by the exchange of DS states. However, there exist important exceptions of observables that are insensitive to UV contact terms and are thus a genuine probe of the dark sector.

Consider, for example, the force between two SM fermions (e.g. nucleons or leptons) measured at some finite distance.  The tree-level exchange of DS states induces a potential that can be tested in a variety of precision experiments operating at different scales, such as torsion balance experiments, Casimir force experiments, neutron scattering and bouncing, atomic and molecular spectroscopy, and nuclear magnetic resonance experiments (see for example Refs.~\cite{Fichet:2017bng,Brax:2017xho,Costantino:2019ixl,Banks:2020gpu}).
The potential from the DS exchange can be computed, in the non-relativistic limit,  from the Fourier transform of the scattering amplitude over the transferred three-momentum. It is thus written as an integral over the two-point DS correlator, which has a non-analytic (in momentum) part encoding the contribution from the dark-sector infrared dynamics, plus polynomial terms due to the UV dynamics whose coefficients are incalculable within the effective field theory. Upon integration, these two contributions map respectively into a long-range potential of the form  $1/r^{2\Delta -1}$ (at distances $r\ll 1/\Lir$), where $\Delta$ is the dimension of the DS operator, and a contact potential given by a delta function $\delta^{3}(\vec r)$ and its derivatives.
Experiments operating at a finite distance, such as torsion balance and Casimir force experiments, are insensitive to the contact term and thus probe exclusively the contribution from the dark sector states.
Molecular spectroscopy experiments also fall in the same class, since they are sensitive to the potential in the finite range of distances where the molecular wave function $\psi$ is non-vanishing. In practice, a potential $V(r)$ generated by the exchange of DS states induces a shift in the energy levels of the molecular system equal to
\begin{equation}
\label{eq:energylevelshift}
\Delta E = \int \! d^3 r \; \psi^*(r) V(r) \psi(r)\, .
\end{equation}
If the wave function vanishes sufficiently fast at the origin, the integral converges and the contribution from contact terms vanishes. For systems of this kind the energy shift is calculable and gives a genuine probe of the DS dynamics. 

Torsion balance experiments and molecular spectroscopy set the most stringent bounds on $1/r^5$ potentials, while molecular spectroscopy is the most effective in the case of $1/r^7$ potentials. Such bounds, however, cannot be used to directly constrain the portals of Eq.~(\ref{eq:portal}), as we now explain.

Let us consider, for example, the $D=6$ portal $J^{DS}_\mu (\kappa_J^{ee} \bar e \gamma^\mu e + \kappa_J^{pp} \bar p \gamma^\mu p + \kappa_J^{nn} \bar n \gamma^\mu n)$ featuring a current of electrons, protons and neutrons. It generates a potential
\begin{equation}
V_{ik}(r) = \frac{c_J \kappa_J^{ii}\kappa_J^{kk}}{32\pi^3} \frac{1}{\Luv^4} \frac{1}{r^5} + \text{contact terms}
\end{equation}
between any two (distinguishable) fermions $i$ and $k$. The corresponding energy level shift induced in a molecule is calculable as long as the molecular wave function vanishes at the origin faster than $r$.~\footnote{The ground state of the hydrogen atom is an example where the integral in Eq.~(\ref{eq:energylevelshift}) diverges, since the wave function is constant at the origin.} This behavior characterizes several molecular systems whose transitional frequencies can be measured accurately with ultra stable lasers. For example, recasting the bounds on large extra dimensions set in Ref.~\cite{Salumbides:2015qwa} by measurements of the energy levels in molecular hydrogen (H$_2$), we obtain
\begin{equation}
\label{eq:moleculaspectroscopy}
\Luv \gtrsim 0.2\,\text{MeV} \,\left( c_J (\kappa_J^{pp})^2 \right)^{1/4}\quad \text{for } \Lir \ll 1\,\text{keV} \, .
\end{equation}
The condition on $\Lir$ stems from the fact that molecular spectroscopy tests distances of order $1 \text{\AA} \sim 1/(1\,\text{keV})$. Other molecular systems also lead to constraints on $\Luv$ in the MeV range~\cite{Brax:2017xho,Banks:2020gpu}. 
Torsion balance experiments operating on distances of order $0.01 - 1\,$mm give slightly stronger bounds, which assume however much smaller IR scales $\Lir \ll 10^{-3}\,$eV. Constraints on $\Luv$ from long-range potentials induced by $D=5$ ${\cal O} H^\dagger H$ portals ($V\sim 1/r^5$) and $T_{DS}^{\mu\nu} O_{\mu\nu}^{DS}$ or $D=6$ ${\cal O} H^\dagger H$ portals ($V\sim 1/r^7$) are much weaker.

The bound on $\Luv$ set by Eq.~(\ref{eq:moleculaspectroscopy})
is below the mass of the nucleon. The effective theory obtained by integrating out the UV dynamics at $\Luv$ is therefore a non-relativistic one, and its expansion must be performed in terms of the nucleon velocity or kinetic energy rather than in powers of 4-dimensional derivatives. The set of effective operators characterizing such non-relativistic effective theory is not in one-to-one correspondence to those, like the portals of Eq.~(\ref{eq:portal}), one would write at higher energies. We thus conclude that, although molecular spectroscopy and fifth-force experiments in general are interesting probes of dark sectors, the corresponding limits belong to a different category compared to those discussed in the previous sections, as they apply to operators (portals) of a different effective field theory.

\subsection{EW precision tests}
\label{sec:EWPT}

Another example of observables where the virtual exchange of DS states can be calculable is electroweak precision tests (EWPT). Calculability in this case requires the dimensionality of the portal to be $D\leq 5$, as already discussed in Sec.~\ref{sec:indirect}. Let us consider, for example, the effects of a Higgs portal on vector boson self energies, in particular we will focus on the corrections to the $\varepsilon_3$ parameter introduced by Altarelli and Barbieri~\cite{Altarelli:1990zd,Altarelli:1991fk}. 

\begin{figure}[t]
\centering
\includegraphics[width=0.95\textwidth]{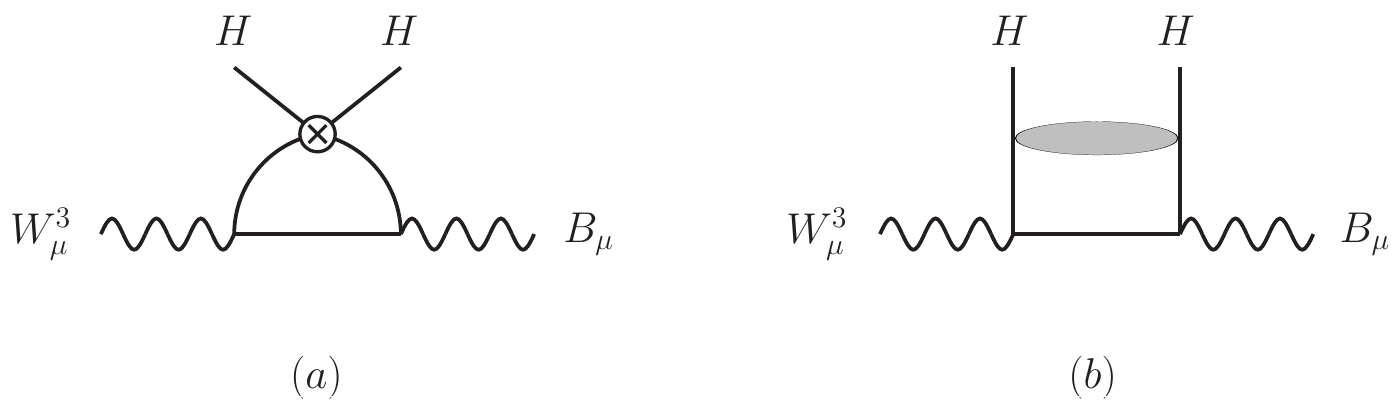}
\caption{\small Diagrams contributing to $\Delta\varepsilon_3$: short-distance contribution from the insertion of $O_H$ (diagram $(a)$); long-distance contribution from the DS exchange (diagram $(b)$). Continuous internal lines correspond to Higgs propagators, the insertion of $O_H$ is denoted by a crossed vertex, and the gray blob represents the DS exchange.}
\label{fig:EWPT}
\end{figure}

A $D=5$ Higgs portal renormalizes the operator $O_H = [\partial_\mu (H^\dagger H)]^2$ via a tree-level diagram with two insertions (see Fig.~\ref{fig:indirect}), implying a coefficient
\begin{equation}
\label{eq:cH}
c_H(\mu) \sim \frac{\kappaO^2 \cO}{16\pi^2} \frac{1}{\Luv^2} \log\frac{\Luv}{\mu}\, .
\end{equation}
The 1-loop diagram of Fig.~\ref{fig:EWPT}a with one $O_H$ insertion, in turn, renormalizes the operators $O_{W} = g D^\mu W^a_{\mu\nu} H^\dagger T^a i \!\overleftrightarrow{D}^{\!\nu} H$ and $O_{B} = g' \partial^\mu B_{\mu\nu} H^\dagger i \!\overleftrightarrow{D}^{\!\nu} H$, which give a short-distance contribution to $\varepsilon_3$. 
We thus estimate
\begin{equation}
\label{de3Deq5}
\Delta \varepsilon_3 = \hat S \sim \frac{m_W^2}{\Luv^2} \frac{\kappaO^2 \cO}{(16\pi^2)^2} \log\frac{\Luv}{\bar \Lambda} \log\frac{\Luv}{m_Z}\, ,
\end{equation}
where $\bar \Lambda = \max(\Lir , m_h)$. Notice that, although it is a short-distance effect due to the UV dynamics, the contributions of Eq.~(\ref{de3Deq5}) is calculable within the effective field theory, since it stems from the RG running of dim-6 operators. Finite contributions are subleading for $D=5$ and have been neglected.

For $4 < D < 5$, the DS exchange leads to a finite correction to $\varepsilon_3$ through the diagram of Fig.~\ref{fig:EWPT}b. If $\Lir > m_Z$, one can integrate out the DS dynamics at $\Lir$ and match to an effective theory with SM fields and higher-dimensional operators. In particular, thresholds at $\Lir$ generate $O_H$ with a coefficient
\begin{equation}
c_H(\Lir) \sim  \frac{\kappaO^2 \cO}{16\pi^2} \frac{1}{\Lir^2} \left(\frac{\Lir}{\Luv}\right)^{2(D-4)} \, .
\end{equation}
The insertion of $O_H$ into the diagram of Fig.~\ref{fig:EWPT}a then gives
\begin{equation}
\label{eq:IRthresholdeps3}
\Delta\epsilon_3 = \hat S \sim \frac{m_W^2}{\Lir^2} \frac{\kappaO^2 \cO}{(16\pi^2)^2} \left(\frac{\Lir}{\Luv}\right)^{2(D-4)} \log\frac{\Lir}{m_Z}\, .
\end{equation}
If instead $\Lir < m_Z$, then the diagram of Fig.~\ref{fig:EWPT}b gives a genuine long-distance correction of order
\begin{equation}
\label{eq:longdistanceeps3}
\Delta\epsilon_3 \sim \frac{m_W^2}{m_h^2} \frac{\kappaO^2 \cO}{(16\pi^2)^2} \left(\frac{m_h}{\Luv}\right)^{2(D-4)} \, .
\end{equation}

For the same value of $\kappaO^2 \cO$, the long-distance effect of Eq.~(\ref{eq:longdistanceeps3}) gives a less suppressed correction compared to those of Eqs.~(\ref{de3Deq5}) and (\ref{eq:IRthresholdeps3}), although it does not have a log enhancement. By requiring $\Delta\varepsilon_3 \lesssim 10^{-3}$, Eq.~(\ref{eq:longdistanceeps3}) implies $\Luv \gtrsim m_h \times (0.02\, \kappaO^2 \cO)^{1/(2D-8)}$, which is a rather weak bound. For example, if one sets $\kappaO$ to its largest value allowed by the naturalness bound of Eq.~(\ref{eq:EWstabilitybound}), it turns into an upper limit $\Lir \lesssim m_h \times (10^3/\cO)^{1/(12-2D)}$, which is easily satisfied (given the initial assumption $\Lir < m_Z$) for not too large $\cO$. We thus conclude that EW precision tests do not set stringent constraints on the DS dynamics.

\section{Summary and Discussion}
\label{sec:Conclusion}

The existence of neutral dark sectors with a low mass scale and irrelevant portal interactions to the visible fields is an intriguing possibility and only apparently an exotic one. Several theoretical extensions of the Standard Model, some of which address one or more of its open issues, predict scenarios of this kind. Neutrinos are an interesting historical precedent. Their existence was proposed by Pauli in 1930 as a solution to the longstanding puzzle of the $\beta$-decay spectrum, but their direct detection came only in 1958 as the culmination of the pioneering experimental efforts of Reines and Cowan. 
The reason why it was so difficult to detect them is because at low energy neutrinos interact feebly with charged particles through $D=6$ portals generated at the weak scale (specifically, a portal of the form $(\bar O_\text{vis}\nu+\text{h.c.})$ mediates $\beta$-decay, whereas $\mu$-decay and neutral-current scatterings proceed through $J^\mu_\text{vis} J_\mu^{(\nu)}$ portals).
Eventually, the properties of neutrinos were uncovered thanks to the possibility of obtaining intense beams from nuclear reactors, as this obviated the huge suppression of signal rates. It was only in 1983 however, more than 50 years after Pauli's original intuition, that the UV mediators responsible for the neutrino portal interactions, the $W$ and $Z$ vector bosons, were produced on shell in the UA1 and UA2 experiments at CERN, and the barrier between dark and visible sector removed forever.

\begin{figure}[tp]
\centering
\includegraphics[width=0.48\textwidth]{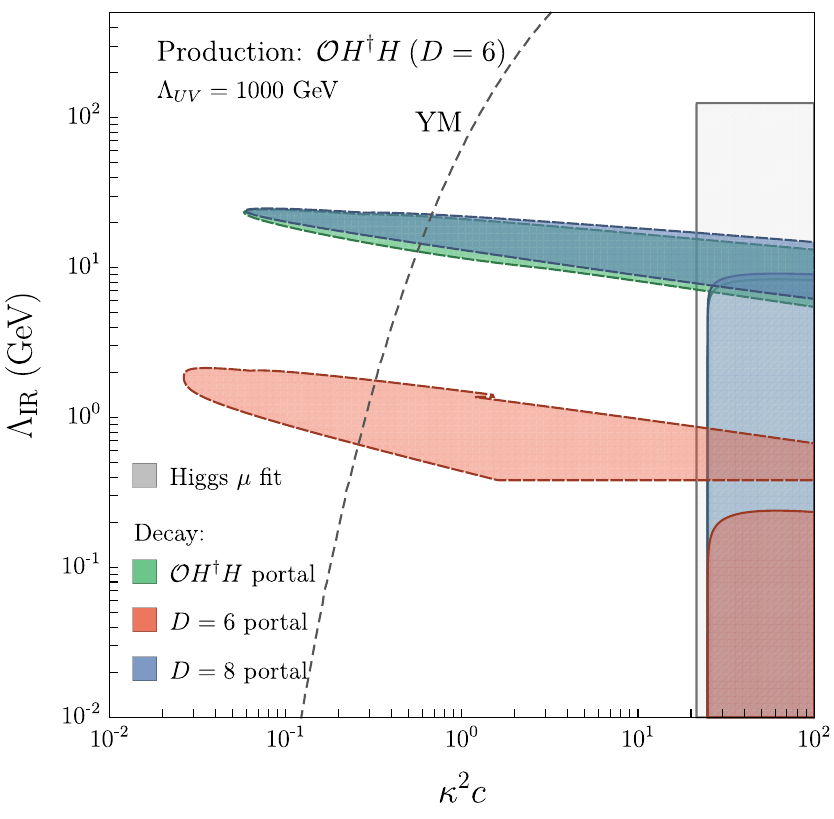}
\includegraphics[width=0.48\textwidth]{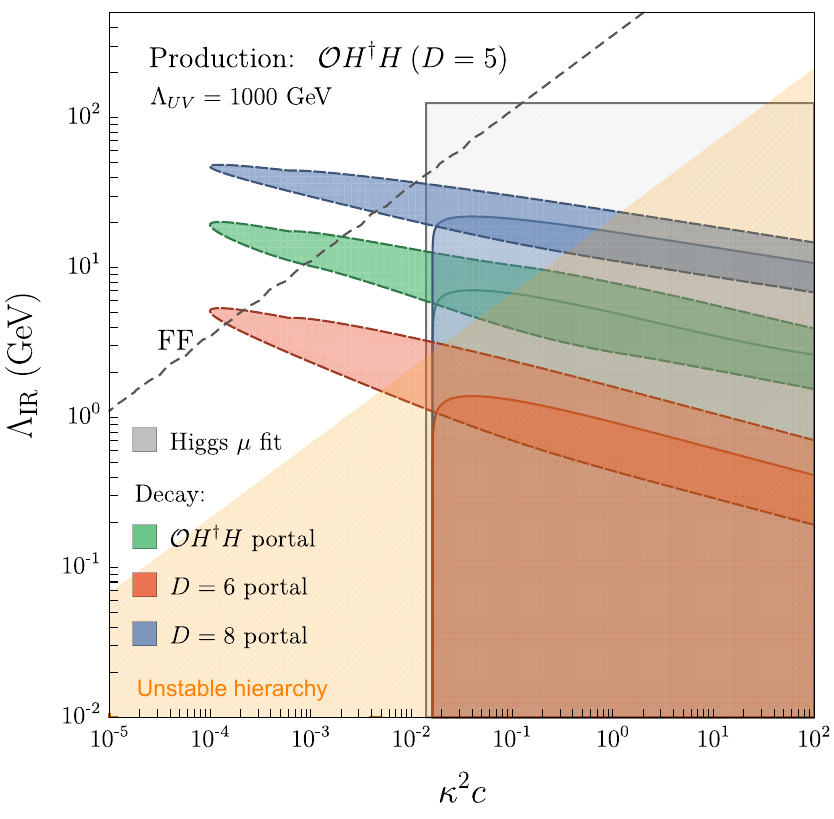}
\includegraphics[width=0.48\textwidth]{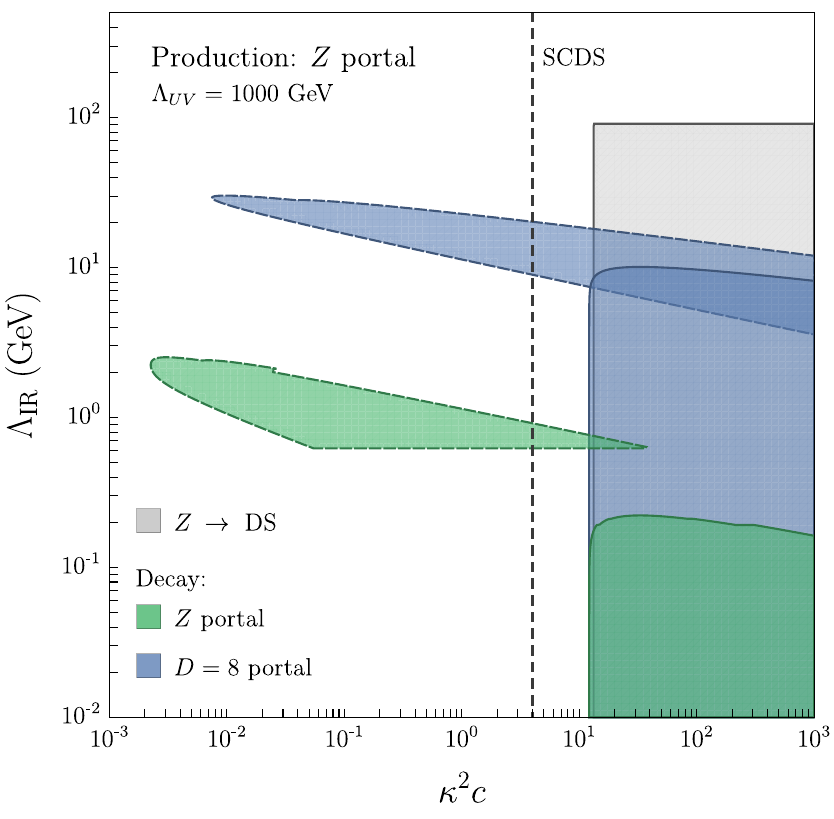}
\includegraphics[width=0.48\textwidth]{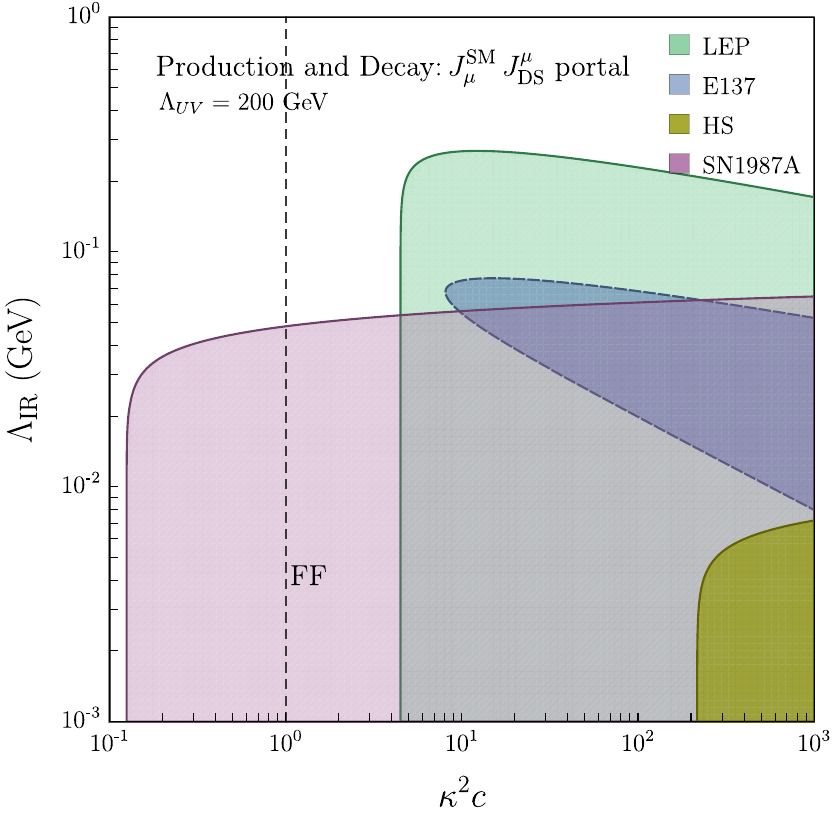}
\caption{\small Exclusions at 95\% probability in the plane $(\kappa^2 c, \Lir)$ for fixed $\Luv$ and various portals. Continuous contours in the upper two panels show the exclusions from the fit to Higgs couplings and the bound on the Higgs invisible branching ratio, while those in the lower left panel arise from the invisible $Z$ decay width and mono-jet searches at the LHC. Dashed contours in these same panels show exclusions from displaced decays at the LHC. The lower right panel shows exclusions from LEP mono-photon searches, E137, SN1987A and stellar evolution (note that the SM current is different for each of these, so they probe different portals. We are taking the simplifying limit of $\kappa$ being the same for all of them). Bounds from other experiments analyzed in the text are too weak to appear in the plots. The dashed curves show the predictions of the benchmark models of Sec.~\ref{sec:Examples} for the following values of the parameters: $y_L=1$, $y_R=0$, $N_{DC}=3$ for the pure Yang-Mills model of Sec.~\ref{sec:YM} (YM) and the strongly-coupled DS model of Sec.~\ref{sec:SCDS}) (SCDS); $\lambda_{HS}=1$ and $y=1$ for the free fermion models of Sec.~\ref{sec:freefermion} (FF).}
\label{fig:summaryplots}
\end{figure}

The current theoretical and experimental landscapes are very different from those of the early decades of the past century, and since then the energy and intensity frontiers have been immensely pushed forward. In light of this, one may ask how a hypothetical elusive dark sector might manifest itself and be discovered at present or future facilities. We have tried to address this question by estimating the relative importance of various effects in Section~\ref{sec:Strategy}. The virtual exchange of UV mediators can be parametrized in terms of $D=6$ effective operators and gives corrections to processes with SM external states that scale as $1/\Luv^2$. The DS contribution to the same processes necessarily involves two insertions of the portals and scales as $1/\Luv^{2(D-4)}$, where $D$ is the dimensionality of the portal. Naively, it is subdominant compared to the UV effect except for $D<5$ or when the experimental observable is sensible only to long-distance contributions and blind to contact ones. Electroweak tests and fifth-force experiments are interesting examples of this kind, and were analyzed respectively in Sections~\ref{sec:EWPT} and~\ref{sec:fifthforce}. Given the current experimental precision, we find that they are not sensitive enough to test portals generated at energies above the EW scale. Direct production of DS states implies signal rates that also scale as $1/\Luv^{2(D-4)}$ but its significance can be competitive with UV virtual effects even for $D>5$. We have analyzed an ample spectrum of processes that are summarized in Table~\ref{tab:processes}. They include searches at high-energy colliders, high-intensity experiments, astrophysical observations (supernova cooling and stellar evolution) and low-energy precision experiments (positronium rare decays). 
We find that the strongest sensitivity on elusive dark sectors is currently obtained at high-energy colliders. The plots in Fig.~\ref{fig:summaryplots} give a summary of our results.
The most stringent constraints can be set on Higgs and $Z$ portals when the DS excitations are produced through the decay of the Higgs or $Z$ bosons, in particular when the lightest DS particles decay back to the SM with displaced vertices. In those cases, UV scales as high as several TeVs are already being probed for $\kappa^2 c$ of order 1 (see Figs.~\ref{fig:resonantHiggs} and~\ref{fig:resonantZ}), where $\kappa$ is the portal coefficient and $c$ measures the multiplicity of DS states. 
As a matter of fact, comparable if not stronger lower bounds on $\Luv$ are set, through their sensitivity to virtual UV effects, by the body of electroweak precision tests performed at LEP, SLC and Tevatron, and by the analysis of Higgs processes at the LHC. Searches for on-shell production of the UV mediators made at colliders, or even DM direct detection experiments (in theories where the DM candidate resides in the UV sector), can also set stringent, though model dependent, limits on $\Luv$. This comparison suggests that, different from the historical neutrino precedent, the first signals of new physics might come this time from the heavy UV dynamics rather than from the light and elusive dark states. For example, in a likely scenario one could first observe deviations in SM precision tests induced by the virtual exchange of UV mediators, and only later on reach the experimental sensitivity to uncover the dark sector. Hence, light and weakly-coupled new physics should not be seen as an alternative to new heavy particles: on the contrary, observing the latter could prelude the discovery of the former.

The above considerations suggest that a future physics programme at a Higgs or $Z$ factory would extend most effectively our sensitivity on Higgs and $Z$ portals thanks to the large statistics of decays. An FCC-ee running at the TeraZ option would be especially beneficial as it would increase the sensitivity on $\Luv$ on two complementary fronts: an order-of-magnitude increase in the precision on electroweak observables~\cite{Baak:2014ora,Fan:2014vta,deBlas:2016ojx} to uncover UV virtual effects, and a sample of $Z$ decays larger by two orders of magnitude compared to the LHC to produce the DS particles. In the longer run, an \mbox{FCC-hh} at $100\,$TeV would produce $\sim 10^{10}$ Higgs bosons, roughly four orders of magnitude larger than the current production at the LHC. This would allow one to probe invisible Higgs decays at the level of $\sim 10^{-4}$~\cite{L.Borgonovi:2642471} and extend considerably the sensitivity on exotic decays. Without looking too much ahead in the future, the approved high-luminosity phase of the LHC will already lead to a substantial increase, by a factor $\sim 30$, of the number of produced Higgs and $Z$ bosons. This corresponds naively to an increase of the lower bound on $\Luv$ by a factor~$\sim 2$ for a $D=6$ portal. In fact, even at future Higgs and $Z$ factories the sheer increase of statistics will imply lower bounds on $\Luv$ larger by at most factors of a few, given that rates scale as $1/\Luv^{2(D-4)}$. For example, a naive rescaling of our results suggests that a GigaZ factory could reach a lower bound on $\Luv$ of order $10\,$TeV in the case of a $Z$ portal with $\kappa_J^2 c_J\sim 1$. Similar conclusions were reached by previous studies, see for example Refs.~\cite{Curtin:2015fna,Cheung:2019qdr,Cheng:2019yai}. Probing higher UV scales will require, for example, to improve our ability to trigger on and reconstruct displaced vertices. 

While portal interactions generated at very large scales will remain elusive, future facilities will be able to extend considerably our reach on low IR scales. 
It is a feature of dark sectors with irrelevant portals that the strength of their interaction with the SM scales with the energy as $\sim \kappa \, (E/\Luv)^{(D-4)} \equiv \alpha_{DS}(E)$. Production rates in the conformal regime are controlled by $\alpha_{DS}(\sqrt{s})$, where $\sqrt{s}$ is the energy characterizing the process, whereas the decay length of the lightest DS particles is determined by $\alpha_{DS}(\Lir)$ and thus crucially depends on the ratio $\Lir/\Luv$.
This has to be contrasted with the case of marginal portals, as in dark photon theories, where the same small parameter (the kinetic mixing) controls both quantities. Future experiments aimed at detecting long-lived particles, e.g. CODEX-b, FASER and MATHUSLA~(see~\cite{Alimena:2019zri} and references therein), will be able to improve the reach on small $\Lir$ by detecting the decays of the lightest DS particles far away from the interaction point. This is especially important since, as illustrated by the plots of Fig.~\ref{fig:summaryplots}, current searches for displaced vertices at the LHC are already sensitive enough to test benchmark models for $\Luv \sim 1\,$TeV, though only in a relatively narrow range of IR scales.

While searches for displaced vertices at high-energy colliders are able to provide the strongest constraints on Higgs and $Z$ portals, it is also important to consider different portals and discovery strategies. Fixed-target and beam-dump experiments making use of very intense beams have been found to be extremely powerful to uncover dark sectors with marginal portals. In particular, simplified dark photon models have been often taken as benchmarks in previous experimental and theoretical studies. We have shown that, at least in the conformal regime, a dark sector coupled through $J^{DS}_\mu J_{SM}^\mu$, where $J_{SM}^\mu$ is an electron or quark current, behaves like a convolution of dark photon theories with a spectrum of masses that depends on the experiment (e.g. on the incoming beam energy and composition of the target). In the case of the NA64 and E137 experiments, such mass spectrum peaks at $\sim 1\,$GeV, see Fig.~\ref{fig:diffxsec-NA64-E137}. In particular, diagrams with DS emission can be obtained from those with an external dark photon field $A^\mu_D$ by replacing $(\varepsilon e) A_D^\mu\to (\kappa_J/\Luv^2) J^\mu_{DS}$. This observation led to Eq.~(\ref{eq:DP-to-DS}) and suggests that simple quantitative estimates for the dark sector can be derived by using the known dark photon results in terms of an effective kinetic mixing parameter $\varepsilon_\text{eff} = (p_{DS}^2/\Luv^2) (\kappa_J^2 c_J)^{1/2}/(4\pi e)$, as a function of the DS invariant mass squared $p_{DS}^2$. Similar considerations were made previously in Ref.~\cite{Cheng:2019yai}. A quick glance to any plot showing the constraints on dark photon theories in the $(\varepsilon, m_{A_D})$ plane, like those in Fig.~6 of Ref.~\cite{Essig:2013lka} and Fig.~20 of Ref.~\cite{Beacham:2019nyx} confirms the hierarchy of effects found by our analysis, namely that the strongest limits come from supernova cooling and beam-dump experiments with extremely intense beams like E137. It also suggests that future experiments, in particular SHiP~\cite{Alekhin:2015byh}, can extend the reach to UV scales of order a few TeV~\cite{Cheng:2019yai}. 
Additional improvement may come if future experimental analyses will be performed so as to optimize their sensitivity to generic dark sectors and not only to benchmark dark photon models. This is especially true for searches, like those performed by \textsc{BaBar} and Belle II, where events are selected by assuming the resonant production of a dark photon.

While the comparison with dark photon theories can be useful for a quick recast of current searches, an experimental programme aimed at the discovery of elusive dark sectors seems justified and would require optimized strategies and analyses. For example, existing high-intensity experiments like those designed for neutrino physics where the detector is placed very far downstream of the target are not particularly effective to detect long-lived particles originating from marginal portals, since very long decay lengths also imply very small production rates. This is not the case for irrelevant portals since, as already mentioned, the decay length of the lightest DS particles can be large as a consequence of a small IR scale.  Besides tailored experimental searches, more in-depth theoretical studies will also be needed to uncover new discovery strategies and thoroughly explore the theoretical landscape of possibilities. The aim of our work was that of making a first step in this direction. We attempted to study elusive dark sectors in a broad perspective and analyzed current experimental results to get insight on how to design a future experimental strategy. We obtained bounds from a large array of experiments by means of a procedure where the validity of the effective field theory used to define the portals is consistently enforced at each step. Our limits are sometimes less stringent than previous ones for this reason. Clearly, much additional work is needed to get a more complete quantitative picture on elusive dark sectors. Information will come not only from laboratory experiments and astrophysical observations, but also from the analysis of the cosmological evolution of these theories.

\section*{Acknowledgements}
We would like to thank D.~Curtin, R.~T.~D'Agnolo, T.~Hahn, Z.~Liu, T.~Okui, G.~Perez, A.~Podo, M.~Redi and F.~Revello for interesting discussions and stimulating questions.
This research was partly supported by the Italian MIUR under contracts 2015P5SBHT 007 (PRIN2015) and 2017FMJFMW (PRIN2017), and by the National Science Foundation under Grant No. NSF PHY-1748958 and NSF PHY-1915071. RKM acknowledges the hospitality of the Kavli Institute for Theoretical Physics, UC Santa Barbara, during the workshops ``Origin of the Vacuum Energy and Electroweak Scales'' and ``From Inflation to the Hot Big Bang'' during which part of this work was completed. 
\appendix

\section{Two-point Dark Sector Correlators}
\label{app:2pt}

We report here the expression of the 2-point correlators of dark sector operators used in our analysis. 

For very large momenta, $p^2\gg \Lir^2$, the form of the 2-point correlators is dictated by conformal invariance, up to an overall normalization constant. We define the latter as follows (in 4D Minkowski space-time):
\begin{align}
\label{eq:2ptOx}
\left<\mathcal{O}(x)\mathcal{O}(0)\right> &= \frac{\cO}{8\pi^4}\frac{1}{(x^{2})^{\Delta_\mathcal{O}}} \\[0.1cm]
\label{eq:2ptJx}
\left<J^{DS}_\mu(x)\,J^{DS}_\nu(0)\right> &= \frac{c_J}{8\pi^4}\frac{1}{(x^2)^3}\left(\eta_{\mu\nu} - 2\frac{x_\mu x_\nu}{x^2}\right)\\[0.1cm]
\label{eq:2ptTx}
\left<T^{DS}_{\mu\nu}(x)\,T^{DS}_{\rho\sigma}(0)\right> &= \frac{c_T}{8\pi^4}\frac{1}{(x^2)^4}\,
\left[ \left(I_{\mu\nu}(x)I_{\rho\sigma}(x) -\frac{1}{4} \eta_{\mu\nu}\eta_{\rho\sigma}\right) + \mu \leftrightarrow \nu \right]\, ,
\end{align}
where $I_{\mu\nu}(x) = \eta_{\mu\nu} - 2x_\mu x_\nu/x^2$.  After Fourier transforming and subtracting the singular terms analytic in momenta, one obtains:
\begin{align}
\label{eq:2pO}
\left<\mathcal{O}(p)\mathcal{O}(-p)\right>
&= \frac{-i\cO}{2\pi^2} \frac{\Gamma(2-\DeltaO)}{4^{\DeltaO-1}\Gamma(\DeltaO)}(-p^2)^{\DeltaO-2} \\[0.1cm]
\label{eq:2ptJ}
\left<J_\mu^{DS}(p)\, J_\nu^{DS}(-p)\right>
&=\frac{-ic_J}{\pi^2} \frac{1}{2^4 3!} \,  p^2 \log (-p^2)  \, P_{\mu\nu} \\[0.1cm]
\label{eq:2ptT}
\left< T^{DS}_{\mu\nu}(p) T^{DS}_{\rho\sigma}(-p) \right>
&=\frac{-ic_T}{2 \pi^2} \frac{1}{2^5 5!} \, p^4 \log(-p^2) \, P_{\mu\nu\rho\sigma}\:, 
\end{align}
for any $p^2$ in the complex plane away from the branch cut on the positive real axis, where the projectors $P_{\mu\nu\rho\sigma}$ and $P_{\mu\nu}$ are defined as
\begin{equation}
\label{P4projector}
P_{\mu\nu\rho\sigma} = 2 P_{\mu\nu} P_{\rho\sigma} - 3 \left( P_{\mu\rho} P_{\nu\sigma} + P_{\mu\sigma} P_{\nu\rho}\right)\, , 
\qquad P_{\mu\nu} = \eta_{\mu\nu} -  \frac{p_\mu p_\nu}{p^2}\, .
\end{equation}
The corresponding imaginary parts, extracted from the discontinuity
across the branch cut, are:
\begin{align}
\label{eq:imO}
\text{Im}\!\left[ i\left<\mathcal{O}(p)\,\mathcal{O}(-p)\right>\right]
&= \frac{\cO}{\pi^{3/2}}\,\frac{\Gamma(\DeltaO + 1/2)}{\Gamma(\DeltaO-1)\Gamma(2\DeltaO)}\,(p^2)^{\DeltaO-2} \\[0.1cm]
\label{eq:imJ}
\text{Im}\!\left[ i\left<J_\mu^{DS}(p)\, J_\nu^{DS}(-p)\right>\right]
&=-\frac{c_J}{\pi} \frac{1}{2^4 3!} \, p^{2}\, P_{\mu\nu} \\[0.1cm]
\text{Im}\!\left[ i \left< T^{DS}_{\mu\nu}(p)T^{DS}_{\rho\sigma}(-p) \right> \right]
&= -\frac{c_T}{2\pi} \frac{1}{2^5 5!} \, p^4  \, P_{\mu\nu\rho\sigma}\, .
\end{align}

The normalization in Eqs.~(\ref{eq:2ptOx}),(\ref{eq:2ptJx}),(\ref{eq:2ptTx}) has been chosen so as to reproduce the following expressions in the case of free canonically-normalized fields (see for example~\cite{Osborn:1993cr, Dolan:2000ut}):
\begin{align}
\label{eq:cO}
&\phantom{T^{DS}_{\mu\nu}} \begin{alignedat}{5}
\mathcal{O} &= \frac12 \left(\partial_\mu \phi\right)^2 & \quad & (\DeltaO = 4),  & \qquad &\cO = 24 \\
\mathcal{O} &= \bar{\psi}\,\gamma^\mu i\overset{\,\leftrightarrow}{\partial_\mu}\psi & \quad & (\DeltaO = 4), & \qquad & \cO = 0 \\
\mathcal{O} &= -\frac14 F_{\mu\nu}^2 & \quad & (\DeltaO = 4), & \qquad & \cO = 24  \\
\mathcal{O} &= \bar{\psi}\psi &\quad & (\DeltaO = 3), &\qquad & \cO = 8
\end{alignedat} \\[0.5cm]
\label{eq:cJ}
&\phantom{\mathcal{O}} \begin{alignedat}{3}
J^{DS}_\mu &= \phi^\dagger i \overset{\,\leftrightarrow}{\partial_\mu}\phi, &\qquad &c_J = 2  \\
J^{DS}_\mu &= \bar{\psi}\gamma^\mu \psi \ \text{or}\ \bar{\psi}\gamma^\mu \gamma^5 \psi & \qquad &c_J = 8 
\end{alignedat} \\[0.5cm]
&\phantom{\mathcal{O}}\begin{alignedat}{3}
T^{DS}_{\mu\nu} &= \partial_\mu\phi\,\partial_\nu\phi -\frac{1}{12}\left(2\partial_\mu\partial_\nu + \eta_{\mu\nu}\,\partial^2\right)\phi^2,\qquad & c_T &= \frac43 \\
T^{DS}_{\mu\nu} &= i\left[\frac12\bar{\psi}\gamma_\mu\partial_\nu \psi - \frac14\partial _\mu(\bar{\psi}\gamma_\nu\psi) + (\mu \rightarrow \nu)\right]
- i\,\eta_{\mu\nu}\,\bar{\psi}\slashed{\partial}\,\psi,\qquad & c_T &= 8 \\
T^{DS}_{\mu\nu} &= F_{\mu\alpha}F_\nu^\alpha - \frac14\eta_{\mu\nu}F_{\alpha\beta}^2, \qquad & c_T &= 16 \, . 
\end{alignedat}
\end{align}
Here $\phi$, $\psi$ and $F_{\mu\nu}$ denote respectively a real scalar, a Dirac fermion and an abelian gauge field strength. In the case of operators made of Majorana fermions, the values of $c_i$ can be obtained by dividing those for Dirac fermions by 2;  values of $c_i$ for operators with non-abelian field strengths are obtained multiplying those of the abelian case by the number of real components of the gauge field.

We end this appendix  by reporting the expression of the 2-point correlators predicted in the benchmark models with a free fermion DS and in the RS model of Sec.~\ref{sec:Examples}. In these models a calculation of the 2-point correlator is possible for values of the momenta down to threshold, i.e. outside of the conformal regime.

In the $B-L$ model of Sec.~\ref{sec:freefermion} the DS consists of three Majorana fermions $\psi_{N_i}$ coupled through the portal (\ref{eq:JJportalforFF}). Using dimensional regularization with minimal subtraction and a 4-component notation, we find
\begin{equation}
\label{eq:JJexact}
\begin{split}
\langle J^{DS}_{\mu}(p) & J^{DS}_{\nu}(-p) \rangle = -\frac{1}{24 \pi^2 } \left( \eta_{\mu\nu}p^2 - p_\mu p_\nu \right) \\
& \times \sum_{i=1}^3\Bigg\{  \left(1+\frac{2 m_{N_i}^2}{p^2}\right) \sqrt{1-\frac{4 m_{N_i}^2}{p^2}} 
\, \log \frac{2 m_{N_i}^2-p^2 + \sqrt{p^4-4 m_{N_i}^2 p^2}}{2 m_{N_i}^2} \\
& \phantom{\times \sum_i\Bigg\{} +\frac{4 m_{N_i}^2}{p^2}+\left(\frac{5}{3}-\gamma_E\right)+\log \frac{\mu ^2}{m_{N_i}^2} 
+ \log 4\pi\Bigg\}\, ,
\end{split}
\end{equation}
where $J_{DS}^{\mu} = (1/2) \sum_i \bar\psi_{N_i}^\dagger \gamma^\mu \gamma^5\psi_{N_i}$ and $\mu$ is the subtraction scale. In the limit $p^2 \gg m_{N_i}^2$ this expression tends to the CFT correlator of Eq.~(\ref{eq:2ptJ}) with $c_J = (1/2) \times 3\times 8$, where the factor $1/2$ appears because the $\psi_{N_i}$ are Majorana fermions (cf. Eq.~(\ref{eq:cJ})). 

In the second model of Sec.~\ref{sec:freefermion} the DS consists of a single Majorana fermion $\chi$ coupled through the portal~(\ref{eq:JJportalforFF2}). The two-point correlator of the current $J_{DS}^{\mu} =\bar\chi \gamma^\mu \gamma^5\chi$ can be obtained by simply keeping the contribution of a single fermion species in Eq.~(\ref{eq:JJexact}) and replacing $m_{N_i}$ with $m_\chi$. Hence, the conformal limit of Eq.~(\ref{eq:2ptJ}) in this case is recovered with $c_J = (1/2) \times 8$.

In the third model of Sec.~\ref{sec:freefermion} the DS consists of one Dirac fermion $\psi$, coupled to the SM through the portal of Eq.~(\ref{eq:JJportalforFF3}). We find
\begin{equation}
\begin{split}
\langle \bar\psi\psi(p) \bar\psi\psi(-p)\rangle =  -\frac{p^2}{8 \pi ^2}\Bigg\{
&\sqrt{1-\frac{4 m_\psi^2}{p^2}} \, \log \frac{2 m_\psi^2-p^2+ \sqrt{p^4-4 p^2 m_\psi^2 }}{2 m_\psi^2} \\
& + \log \frac{\mu ^2}{m_\psi^2} + \log 4\pi   -\gamma_E +2 \Bigg\} + \dots \, ,
\end{split}
\end{equation}
where the dots stand for terms independent of $p^2$. In the limit $p^2 \gg m_\psi^2$ this expression tends to the CFT correlator of Eq.~(\ref{eq:2pO})  with $\DeltaO = 3$ and $\cO =8$ (cf. Eq.~(\ref{eq:cO})).

Finally, let us consider the RS model of Sec.~\ref{sec:5DRS}. In that case the DS consists of the dynamics in the bulk and on the IR brane, coupled to the elementary SM sector through the portal (\ref{eq:dim8RSportal}). Despite the DS being strongly coupled in the infrared (and up to the $\Luv$ scale), the 2-point correlator of $T_{\mu\nu}^{DS}$ can be computed thanks to holography. Indeed, it can be extracted from the UV brane-to-brane graviton propagator by sending the UV brane to the AdS boundary; in Minkowski space-time one finds~\cite{Rattazzi:2000hs}:
\begin{equation}
\label{eq:TTRScorrelator}
\begin{gathered}
\langle T^{DS}_{\mu\nu}(p) T^{DS}_{\rho\sigma}(-p) \rangle  = \frac{(M_5/k)^3}{12} p^4 F(p^2) P_{\mu\nu\rho\sigma}\, , \\[0.4cm]
F(p^2) \equiv \log\frac{p^2}{4k^2} - \pi \frac{Y_1(\sqrt{p^2}/\Lir)}{J_1(\sqrt{p^2}/\Lir)}\, ,
\end{gathered}
\end{equation}
where $\Lir \equiv k e^{-\pi R k}$ and the  transverse and traceless projector $P_{\mu\nu\rho\sigma}$ is defined in Eq.~(\ref{P4projector}). In absence of an explicit breaking of conformal symmetry, the 2-point correlator has a massless pole corresponding to the dilaton (i.e.~the radion of the 5D theory): $F(p^2) \simeq -4\Lir^2/p^2$ for $\sqrt{p^2}\ll \Lir$. The radion acquires a mass through the mechanism that stabilizes the extra dimension.
In using the expression of $\langle T_{\mu\nu}^{DS}T_{\rho\sigma}^{DS}\rangle$ in Sec.~\ref{sec:openproduction}, we have captured this effect by modifying the IR behavior of the form factor as follows:
\begin{equation}
\label{eq:modifiedFF}
F(p^2) \to \hat F(p^2) = F(p^2) +\frac{4\Lir^2}{p^2} -\frac{4\Lir^2}{(p^2 - m_\phi^2)}
\end{equation}
where $m_\phi$ is the dilaton (radion) mass (taken to be $\Lir$ in Sec.~\ref{sec:Strategy}). Notice that upon breaking explicitly the conformal invariance, the 2-point correlator acquires an additional Lorentz structure that is not traceless (see for example Ref.~\cite{Rattazzi:2000hs}). We neglect this effect for simplicity. For $\sqrt{p^2} \gg \Lir$, the expression of $\langle T_{\mu\nu}^{DS}T_{\rho\sigma}^{DS}\rangle$ tends to the pure CFT result provided the limit is taken in the correct way, see the discussion in the next Appendix.

\section{Further Discussion on the 5D Randall-Sundrum Dark Sector}
\label{app:RSmodel}

In this Appendix we analyze a few additional aspects of the 5D Randall-Sundrum dark sector theory that are worth discussing.
Let us first set our notation and derive some useful formulas. 
We take the bulk metric to be 
\begin{equation}
ds^2= e^{-2ky} g_{\mu\nu}(x,y) dx^\mu dx^\nu +dy^2\, ,
\end{equation}
and locate the UV and IR branes respectively at $y=0$ and $y=\pi R$. 
The value of the 4D Planck mass can be computed by taking the low-energy limit of the 5D action, including the localized kinetic term of Eq.~(\ref{eq:UVbraneaction}). One has:~\footnote{We define $M_5$ and $\MPl$ as in Ref.~\cite{Rattazzi:2000hs}.}
\begin{equation}
\MPl^2 = \frac{M_5^3}{k} \left( 1 - e^{-2\pi Rk} \right)+ M_0^2\, .
\end{equation}
This equation can be used to express the value of $M_0$ in terms of the other parameters. 
Performing a Kaluza-Klein (KK) decomposition of the graviton field and neglecting at first order the effect of the second term of Eq.~(\ref{eq:UVbraneaction}), the wave function of the $n$-th KK mode has the standard expression 
\begin{equation}
\label{eq:KK_grav_profile}
f_n(y) = N_n \, e^{2k y}\left[J_2\big(x_n e^{k y}\big) + b_n \, Y_2\big(x_n e^{k y}\big) \right] \, ,
\end{equation}
where $x_n = m_n/k$, $m_n$ is the KK mass, $N_n$ is a normalization factor and 
\begin{equation}
b_n=-\frac{J_1(x_n) - r_0  \,x_n J_2(x_n)}{Y_1(x_n) - r_0  \,x_n Y_2(x_n)}\, , \qquad\quad r_0 = \frac{kM_0^2}{M_5^3}\, .
\end{equation}

We are thus ready to make our considerations about this model.
First of all, we would like to justify our claim that the UV-localized interaction of Eq.~(\ref{eq:UVbraneaction}) corresponds, in the 4-dimensional holographic theory, to the dim-8 portal~(\ref{eq:dim8RSportal}). We do so by considering the interaction between the SM energy-momentum tensor and the $n$-th KK mode; from the 5D Lagrangian, after the KK decomposition, one has
\begin{equation}
\frac{1}{M_5^{3/2}}  \left( f_n(0) + \frac{1}{\Luv^2}  \tilde f_n(0) \right)  h_{\mu\nu}^{(n)}(x) \,T^{\mu\nu}_{SM}(x) \, ,
\end{equation}
where $\tilde f_n(0) \equiv (-2k \, \partial_y+\partial_y^2) \, f_n(y)\big|_{y=0}$. The first term in parenthesis originates from the minimal coupling between gravity and matter, while the second term is due to the non-minimal interaction of Eq.~(\ref{eq:UVbraneaction}).
By using the solution~(\ref{eq:KK_grav_profile}) and expanding for $1\ll r_0 x_n^2 \sim \MPl^2 \Lir^2/\Luv^4$, we find:
\begin{equation}
\label{eq:KKTSM}
\frac{\pi}{2\sqrt{2}} \bar x_n |Y_1(\bar x_n)| \Lir \left( \frac{M_5^{3/2}}{k^{3/2}} \frac{1}{\MPl^2} - \bar x_n^2 \frac{k^{3/2}}{M_5^{3/2}} \frac{1}{k^2 \Luv^2}\right) h_{\mu\nu}^{(n)}(x) \,T^{\mu\nu}_{SM}(x)\, , 
\end{equation}
where, we recall, $\Lir \equiv k e^{-\pi Rk}$, and we have defined $\bar x_n \equiv x_n e^{\pi R k}$, so that $\bar x_n \sim O(1)$. Notice that $(k/M_5)^{3/2}$ has $\hbar$ dimension of a coupling. The form of Eq.~(\ref{eq:KKTSM}) matches the behavior expected from the 4D holographic theory where
\begin{equation}
\label{eq:holoL}
{\cal L}_{holo} \supset \frac{1}{\MPl} h_{\mu\nu} T^{\mu\nu}_{SM} + \frac{1}{\MPl} h_{\mu\nu} T^{\mu\nu}_{DS} + \frac{\kappa_T}{\Luv^4}  T^{\mu\nu}_{SM} T_{\mu\nu}^{DS}\, .
\end{equation}
Indeed, as a consequence of the second term above, the elementary graviton mixes with the tower of composites spin-2 states once conformal invariance is broken in the infrared. This implies that $h_{\mu\nu}$ in the first term in Eq.~(\ref{eq:holoL}) will have some component of the spin-2 massive eigenstate. This leads to the Planck-suppressed contribution in Eq.~(\ref{eq:KKTSM}) (first term in parenthesis). The non-minimal interaction of the holographic theory, on the other hand, is expected to give a contribution that is not suppressed by the Planck scale. That is exactly the second term in parenthesis in Eq.~(\ref{eq:KKTSM}), from which we infer $\kappa_T \sim (k/M_5)^{3}$. The exact expression of $\kappa_T$ can be extracted from Eq.~(\ref{eq:KKTSM}) if one knows the matrix element between the spin-2 bound states and the energy-momentum tensor in the holographic theory. Such matrix element, in turn, can be derived from the residues of the poles in the $\langle T_{\mu\nu}^{DS} T_{\rho\sigma}^{DS}\rangle$ correlator.

The other aspect that we would like to discuss about the RS dark sector model concerns the high-energy limit of Eq.~(\ref{eq:TTRScorrelator}) (similar considerations appeared previously in the literature, see also the related arguments of Refs.~\cite{Fichet:2019hkg,Costantino:2020msc}). We expect that for $p^2\gg \Lir^2$ the expression of $\langle T_{\mu\nu}^{DS} T_{\rho\sigma}^{DS}\rangle$ tends to the result valid for a CFT dynamics, see Eq.~(\ref{eq:2ptT}). However, Eq.~(\ref{eq:TTRScorrelator}) has been obtained from a tree-level calculation in the 5D theory, which, on the 4D holographic side, corresponds to the leading order in $1/N_{CFT}$, where $N_{CFT}$ is the number of colors of the CFT dynamics. Correspondingly, the form factor $F(p^2)$ has an infinite series of poles on the real axis, interpreted as due to the exchange of non-interacting, stable bound states in 4D. The corresponding spectral function, computed by taking the imaginary part of $F(p^2)$, is an infinite sum of delta functions. It is thus clear that when taking the limit $p^2\gg \Lir^2$, $F(p^2)$ does not lead to the logarithm predicted by a CFT.
The solution to this apparent paradox comes by noticing that after including 1-loop corrections in the UV brane-to-brane calculation, the poles of $F(p^2)$ acquire an imaginary part and move above the real axis.~\footnote{The 1-loop corrections also introduce a multiparticle branch cut on the real axis. This effect has been neglected in deriving Eq.~(\ref{eq:TTRScorrelator}).} This fact has a simple interpretation in the 4D holographic theory: the finite width of the resonances comes in only at next order in $1/N_{CFT}$. Including such subleading effect is crucial to recover the correct conformal limit. Doing so, indeed, corresponds to first evaluate the form factor at $p^2 \to p^2(1+i \epsilon)$, where $\epsilon = \Gamma_n/m_n$ and $\Gamma_n$ is the resonance's width. Taking the limit $p^2\gg \Lir^2$ then gives the correct result for Eq.~(\ref{eq:TTRScorrelator}), since $\lim_{x\to \infty} Y_1(x(1+i \zeta)/J_1(x(1+i \zeta)) = i$, for real and finite $\zeta$. The reason why a finite width of the resonances is crucial to recover the CFT result is also clear from the `quark-hadron' duality viewpoint: the `quark' behavior is obtained only by resumming over the contribution of an infinite number of `hadrons'. Increasing $\epsilon$ implies that the tails of a larger number of resonances will enter a given interval in $p^2$. Conversely, for fixed and finite $\epsilon$, the number of resonances effectively contributing into a $p^2$ interval of given length increases as $p^2\to \infty$.

Comparing the high-energy limit of Eq.~(\ref{eq:TTRScorrelator}) with the CFT result of Eq.~(\ref{eq:2ptT}) we find the value of $c_T$ in the RS model:
\begin{equation}
\label{eq:cTholo}
c_T = 640 \pi^2 \frac{M_5^3}{k^3} = 40 (N_{CFT}^2-1)\, .
\end{equation}
The last equality follows by using the standard holographic dictionary where $16\pi^2 (M_5/k)^3 = N_{CFT}^2-1$.

\section{Probabilities for Displaced Decays}
\label{sec:probabilities}

In this Appendix we describe how we modelled the probability for a signal event to pass the selections made by ATLAS in the searches for displaced jets of Refs.~\cite{Aaboud:2018aqj, Aad:2019xav}. The simplest search of Ref.~\cite{Aaboud:2018aqj} selects events with at least two displaced hadronic vertices in the MS, while Ref.~\cite{Aad:2019xav} considers events with one decay in the MS and one in the ID. Let 
\begin{equation}
P_{ij} = \exp\left( - \frac{L_i}{c\tau_\psi \gamma} \right) - \exp\left( - \frac{L_j}{c\tau_\psi \gamma} \right)
\end{equation}
be the probability for a single LDSP with boost $\gamma$ to decay within distances $L_i$ and $L_j$ from the primary vertex (we assume for simplicity that the LDSP is produced promptly after the hard collision). Then, for a signal event with $n$ LDSPs, the probability to have at least two decays within distances $L_1$ and $L_2$ is:
\begin{equation}
\label{eq:2inL1L2}
P_{\geq 2 \text{ in } [L_1,L_2]} = 1 - \left(1-P_{12}\right)^n - n\, P_{12} \left( 1-P_{12}\right)^{n-1}\, .
\end{equation}
The probability to have at least one decay in $[L_1,L_2]$ and at least one in $[L_3,L_4]$ is instead:
\begin{equation}
\label{eq:1inL1L1and1inL3L4}
P_{\stackrel{\scriptstyle\geq 1 \text{ in } [L_1,L_2]}{\geq 1 \text{ in } [L_3,L_4]}} = 1 - \left\{
\left(1-P_{12}\right)^n + \left(1-P_{34}\right)^n - \left( 1-P_{12}-P_{34} \right)^{n} \right\}\, .
\end{equation}
We assess the signal yield by setting $n$ to equal the average values $\langle n\rangle =2$ and that in Eq.~(\ref{eq:naverage}) to characterize the behavior of respectively weakly-coupled and strongly-coupled dark dynamics. The boost factor $\gamma$ is set to its average value of Eq.~(\ref{eq:gammaaverage}). We then recast the results of Ref.~\cite{Aaboud:2018aqj} by assigning each event a weight given by Eq.~(\ref{eq:2inL1L2}) with $L_1 =4\,$m and $L_2 =13\,$m, where these distances correspond to the region where the efficiency of the Muon RoI Cluster trigger of ATLAS is largest (see Fig.~2 of Ref~\cite{Aaboud:2018aqj}). Similarly, we recast the results of Ref.~\cite{Aad:2019xav} by assigning each event a weight given by Eq.~(\ref{eq:1inL1L1and1inL3L4}), with $L_1$, $L_2$ as above and $L_3 =4\,$mm and $L_4 =300\,$mm. The values chosen for $L_3$, $L_4$ correspond to the region where the efficiency to select the hadronic vertex in the ID is largest (see Tab.~4 and Fig.~3 of Ref.~\cite{Aad:2019xav}).

\bibliographystyle{utphys}
\bibliography{references}
 
\end{document}